\documentclass[twoside,12pt]{article}
\usepackage{epsfig}

\def\Journal#1#2#3#4{{#1} {#2} (#4) #3 }

\def\NPA{{\em Nucl. Phys.} A}

\def\PLB{{\em Phys. Lett.} B}

\def\PRL{\em Phys. Rev. Lett.}

\def\PREP{\em Phys. Rep.}

\def\PRC{{\em Phys. Rev.} C}

\newcommand{\be}{\begin{equation}}
\newcommand{\ee}{\end{equation}}
\newcommand{\bea}{\begin{eqnarray}}
\newcommand{\eea}{\end{eqnarray}}

\begin{document}

\title{ \vspace{1cm} Collision dynamics at medium and relativistic energies}
\author{M.\ Colonna$^1$\\
\\
$^1$INFN, Laboratori Nazionali del Sud, I-95123 Catania, Italy}

\maketitle
\begin{abstract} 
Recent results connected to nuclear collision dynamics, 
from low up to relativistic energies, are reviewed.
Heavy ion reactions offer the unique opportunity to 
probe the complex nuclear many-body dynamics and to explore, in laboratory 
experiments, transient states of nuclear matter under several conditions
of density, temperature and charge asymmetry.
From the theoretical point of view, 
transport models are an essential tool to undertake these investigations
and make a connection betwen the nuclear effective interaction
and sensitive observables of experimental interest.
In this article, we mainly focus on the description of results 
of transport models for a selection of reaction mechanisms, 
also considering comparisons of predictions of different approaches. 
This analysis can help understanding the impact of the interplay
between mean-field and correlation effects, as well as of in-medium
effects, on reaction observables, which is an essential point also for 
extracting information on the nuclear Equation of State. 
A special emphasis will be given to the review of recent studies aimed
at constraining the density 
behavior of the nuclear symmetry energy. 
For reactions at medium (Fermi) energies, we will describe light particle and
fragment emission mechanisms, together with isospin transport effects. 
Collective effects characterizing nuclear collision dynamics, such as 
transverse and elliptic flows, will be discussed for relativistic
heavy ion reactions, together with meson production and isotopic ratios.
 
%This discussion will be framed also within the context of the present
%theoretical and experimental efforts aimed at constraining the density 
%behavior of the symmetry energy.   
%Heavy ion collisions offer the possibility to explore, in laboratory 
%experiments, transient states of nuclear matter under several conditions
%of density, temperature and charge asymmetry. Transport models are an essential tool to describe the reaction
%dynamics and make a connection betwen the nuclear effective interaction
%employed in the calculations and sensitive observables of experimental interest.
%We review here recent analyses which allow to constrain, in particular,
%the controversial density behavior of the symmetry energy of the nuclear EOS, 
%also discussing future perspectives.
\end{abstract}
%\eject
\tableofcontents
\section{Introduction}
%The nuclear matter Equation of State (EOS) is a key ingredient 
%for the understanding of 
%many phenomena involving nuclear systems and astrophysical compact objects. 
%The behavior of nuclear matter under several conditions of density
%and temperature is of crucial importance for the understanding of a large variety of phenomena, ranging from the structure of nuclei and their decay modes, up to the life and the properties of massive stars. 
%A large variety of phenomena, ranging from the structure of nuclei and their decay modes, up to the life and the properties of massive stars are 
%connected to the behavior of nuclear matter under several conditions of density
%and temperature. Indeed,
%the concept of the nuclear Equation of State (EOS) links 
%mechanisms involving an enormous range of scales in size, characteristic time and energy, but all based on nuclear processes at fundamental level.
%, are actually linked by the concept of the nuclear Equation of State (EOS).
%In terrestrial laboratories, new insights into the features of the nuclear EOS can be gained by mean 
%This information can be accessed by mean 
%of heavy ion collision experiments, where transient 
%states of nuclear matter (NM) spanning a wide spectrum of regimes can be created. 
%Transient states of nuclear matter far from normal conditions can be created in terrestrial laboratories via Heavy Ion Collisions (HIC).  
Many experimental and theoretical efforts have been devoted, over the past 40 years or so, to the study of nuclear reactions from low to relativistic energies, %as a possible tool to 
to probe new aspects of collision dynamics and 
scrutinize relevant properties of the nuclear medium.
Indeed, heavy ion collision experiments, where transient 
states of nuclear matter (NM), spanning a wide spectrum of regimes, are 
created, provide crucial insights into the features of the nuclear Equation of State (EOS).  The latter is a rather important object, which influences 
 a large variety of phenomena, ranging from the structure of nuclei and their 
decay modes, up to the life and the properties of massive stars. 
In particular, the understanding of the properties of exotic nuclei, as well as neutron stars and supernova dynamics, entails the knowledge of the 
behavior of nuclear 
symmetry energy.

Measurements of experimental observables linked to
isoscalar collective vibrations in nuclei, 
collective flows and meson production in nuclear reactions, have contributed to constrain the EOS of charge symmetric matter 
for densities up to five time the saturation value \cite{Dan2002}.
%More recently, the availability of neutron-rich and exotic beams has opened the way to investigate, in laboratory conditions, new aspects of nuclear structure and dynamics up to extreme ratios of neutron to proton numbers. Thus it has become possible to explore the behavior of nuclear matter along a new degree of freedom, the asymmetry I= (N−Z)/(N+Z), aiming at constraining the density and/or temperature dependence of the symmetry energy (Iso−EoS).
More recently, the availability of exotic beams has made it 
possible to explore, in laboratory conditions, new aspects of nuclear structure and dynamics up 
to extreme ratios of neutron (N) to proton (Z) numbers,  thus giving a
strong boost to the investigation of the EOS of
asymmetric matter. %, which  has %, comparatively to the symmetric EOS, 
%few experimental constraints. % yet.  
%Indeed, the isovector part of the nuclear effective interaction and the corresponding
%symmetry energy of the EOS (Asy-EOS) are largely unknown as soon as we move away from normal density. 

%Nevertheless, this information is essential in the astrophysical context, for the understanding of the 
%properties of compact objects such as neutron stars, whose crust behaves as low-density 
%asymmetric nuclear matter \cite{Lattimer,bur1} and whose core may touch extreme
%values of density and asymmetry. 
%Moreover, the low-density behavior of the symmetry energy also affects
%the structure of exotic nuclei and 
%%features such as the appearance of 
%%new collective modes involving the neutron skin. 
%the appearance of new features involving the neutron skin,
%whish are currently under intense investigation (see \cite{Burrello2019}
%and refs. therein). 
%which describes the difference between the binding energy of symmetric matter (with equal proton and neutron numbers, N=Z), and that of pure neutron matter.  
%We recall that the knowledge of the EoS of asymmetric matter is very important at low densities, 
The low-density regime has an impact on reaction dynamics at Fermi energy (isotopic features in fragmentation \cite{Baran2002}, 
charge equilibration \cite{tsang2004,betty2009}), as well as 
on nuclear structure (neutron skins, pygmy resonances \cite{roca2018}) 
and in the astrophysical context (neutron star formation and crust
\cite{Lattimer,Burrello2015,Horo}). 
On the other hand,  
%as well as at high densities in 
relativistic heavy ion reactions (isospin flows, 
meson production \cite{Xiao2009,Russotto2016}) and compact star features (neutron star mass-radius relation, cooling, hybrid structure, formation of black holes \cite{Lattimer,Baldo2016}) 
are strongly influenced by the high density behavior of the symmetry energy. 

In this article, we will discuss heavy ion reactions 
in the beam energy domain mentioned above, 
ranging from few tens of MeV/nucleon (Fermi energy domain) up
to several hundred MeV/nucleon (relativistic energy regime). 
The overall reaction dynamics appears characterized by an initial compression phase, 
where, depending on beam energy and reaction centrality, densities up to
two-three times the saturation density can be reached.  
A significant degree of thermalization, together with ``pre-equilibrium'' light particle emission, is expected, especially in central reactions. 
During the following expansion phase (on characteristic time scales of
the order of 10-100 fm/c), 
several clusters and nuclear fragments are produced from a hot
source whose excitation energy is typically comparable to the binding energy (per
nucleon) of a nucleus \cite{Hudan2003,Reisdorf1997,Reisdorf2010}. 
One also observes the development of collective flows,
such as radial flow of the initially compressed matter and/or transverse flow of the spectator matter.

From the theoretical point of view, 
understanding the features and the reaction mechanisms involving 
complex systems, such as nuclei,  in terms of
their constituent particles and the interaction among them is a true challenge. The
original quantal many-body problem, is often approached adopting the mean-field
approximation (or suitable extensions), yielding a so-called effective interaction  \cite{Neg,Dec,Rein,Bend}.
%learn about the behavior of nuclear matter and its EoS.  
%Relevant conclusions have been reached concerning the EoS of symmetric matter for densities up to five time the saturation value [22]. 
Hence, one can say that the collision dynamics is governed to a large extent by the details
%Actually this study allows one to learn about the corresponding behavior 
of the nuclear effective interaction, which provides the nuclear EOS  in the equilibrium limit.  
%As anticipated above, 
%from the theoretical point of view, 
%the study of heavy ion collisions is a rather intricate 
%quantum many-body problem. 
However, many-body correlations, beyond the mean-field
picture, are important and certainly influence the dissassembly of the
composite nuclear system and its re-aggregation in new configurations
along the reaction path.  In general, it is more appropriate to state that the reaction dynamics is
ruled by the delicate interplay between mean-field effects and many-body correlations.    Hence the investigation of heavy ion reactions 
presents a twofold interest:
to unveal new aspects of the complex nuclear dynamics and, at the same time, 
to probe relevant features of the nuclear effective interaction and
EOS. It is clear that the two goals are intertwined: the extraction of 
robust constraints on the EOS relies on suitable descriptions of the
reaction dynamics.  

Several extensions of mean-field models have to be introduced to take explicitly
into account the effects of relevant interparticle correlations. Focusing on nuclear
dynamics, an intense theoretical work on correlations and density fluctuations has
started in the past years, fostered by the availability of large amounts of
experimental data on fragment formation and light cluster production 
in intermediate energy heavy ion collisions \cite{Frankland,Moretto,Dagostino,EPJA_Tab,chomaz2004,Ono1999}, also in connection with 
the possibility to observe volume (spinodal) instabilities 
%related to the occurrence of 
and liquid-gas phase transitions in nuclei. 
%thus assimilating the
%nuclear disassembling to the occurrence of liquid-gas phase transitions.8–13
In passing, we note that
low-density clustering is of interest also in other contexts where
nuclear matter at subsaturation densities can appear,  
%Nuclear clusters are predicted to appear also in equilibrated
%nuclear matter below saturation density, 
%i.e. in 
such as the conditions encountered in the crust of neutron stars 
and/or along supernova explosion processes \cite{Lattimer}.
%The low-density nuclear matter is also important for astrophysical problems such
%as the core-collapse supernovae. 
Sophisticated thermodynamical approaches have been formulated to evaluate
the EOS of clustered matter, 
which can be employed in astrophysical applications \cite{Typel2014}. 
For instance, the appearance of clusters is expected to influence neutrino
transport, thus modifying the cooling mechanism by neutrino emission of protoneutron stars \cite{Reddy2000,Margueron2004,Horowitz2015}.  
%The equation of state (EOS) for astrophysical applications has been calculated recently with
%elaborated thermodynamical 
%approaches allowing the existence of clusters at low densities. 
%The composition of clusters may be important for
%some phenomena in compact stars such as the neutrino transport in core-collapse supernovae.

%The dynamics of nuclear collisions at intermediate
%energy is often investigated within the framework of 
A commonly employed scheme to deal with the dynamics of nuclear collisions
at intermediate energy is represented by 
semi-classical transport theories, such as the Nordheim approach, in which the Vlasov equation for the
one-body phase space density, $f({\bf r}, {\bf p},t)$, 
is extended by the introduction of a Pauli-blocked
Boltzmann collision term \cite{bertsch1988,Aldo_rep},
which accounts for the average effect of the two-body residual interaction. 
%(quantal correlations are not included).  
%%%%%%%%%%%%%%%%%%%%%%   add references
%The basic ingredients that enter 
Thus the resulting transport 
equation, often called  Boltzmann-Uehling-Uhlenbeck (BUU) 
%or Boltzmann-Nordheim-Vlasov (BNV)
equation, contains two basic ingredients:  
the self-consistent mean-field potential and the two-body scattering cross sections.
%These transport models hence describe the time evolution of the reduced
%one-body density in phase-space and, 
%consequently, they are suited for
%the description of one-body observables, such as inclusive particle spectra
%in nuclear collisions, average collective flows and excitations.
%However, they cannot provide a reliable description of fluctuation phenomena,
%such as clustering aspects and multi-fragmentation processes, i.e. the break-up of excited nuclear
%systems into many pieces. In fact, 
%neither fluctuations of one-body observables nor many-body correlations can be 
%addressed with this class of mean-field models. 
%%Hence suitable extensions, including fluctuations of the one-body density, have to
%%be considered.
In order to introduce fluctuations and further (many-body) correlations in transport theories, 
a number of different avenues have been taken, that can
be essentially reconducted to two different 
classes of models
(see Ref.\cite{Xu2019} for a recent review). 
%{\bf *** NB:no a capo}
One is the class of molecular dynamics (MD) models \cite{Ono1999,Aich1991,
feldmeier1990,FFMD,ono1992prl,ono1992,Papa2001,Zhang2006} while the other
kind is represented by stochastic mean-field approaches 
\cite{Ayik,Ayik1,Randrup,chomaz2004}.
In the latter approaches, 
fluctuations of the one-particle
density,  which should account for the effect of the neglected many-body
correlations, are introduced by adding to the transport equation
a stochastic term, representing the 
fluctuating part of the collision integral \cite{Ayik,Ayik1,Randrup}, in close analogy with the Langevin equation for a Brownian motion.
This extension leads to the Boltzmann-Langevin (BL) equation, which can be 
derived as the next-order correction, in the equation describing
the time evolution of $f$, with respect to the standard 
average collision integral. %leading to the Boltzmann-Langevin (BL) equation. 
Within such a description, though
the system is still described solely in terms of the reduced one-body density
$f$, this function may experience a stochastic time evolution in response
to the random effect of the fluctuating collision term.
%In this way density fluctuations, resulting from many-body correlations, 
% are introduced, that are amplified when 
The effects of the fluctuations introduced are particularly important when
instabilities or bifurcations occur in the dynamics.  
Indeed this procedure is suitable for addressing multifragmentation phenomena where clusters 
%emerge from the growth, driven by the unstable mean-field, of density inhomogeneities. 
emerge from the growth of density inhomogeneities, driven by the unstable mean-field. %PN2[reformulated]     
%since fragments can be associated with the regions where the spacial density
%becomes larger, which finally can be reconstructed by sampling the 
%one-body distribution function.
However, the fluctuations introduced in this way are not strong enough to
fully account for the production of 
light clusters, which are loosely bound by the mean-field and would require
stronger nucleon correlations. %of the nucleon wave packet (as in MD approaches).
%In the mean-field class of descriptions, the dynamical state of
%the nuclear system is characterised by the reduced one-body density
%in phase space, $f({\bf r},{\bf p},t)$, the classical analogue of the
%Wigner transform of the one-particle density matrix.
%At low energies, the time evolution of the one-body density is governed by the
%Vlasov equation, which can be regarded as the semi-classical approximation 
%to the time-dependent Hartree-Fock theory. The residual direct collisions
%between the constituent nucleons are incorporated by means of a Pauli-blocked
%collision integral, leading to the so-called Boltzmann-Uehling-Uhlenbeck (BUU) 
%or Boltzmann-Nordheim-Vlasov (BNV) approaches \cite{Bert,Aldo}. 

In molecular dynamics models the many-body state is represented 
by a simple product wave function, with or without antisymmetrization.
%,similarly to what is done in transport theories. %, such as BUU. 
%However,
The single particle wave functions are usually 
assumed to have a fixed Gaussian shape.
In this way, though nucleon wave functions are supposed to be
independent (mean-field approximation), the use of localised wave packets
induces many-body correlations both in mean-field propagation 
and hard two body scattering (collision integral), which 
is treated stochastically.
Hence this way to introduce many-body correlations 
and produce a possible trajectory branching
%, leading to a variety of clustered configurations,
 is essentially based on the use of localized nucleon 
wave packets. 
%If wave functions were allowed to assume any shape, the method would
%become identical to standard mean-field descriptions.

These approaches (in particular, the AMD approach \cite{Ono1999})
%This localisation of the nucleon wave packet 
have shown to be quite successful in
describing the clustered structures characterising the ground state 
of several light nuclei \cite{kanada2012}.    %PN[Kanada2012]
%This can be surely considered as an efficient way to introduce many-body
%correlations into the dynamics, however this mean of producing trajectory branching
%lacks a rigorous derivation {\bf is it true ???}. 
%Moreover, 
Moreover, as far as nuclear dynamics is concerned, %PN[as far AS]
the wave function localisation appears quite appropriate
 to describe final fragmentation channels,  
%for fragmentation reactions,
where each single particle wave function should be localised within a
fragment.
However, one should notice that 
the use of fixed shape localised wave packets in the full dynamics 
%is rather questionable since it 
could affect the correct description of  
one-body effects, such as spinodal instabilities and zero sound 
propagation \cite{FFMD,ColonnaPRC2010}. 
%are not as precisely described as in
%mean-field models. 
%From the point of view of stochastic mean-field models, the philosophy of
%molecular dynamics would be to introduce a special kind of fluctuations
%by stochastically localising the single particle wave functions.
A simplistic description of the differences between the two classes of models
would be to say that (extended) mean-field models may lack many-body 
correlations,  whereas molecular dynamics models are mainly 
classical models.   
%Naive statements about the defects of these
%models are that the mean-field models lack many-body correlations and that the %molecular dynamics models are classical,
%though the actual situation is not so simple. 

%%%%%%%%%%%%%%%%%%%%%%%%  Mettere dopo
%Some of the mean-field and molecular dynamics models have been improved
%e.g. by considering light clusters explicitly, which can be regarded as incorporating
%quantum mechanical features mentioned above in practical manners. 

%These attempts of transport models will be reviewed
%in this article with emphasis on the description of the formation of fragments and clusters.
The approaches discussed so far are mostly based on non-relativistic formalisms, in which non-nucleonic degrees of freedom are integrated out giving nucleon-nucleon potentials. Then nuclear matter is described as a collection of quantum non-relativistic nucleons interacting through an instantaneous effective potential.
This prescription is mainly employed in the medium-energy collisional domain.

On the other hand, reactions at relativistic energies are more properly 
described via a fully covariant transport approach, related to an effective field exchange model, where the relevant degrees of freedom of the nuclear dynamics are accounted for \cite{SerotIJMPE6,TypelNPA656}. This leads to a propagation of particles (nucleons and mesons) 
suitably dressed by self-energies that will influence particle emission, collective flows and in medium nucleon-nucleon inelastic cross sections. 
%The construction of an Hadron−EoS at high baryon and isospin densities will finally allow the possibility of developing a model of a hadron-deconfinement transition at high density for an asymmetric matter[21]. 

One of the goals of this review is to try to get a deeper insight into 
the interplay between mean-field effects and
many-body correlations in light particle and fragment emission mechanisms, 
as well as on collective dynamics.
%, through the comparison of the results of selected transport models.   
%At the same time, such systems may be closely linked with the equilibrium properties of nuclear matter at finite
%temperatures below the saturation density $\rho_0$. In fact, e.g., fragment observables have often been well explained by statistical
%models in many cases. The relation of observables to nuclear liquid–gas phase transition has also been indicated based on
%experimental data (see e.g. Refs. [4,5] for reviews).
A crucial point to be understood would be the extent of the impact of many-body
correlations on the global reaction dynamics (such as compression-expansion,
development of collective flows, thermalization, charge equilibration) and 
on corresponding reaction observables which are 
expected to reflect specific ingredients of the nuclear effective interaction. 
% An important question is whether the correlations to form clusters and fragments have only a small effect on top of
%the global dynamics of collisions, such as compression, expansion and collective flows, or they have large reverse impacts on
%the global dynamics. 
Sizeable effects could be expected on the basis of the different energetics
and of the reduction of the degrees of freedom induced 
by the formation of clusters and fragments.
% We will see some examples in theoretical results reviewed here.

For reactions at relativistic energies, an important aspect to be discussed will be the
impact of (isospin-dependent) in-medium effects, such as the modification of particle self-energies, on nucleon and meson emission.

%The traditional approach to nuclear physics starts from non-relativistic formalisms in which non-nucleonic degrees of freedom are integrated out giving nucleon–nucleon potentials. Then nuclear matter is described as a collection of quantum non-relativistic nucleons interacting through an instantaneous effective potential. Although this approach has had a great success, a more appropriate set of degrees of freedom consists of strongly interacting effective hadron fields, mesons and baryons. These variables are the most efficient in a wide range of densities and temperatures and they are the degrees of freedom actually observed in experiments, in particular in heavy-ion collisions at intermediate energies. Moreover, this framework appears in any case a fundamental “doorway step” toward a more microscopic understanding of the nuclear matter. Relativistic contributions to the isospin physics for static properties and reaction dynamics will be discussed in the second part of the report, 

%After a brief description of  
%transport models, %widely employed to model heavy ion reactions,   
We will review a selection of recent results on dissipative collisions in a wide range of beam 
energies, from reactions at Fermi energies up to the $AGeV$ range.
%on the basis of transport theories of the Stochastic Mean Field (SMF) type.  
%Isospin effects
%on the chiral/deconfinement transition at high baryon density will be also
%discussed.
Several observables, which are sensitive in particular to the isovector sector
of the nuclear effective interaction and corresponding EOS terms (asyEOS) 
have been suggested \cite{baranPR,bao-an2008,EPJA_Colonna,Horo}.
Fermi energies
bring information on the EOS features (and symmetry energy term) around or 
below normal density, 
whereas intermediate energies probe higher density regions.
We mainly focus on the description of results of transport models, trying to 
compare predictions of different models and to probe the impact of the interplay
between mean-field and correlation effects, as well as of in-medium
effects, on reaction observables.
This discussion will be framed also in the context of the present
theoretical and experimental efforts aimed at constraining the density 
behavior of the symmetry energy.   

%%%%%%%%%%%%%%%%%%%%%%%%%%%%%%%%%%%%%%  da PPNP Akira
%Here, we review the field trying to pin down the most interesting theory questions and eventually the related key observables.

The article is organized as it follows: A brief description of the transport
models commonly employed in the treatment of collision dynamics at 
medium and relativistic energies is given in Section 2. 
Section 3 is devoted to the description of effective interactions 
used in transport models. A survey of results relative to the Fermi energy
domain for central and semi-peripheral collisions is given in Sections 
4 and 5, respectively.  Section 6 is devoted to results of collision dynamics
at relativistic energies.  Finally conclusions and perspectives are drawn 
in Section 7. 

%%%%%%%%%%%%%%%%%%%%%%%%%%%%%%%  da JPG

%%%%%%%%%%%%%%%%%%%%%%%%%%%%%%%%%%%%%%%%%%%  da Akira

%\section{\label{sec:basicmodels}Basic transport models}
\section{\label{theo:intro}Theoretical description of collision dynamics}

Heavy ion collisions are rather intricate processes whose understanding would
imply to solve the complex many-body problem. 
% To understand the evolution from the early phase of compression to the late phase of collective expansion and fragment formation, we need to solve the time evolution of a quantum system in some way.  We should prepare for the emergence of so many reaction channels, each of which corresponds to a configuration of decomposing the whole system into fragments.
%Even if we may know the initial many-body state $|\Psi(t=0)\rangle$ completely,
In principle, this task should be tackled by 
solving the many-body Shr\"odinger equation, 
to obtain the many-body state $|\Psi(t)\rangle$ at a time $t$.   
However, this is presently anaffordable in the general case; 
an important simplification derives from 
the fact that, for several practical purposes, it is enough to
know, with sufficient accuracy, the corresponding solution for the system 
one-body 
density, thus reducing the
problem to single-particle dynamics.

%Traditionally several types of transport models have been proposed and practically applied to heavy-ion collisions.  These calculations are actually important and useful to understand collision dynamics and to extract physical information from heavy-ion collision data such the nuclear EOS at various densities.  We will review these transport models below.  We will not try to rigorously derive or justify these models here but will put more emphasis on their achievements, advantages and disadvantages, which is useful in improving models and also in understanding important physics in heavy-ion collisions.  Although most of practical transport models are classical or semiclassical, the link with quantum mechanics will be still essential.

%Practically all the transport models try to solve the time evolution of 
The one-body density operator $\hat{\rho}$ for a system containing
$A$ particles is defined
as
\begin{equation}
\hat{\rho} = A\mathop{\mathrm{Tr}}_{2,3,\ldots,A}|\Psi\rangle\langle\Psi|
\end{equation}
%by taking the trace of the many-body density operator $|\Psi\rangle\langle\Psi|$ with respect to all the particles excluding one.  
(The state $|\Psi\rangle$ is assumed to be normalized).  
From the knowledge of the one-body density operator $\hat{\rho}(t)$, 
it becomes possible to evaluate 
the expectation value of any one-body observable, thus allowing one to
make predictions also for observables of experimental interest, in view of 
a comparison to data.

In presence of only two-body interactions  $\sum_{i<j}v_{ij}$,  
%The aim is to have a closed equation of the time evolution for $\hat{\rho}(t)$.  Generally speaking, however, one should not assume that it is possible.  If there are two-nucleon interactions $\sum_{i<j}v_{ij}$, 
one can deduce, for the time derivative of the one-body density 
$\frac{d}{dt}\hat{\rho}$, the following equation:
\begin{equation}
\label{eq:drhodt-general}
i\hbar\frac{\partial}{\partial t}\hat{\rho}=
[\frac{1}{2m}{\bf p}^2,\ \hat{\rho}]
+\mathop{\mathrm{Tr}}_2[v,\ \hat{\rho}^{(2)}],
\end{equation}
where $m$ is the particle mass and the two-body 
density operator $\hat{\rho}^{(2)}$, defined as
\begin{equation}
\hat{\rho}^{(2)} = A(A-1)\mathop{\mathrm{Tr}}_{3,\ldots,A}
|\Psi\rangle\langle\Psi|,
\end{equation}
has been introduced. 
The equation above shows that 
the knowledge of the two-body density operator is required to solve the exact
time evolution of the one-body density.    In turn, 
deriving the equation for the
time evolution of the two-body density, one would see %realize
that it contains the
three-body density operator, and so on. This actually corresponds to the
Martin-Schwinger hierarchy set of equations (see e.g.\ Ref.~\cite{wang1985}). 
In order to get a closed equation for the time evolution of $\hat{\rho}(t)$, 
it is necessary to resort to some kind of approximation, i.e. to truncate the hierarchy at a given level. 

The simplest approximation corresponds to the independent particle picture, 
according to which the two-body density operator can be written as an anti-symmetrized product of one-body densities: 
\begin{equation}
\label{eq:rho2-hf}
\hat{\rho}^{(2)}_{12}=\hat{A}_{12}\hat{\rho}_1\hat{\rho}_2,
\end{equation}
where $\hat{A}_{12}$ is the antisymmetrization operator acting on a two-body state.  
%The subscript 1 or 2 attached to a one- or two-body operator indicates on which particle it operates in a two-body state.  More explicitly, $\hat{\rho}^{(2)}(|\psi\rangle\otimes|\phi\rangle) = \hat{\rho}|\psi\rangle\otimes\hat{\rho}|\phi\rangle - \hat{\rho}|\phi\rangle\otimes\hat{\rho}|\psi\rangle$.

%and the equation for $\frac{d}{dt}\hat{\rho}^{(2)}$ contains the three-body 
%density operator, and so on (see e.g.\ Ref.~\cite{wang1985}).  The equations will not be closed until the full many-body state $|\Psi(t)\rangle$ is included.

%The time-dependent Hartree-Fock (TDHF) theory assumes a single 
%Slater determinant as the many-body state $|\Psi\rangle$.  An idempotent one-body density matrix that satisfies $\hat{\rho}^2=\hat{\rho}$ with $\mathop{\mathrm{Tr}}\hat{\rho}=A$ uniquely corresponds to a many-body state which is a Slater determinant, and therefore the two-body density operator is written with $\hat{\rho}$ as

Then Eq.~(\ref{eq:drhodt-general}) can be reformulated as:
\begin{equation}
\label{eq:drhodt-tdhf}
i\hbar\frac{\partial}{\partial t}\hat{\rho}=
\Bigl[\frac{1}{2m}{\bf p}^2+U[\hat{\rho}],\ \hat{\rho}\Bigr]
\equiv\Bigl[H_0[\hat{\rho}],\ \hat{\rho}\Bigr],
\end{equation}
where the mean-field potential $U[\hat{\rho}]$ has been introduced, that is
defined as 
\begin{equation}
\label{eq:meanfield-of-rho}
U_1[\hat{\rho}]=\mathop{\mathrm{Tr}}_2\hat{A}_{12}v_{12}\hat{\rho}_2.
\end{equation}

A particular case, within the independent particle picture, would be to
consider a Slater determinant for the many-body state $|\Psi\rangle$. 
This approximation corresponds the the time-dependent Hartree Fock (TDHF) approach.
In practical applications, effective interactions are used for the mean-field 
potential, to account for
the neglected correlations beyond the mean-field picture. 

Extensions towards the inclusion of two-body correlations can be developed 
by including a correlated part in the two-body density operator, i.e.
\begin{equation}
\label{eq:rho2-hf}
\hat{\rho}^{(2)}_{12}=\hat{A}_{12}\hat{\rho}_1\hat{\rho}_2 + \hat{\sigma}_{12},
\end{equation}   
where the quantity $\hat{\sigma}_{12}$ encloses the explicit
two-body correlation effects. 
Accordingly, the system Hamiltonian can be written as the sum of the term 
treated and the one-body level ($H_0$) and a residual
two-body interactions: ${H}$ =  ${H_0} + {\nu}_{12}$. 
Then neglecting three-body correlations and retaining only terms 
up to leading order in the residual interaction and/or 
the explicit two-body correlations, 
Eq.(\ref{eq:drhodt-tdhf}) will be extended as it follows:
\begin{equation}
\label{eq:tdhf-ext}
i\hbar\frac{\partial}{\partial t}\hat{\rho}=
\Bigl[H_0[\hat{\rho}],\ \hat{\rho}\Bigr] + K\Bigl[\hat{\rho}, {\nu_{12}}
\Bigr] 
+ \delta K\Bigl[{\nu_{12}},\hat{\sigma}_{12} \Bigr]
\end{equation}
The second term of the (r.h.s.) of Eq.(\ref{eq:tdhf-ext}) represents the 
average effect of the residual interaction, i.e. of the interaction which is
not accounted for by the mean-field description, whereas the third term contains
the explicit correlations $\hat{\sigma}_{12}$. 
In other words, the $K$ term accounts for explicit two-body interactions,
%(2p-2h) excitations, 
whereas
the fluctuating term $\delta K$ may even account for higher order correlations.
 
Fluctuations with respect to the mean-field trajectory might originate from the initial conditions, as well as from 
the dynamical evolution. 
We will come back to this point in deeper detail later, when we 
will discuss the semi-classical approximation. 
	It may be noted that the equation above 
is similar to Stochastic TDHF (STDHF)~\cite{Reinhard1992}, and it
transforms into the extended TDHF (ETDHF) theory~\cite{Wong1978,Wong1979,Lacroix2004} if the fluctuating term $\delta K$ is suppressed;
	ETDHF can in fact efficiently describe the behavior %widening 
of some observables %spread widths 
related to dissipative processes, but it can not follow possible bifurcation paths deviating from the mean trajectory.

New stochastic extensions have been recently proposed for time-dependent 
quantum approaches \cite{Lacroix2016,Simenel2012}. 
A possibility is to inject quantum and/or thermal fluctuations just in the
initial conditions of TDHF calculations, leading to a spread of dynamical trajectories and
corresponding variances of physical observables.  Interesting results have been
obtained for the description of the spontaneous fission of 
superheavy systems \cite{Tanimura2017}. 

%so that the equation for $\hat{\rho}$ is now closed.  This TDHF equation  (\ref{eq:drhodt-tdhf}) conserves the idempotency $\hat{\rho}^2=\hat{\rho}$ as well as the normalization $\mathop{\mathrm{Tr}}\hat{\rho}=A$.

%The TDHF equation for $\hat{\rho}$ may be more general than the assumption of a Slater determinant because the assumption of the non-correlated two-body density operator by Eq.~(\ref{eq:rho2-hf}) can be introduced for a general $\hat{\rho}$ that does not correspond to a Slater determinant, i.e., in the case of $\hat{\rho}^2\ne\hat{\rho}$.  However, a benefit of restricting to the case of $\hat{\rho}^2=\hat{\rho}$ is that we then have a uniquely corresponding many-body wave function which is easily constructed from $\hat{\rho}$.  We will come back to this point later.

\subsection{\label{BUU}Semi-classical approximation and BUU models}

For the description of heavy ion reaction dynamics in the Fermi and 
intermediate energy regimes,  Eq.(\ref{eq:tdhf-ext}) is often solved in 
the semi-classical approximation. 
Within such a scheme, the goal is to derive an equation for the time evolution
of the one-body distribution function in phase space, $f({\bf r},{\bf p},t)$, 
which is nothing but
the semi-classical analog of the Wigner transform of the one body density matrix. 
The Boltzmann-Uehling-Uhlenbeck (BUU) theory can be 
considered as the semi-classical analog of Eq.(\ref{eq:tdhf-ext}), where the
$\delta K$  term is neglected.
The BUU equation can also be directly 
derived from the Born-Bogoliubov-Green-Kirkwood-Yvon (BBGKY) hierarchy or,  
within the real-time Green's-function formalism, from the Kadanoff-Baym equations \cite{danielewicz1984,botermans1990,buss2012}.  
%  Then two nucleons collide locally and instantaneously in the macroscopic scale, as an energy-conserving scattering of two nucleons of definite incoming momenta.  There are similar but different ways of arriving at similar conclusions, e.g.\ Refs.~\cite{reinhard1992stdhf,aichelin1991,kawai1992}.

Considering, for the sake of simplicity, a mean-field potential depending
only on the local density $\rho({\bf r})$, in non-relativistic kinematics, the BUU 
equation reads: 
\begin{equation}
\label{eq:buu}
\frac{\partial f({\bf r}, {\bf p},t)}{\partial t}
+\frac{{\bf p}}{m}\cdot\frac{\partial f}{\partial {\bf r}}
-\frac{\partial U[\rho]}{\partial{\bf r}}
\cdot\frac{\partial f}{\partial{\bf p}}
=\bar{I}_{coll}[f]%(\bf{r},\bf{p}),
\end{equation}
where the collision term $\bar{I}_{coll}$ has been introduced, which represents
the semi-classical analog of the $K$ term on the r.h.s. of 
Eq.(\ref{eq:tdhf-ext}).
The latter is written as: 
\begin{equation}
\label{eq:icoll-buu}
\bar{I}_{coll}[f]({\bf r}, {\bf p})
= g\int\frac{d{\bf p}_1}{(2\pi\hbar)^3}d\Omega\
v_{rel}\frac{d\sigma}{d\Omega}
[f'f_1'(1-f)(1-f_1)-ff_1(1-f')(1-f_1')],
\end{equation}
where $g$ is the degeneracy factor and
the coordinates of isospin are not shown for brevity. 

The distribution function entering the above integral is evaluated at the
coordinate ${\bf r}$ 
and at four locations in momentum space, with the final momenta  ${\bf p'}$  and 
${\bf p'_1}$ connected to the momenta   
 ${\bf p}$  and ${\bf p_1}$  
%the abbreviations mean $f=f(\bm{r},\bm{p},t)$, $f_1=f(\bm{r},\bm{p}_1,t)$, $f'=f(\bm{r},\bm{p}',t)$ and $f_1'=f(\bm{r},\bm{p}_1',t)$, with $\bm{p}'$ and $\bm{p}'_1$ related to $\bm{p}$ and $\bm{p}_1$ 
by the scattering angle $\Omega$ and the energy and momentum conservation.
Thus, within a classical picture, the effect of the two-body residual 
interaction is interpreted in terms of hard two-body scattering between
nucleons. In Eq.(\ref{eq:icoll-buu}), the quantity $v_{rel}$ denotes the 
relative velocity between the two initial phase-space portions and $d\sigma / d\Omega$
represents the differential nucleon-nucleon (n-n) cross section, accounting for the
residual interaction. 
In simulations of heavy ion collisions at intermediate energy, a density
dependent screened value of the n-n cross section, accounting for
in-medium effects, is often adopted (see, for instance, 
Refs.\cite{Dancross,Daniel}).
%P. Danielewicz, Hadronic transport models, Acta Phys. Pol.B33, 45 (2002)
The fermionic
nature of the system is preserved by the Pauli-blocking factors (the terms
like $(1-f)$ ).%in Eq.\ref{eq:icoll-buu}). 

Transport equations are usually solved numerically, adopting the test particle
method \cite{wong1982}, i.e. the system is sampled by $N_{test}$ test-particle per nucleon. 

\subsection{\label{stochastic}Semi-classical stochastic models}

\begin{figure}
\centering
\includegraphics[scale=1.2]{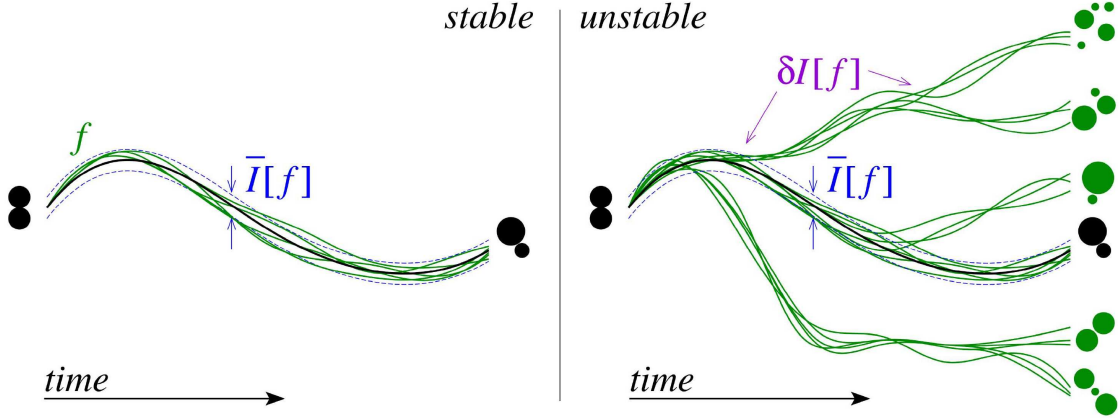}
\caption{\label{fig_1}
Illustration of the effect of fluctuations on a dynamical trajectory.
In stable conditions a moderate spread of trajectories, around
the average (associated with the average collision term ${\bar I}_{coll}[f]$)
is observed (left panel).  The right panel shows that in presence of 
instabilities the fluctuating BL term leads to bifurcation of trajectories. 
Taken from \cite{Paolo_fig1}.}
\end{figure}

	As anticipated above, if a suited stochastic approach is adopted, simplified higher-order contributions to the dynamics are taken into account even though not explicitly implemented.
	Already within a first-order-truncation scheme, 
%in a low-energy framework, 
it was found that a stochastic approach, including fluctuations only 
in the initial 
conditions,  can be used to restore all the BBGKY missing orders approximately~\cite{Lacroix2016}, and generate large-amplitude fluctuations; in this case, a coherent ensemble of mean-field states is propagated along different trajectories from an initial stochastic distribution.
	Such scheme, which appears appropriate in low-energy framework, 
is however insufficient to address highly dissipative regimes.
We will discuss in the following some approaches which attempt to solve
Eq.(\ref{eq:tdhf-ext}), but within the semi-classical approximation. 
Thus we introduce the so-called Boltzman-Langevin equation (BLE) \cite{Ayik1,chomaz2004}:
\begin{equation}
{{df}\over{dt}} = {{\partial f}\over{\partial t}} + \{f,H\} = {\bar I}_{coll}[f] 
+ \delta I[f].
\label{BL_equation}
\end{equation}
%\cite{Ayik1990}%from which kinetic equations are obtained~\cite{Balescu1976,Balian1991,Cassing1992,Bonasera1994}.
	Initial quantum fluctuations are neglected in this case and 
the stochastic treatment, mainly associated with hard
two-body scattering,  
%introduced in the initial state 
%obtained %from exploiting a distribution of initial states 
%progressing from a single initial state and 
acts intermittently all along the temporal evolution, producing successive splits of a given mean-field trajectory into subensembles, as illustrated in Fig.(\ref{fig_1}).
Within this framework, 
the system is still described in terms of the one-body distribution function $f$, but this function
may experience a stochastic evolution in response to the action of the fluctuating term. 
When instabilities are encountered along the reaction path, 
the evolution of the fluctuation "seeds" introduced by the BL approach is then
driven by the dissipative dynamics of the BUU evolution, 
%Thus it does not rely on a knowledge of the unstable modes but rather 
allowing 
the system to choose its trajectory through possible bifurcations leading
to different fragmentation paths. In this way a series of
"events" are created in a heavy-ion collision, which can then be analyzed and sampled 
in various ways.

Assuming that fluctuations are of statistical nature, the term $\delta I$ is 
simply interpreted as the fluctuating part of the collision
integral, with a vanishing mean value and a 
variance which equals the average collision integral.

\subsection{\label{BLOB}Fluctuations in full phase space 
and the BLOB model}

The latter procedure can be implemented  by replacing 
the residual terms $({\bar I}_{coll} + \delta I )$ by a similar 
Uehling Uhlenbeck (UU) - like term, involving nucleon packets, which respects the Fermi statistics both for the occupancy mean value and for the occupancy variance.
%\cite{Bauer1987}.

%	A natural solution to satisfy such requirement is to rewrite the residual contribution in the form of 
Thus one may consider a rescaled UU collision term where a single binary collision involves extended phase-space portions of 
nucleon distribution of the same type (neutrons or protons),
$A$, $B$, to simulate nucleon wave packets, and Pauli-blocking factors act on the corresponding final states $C$, $D$, also treated as extended phase-space portions.
	The choice of defining each phase-space portion $A$, $B$, $C$ and $D$ so that the isospin number is either $1$ or $-1$ is necessary to preserve the Fermi statistics for both neutrons and protons,
%to produce isovector fluctuations, 
and it imposes that blocking factors are defined accordingly in phase-space cells for the given isospin species.
	The above prescriptions lead to the Boltzmann Langevin One Body
(BLOB) equations~\cite{Napolitani2013}:
\begin{eqnarray}
	\frac{\partial f}{\partial t} + \{f , H\} 
		= {\bar I}_{coll} + \delta I  = \nonumber\\
	= g\int\frac{d{\bf p}_b}{h^3}\,
	\int
	W({AB\leftrightarrow CD})\;
	F({AB \rightarrow CD})\;
	d\Omega\;.
\label{eq:BLOB}
\end{eqnarray}
%where $g$ is the degeneracy factor. 
	In the above equation, $W$ is the transition rate, in terms of relative velocity between the two colliding phase-space portions and differential nucleon-nucleon cross section:
\begin{equation}
	W({AB \leftrightarrow CD}) = |v_A\!-\!v_B| \frac{d\sigma}{d\Omega}\;.
\label{eq:transition_rate}
\end{equation}
	The term $F$ contains the products of occupancies and vacancies of initial and final states over their full phase-space extensions:
\begin{equation}
	F({AB\rightarrow CD}) =
	\Big[(1\!\!-\!\!{f}_A)(1\!\!-\!\!{f}_B) f_C f_D - f_A f_B (1\!\!-\!\!{f}_C)(1\!\!-\!\!{f}_D)\Big]\;.
\label{eq:Pauli_scaled}
\end{equation}
%	Details on the implementation of BLOB are given in appendix~\ref{sec_appendix_metrics}.
	In practice, if the test-particle method is employed, 
%so that the system is sampled by $N_{test}$ test-particle per nucleon, 
the phase-space portions $A$, $B$, $C$ and $D$ should be agglomerates of $N_{test}$ test-particles each, and the nucleon-nucleon cross section used in Eq.~(\ref{eq:transition_rate}) should be scaled by the same amount $N_{test}$, considering that each nucleon packet is associated with $N_{test}$ possible samplings.
  
\begin{figure}
\centering
\includegraphics{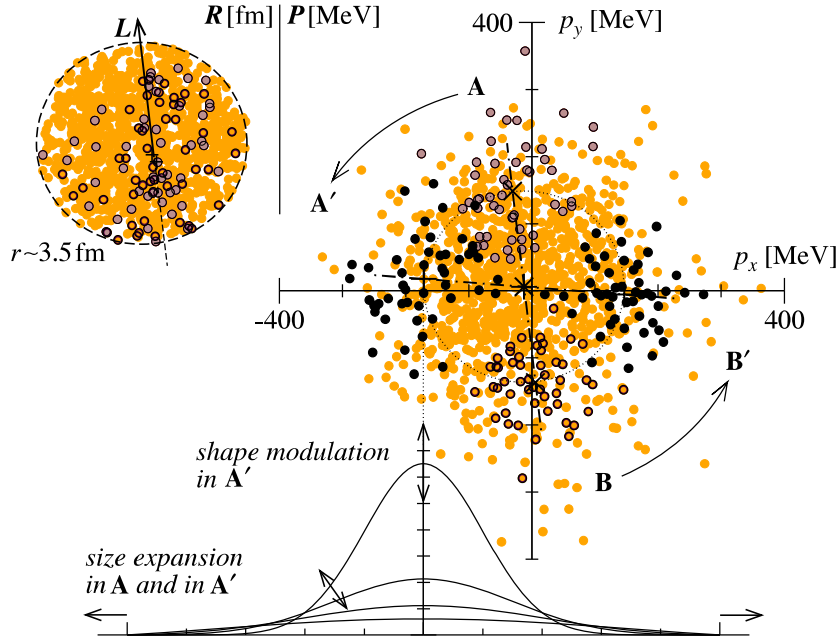}
\caption{\label{fig_2}
Example of one collision event in BLOB. Two nucleons are represented by two
agglomerates of test particles A and B (black points) which share the same
volume in coordinate space $R$.  In the momentum space $P$ the collision
process induces a rotation, according to a given set of 
scattering angles ($\theta$, $\phi$), to the destination sites A' and B', 
where the test particles are distributed according to Pauli-blocking
and energy conservation constraints.  The latter are enforced by modulating 
the shape and the size of the nucleon packet (see the bottom part of the figure). Taken from \cite{Paolo_fig2}.
}
\end{figure}

	By this procedure, it can be shown that 
for a free Fermi gas, 
the occupancy variance at equilibrium equals ${\bar f}(1-
{\bar f})$ in a phase-space cell $h^3$, resulting from  
the movement of extended portions of phase space which have the size of a nucleon. Thus the residual term carries nucleon-nucleon correlations which fulfill
the analytical predictions of an equilibrated fermionic system \cite{Rizzo2008}.    
	Hence, the BLOB approach exploits the stochastic term of Eq.~(\ref{eq:BLOB}), recovering correlation orders higher than the 
second order %$k\!=\!2$ 
truncation (which would correspond to the average collision integral), and inducing the BL fluctuation-bifurcation scheme.
The method is schematically illustrated in Fig.\ref{fig_2}.

%%%%%%%%%%%%%%%%%%%%%%%%%%%%%%%%%%  mettere un commento su nuovi approcci
%%%%%%%%%%%%%%%%%%%%%%%%%%stocastici

A new framework to treat the dissipation and fluctuation dynamics associated with n-n scattering in heavy-ion collisions has been recently introduced in \cite{Hao2019}. Two-body collisions are effectively described in terms of the diffusion of nucleons in the viscous nuclear medium, according to a set of Langevin equations in momentum space. The new framework, combined with the usual mean-field dynamics, has been shown to be suited to simulate heavy-ion collisions at intermediate energies. Applications of the method, as well as the comparison
with other transport models,  are presently in progress.

\subsection{\label{SMF}Fluctuation projection and simplified stochastic approaches}
%
%	The stochastic term in Eq.~(\ref{eq:SMF}) can be kept separate and treated as a stochastic force related to an external potential $U'$, like in the corresponding semi-classical stochastic mean field (SMF) model~\cite{Colonna1998}. This leads to treatments where fluctuations are implemented only in the coordinate space, i.e. they are projected on a suited coordinate, like the density space, so that $\delta I[f]=\partial_{\vecr}\,U'\;\partial_{\vecp}\,f$.
%	Eq.~(\ref{eq:BLOB}) may be confused with the early approach of Bauer and Bertsch~\cite{Bauer1987}. There is however a very fundamental difference: in Bauer-and-Bertsch approach the Pauli-blocking term is not applied to the involved portions of phase space which are actually interested by the scattering at a given time $t$, but it is applied in average. Such approximation has the effect of loosing the Fermi statistics because the variance $f^{(n)}(1-f^{(n)})$ is not respected~\cite{Chapelle}.
%
%	--- FIGURE collisionvariance
%

%
Several approximate treatments of the BL approach have been introduced and are
still employed to deal with the description of collision dynamics. 
For instance, at variance with the above description, the stochastic term in 
the r.h.s. of Eq.(\ref{BL_equation}) can be kept separate and treated as a stochastic force related to an external potential $U'$~\cite{BOB}.
%	This leads to treatments where fluctuations are implemented only in the coordinate space, i.e. they are projected on the spacial density.	
	More generally, %the difference between Eq.~(\ref{eq:BLOB}) and 
approximated stochastic mean-field approaches 
%is that those latter 
build fluctuations from 
suitable projections \cite{Xie2013,Maccarone1998} or 
introducing a well adapted external force or a distribution of initial conditions which should be accurately prepared in advance \cite{chomaz2004}.
	On the contrary, Eq.~(\ref{eq:BLOB}) introduces fluctuations in full phase space and let them develop spontaneously and continuously over time in a dynamical process.

In the Stochastic Mean Field (SMF) treatment \cite{Maccarone1998}
the fluctuations of the distribution
function generated by the stochastic collision integral are projected 
on the coordinate space. Thus only local density fluctuations, 
which could be implemented
as such in numerical calculations, are considered. 
The further
assumption of local thermal
equilibrium is made, with the possibility to derive  analytical
expressions for the density fluctuations.  This implies that fluctuations
can be implemented only when, after the first collision instants, the phase
space occupancy is locally thermalized. 
In principle, thermal fluctuations could be introduced directly in
the phase
space, i.e. implementing $\sigma^2_f = {\bar f}(1 - {\bar f}) $
locally in the phase space. %, where ${\bar f}$ denotes the mean value of the distribution function.  
However,
this is a difficult
numerical tack because of the high dimension of the phase space. In the
standard application
of the SMF method therefore one considers density fluctuations 
in a volume $V$:
\begin{equation}
\sigma^2_\rho({\bf r}, t) = {1 \over V} \int {d{\bf p} \over h^3/4} \sigma^2_f
({\bf r},{\bf p},t).
\label{sigma_ro}
\end{equation}

This variance could be directly calculated from the value of the
average one-body distribution function provided by the BUU simulation, and
fluctuations
be introduced accordingly. However it is more practical to have explicit
analytical expressions
for the density fluctuations. 
Within the assumption of local thermal equilibrium, 
the mean distribution function can be
parametrized by the expression
${f} ( {\bf r} , {\bf p} , t ) = 1/(1 + exp(e - \mu( {\bf r} , t ) ) 
/ T ( {\bf r} , t ) ) $
with a local chemical potential and temperature $\mu({\bf r}, t)$ 
and $T({\bf r}, t)$,
respectively, and
with $e = p^2/2m$. 
%The determination of the temperature will be discussed below.
Introducing the expression for the fluctuation variance
 into Eq.(\ref{sigma_ro}) 
%and converting the
%$\vecp$-integration into an $\epsilon$-integration, 
%%expressing f(1 - f) by a derivative of Eq. (10) with
%respect to $\epsilon$ and
%integrating by parts, 
one obtains, after some algebra:
\begin{equation}
\sigma^2_\rho =  {1 \over V} {{2\pi m \sqrt{(2m)}~T}  \over {h^3/4}}
\int {1 \over {\sqrt e}} {1 \over {1+exp(e-\mu)/T}} de 
\label{sigma_ro_final}
\end{equation}
We note that Eq.(\ref{sigma_ro_final}) is consistent with the thermodynamical relation
for the variance
of the particle number in a given volume.
 To obtain
a more explicit expression, and to eliminate the chemical potential we can
use the
Sommerfeld expansion, for the function ${f}$ around $e =\mu$ 
for small $T/\epsilon_F$ ratio, being $\epsilon_F$ the Fermi energy at the considered density.  We then obtain
\begin{equation}
\sigma^2_\rho ={{16\pi m \sqrt{(2m)}} \over {h^3 V}} \sqrt{\epsilon_F} T 
[1 - {\pi^2 \over 12} ({T \over \epsilon_F})^2 + . . . ].
\end{equation}
The procedure can be considered and implemented separately for neutrons
and protons.

%%%%%%%%%%%%%  check equation number !!
%%%%%%%%%%%%%%%%%%%%%%%%%%%%%%%%%%%%%%%%%%%%%%%%%%%%%%%%%%%%%%%%%%%%%%%%%%%
%%%%%  mettere figura sulla biforcazione e sul metodo BLOB

\subsection{\label{QMD}QMD models}

Molecular dynamics (QMD) approaches are widely employed to describe the 
nuclear reaction dynamics (several examples can be found in Refs.\cite{Aich1991,Papa2001,Zhang2006}). 
Within such a family of models, the nuclear 
many-body state is written as a product 
%(anti-symmetrized in the case
%of the AMD \cite{Ono1999,Ikeno2016} approach) 
of single particle states, usually represented by Gaussian
wave packets. Two-body collisions between nucleon packets are treated stochastically, by choosing 
randomly the scattering angle associated with the n-n scattering process. 
Moreover, the localization of the wave packets induces some additional correlations also in the mean-field propagation. 
Then these approaches are very well suited to  enable the formation of clusters and fragments.
%Besides the BUU models, there are quantum molecular dynamics (QMD) models which are also actively employed in the studies of heavy-ion collisions.  Regardless of its name and its historical origin \cite{aichelin1991}, the QMD approach can be understood as a method to avoid the problems of the BUU equation discussed at the end of the previous subsection, i.e., to handle the propagation of correlations induced by two-nucleon collisions and to avoid too smooth mean-field potential, in order to enable the formation of clusters and fragments.

In QMD, the one-body distribution function is obtained as a sum of Gaussian wave packets:
\begin{equation}
\label{eq:f-qmd}
f_{QMD}({\bf r},{\bf p},t)
=\sum_{k=1}^A \Bigl(\frac{2\nu}{\pi}\Bigr)^{3/2}e^{-2\nu({\bf r}-{\bf R}_k(t))^2}
(2\pi\hbar)^3\delta({\bf p}-{\bf P}_k(t)),
\end{equation}
where $\nu$, or $\Delta x=(4\nu)^{-1/2}$, is a fixed parameter representing the width of the wave packet of each nucleon.  
The centroids of the wave packets (${\bf R}_k$ and ${\bf P}_k$) are propagated according to classical 
equations of motion (Hamilton equations), with the Hamiltonian evaluated 
starting from the effective interaction considered and the nucleon 
distribution given by Eq.(\ref{eq:f-qmd}).

As far as the collision integral is concerned, 
similarly to the 
approached introduced before,
%two-nucleon collisions in QMD models 
%are governed by 
the in-medium n-n 
cross sections $d\sigma/d\Omega$ is employed to evaluate the probability
for a nucleon-nucleon collision to occur.  
More specifically, a pair of nucleons 
represented by 
$({\bf R}_i, {\bf P}_i)$ and $({\bf R}_j, {\bf P}_j)$ 
will collide when the radial distance among them 
becomes minimum during the considered time step interval $\Delta t$ and this closest distance is less than $\sqrt{\sigma/\pi}$, in analogy with the geometrical prescription by Bertsch and Das Gupta \cite{bertsch1988}.
This implies that nucleons collide at a given finite relative distance, 
whereas in the BUU equation, distribution functions are evaluated at the
same point ${\bf r}$ in configuration space, i.e. collisions are local
in space.   

The main general drawback of QMD approaches is that 
the one-body distribution $f_{QMD}$ of Eq.(\ref{eq:f-qmd}) 
%or $f_{\text{W}}$ of Eq.~(\ref{eq:f-qmd-w}) 
does not preserve the Pauli principle along the dynamical evolution, though 
it can be enforced
in the initial conditions. 
%i.e., it does not correspond to a one-body density operator $\hat{\rho}$ with all the eigenvalues between 0 and 1.  
%Furthermore, the Heisenberg's uncertainty principle is not respected by $f_{QMD}$.  
As a consequence, %the dynamical evolution may drive quickly the system
the system can be pushed
towards a classical type of behavior. 
Some methods have been proposed to cure this problem \cite{Papa2001}. 

From the above discussion, we expect mean-field (BUU-like and stochastic) theories to better
describe genuine mean-field effects, owing to the better mapping of the
one-body distribution function in phase space.  On the other hand, the effect of correlations is emphasized in QMD-like models, but the fermionic nature of the nuclear systems under study could be quickly lost along 
the dynamical evolution. 
%In the following we will discuss results obtained with both classes of models.

\subsection{\label{AMD}Antisymmetrized Molecular Dynamics}

%As already stressed above, the fermionic nature of nucleons most probably plays important roles in fragmentation reactions, as well as in many other problems of reaction dynamics.

In  the antisymmetrized molecular dynamics (AMD) approach, a system of  
$A$-nucleon is represented by a Slater determinant of Gaussian wave packets,
\begin{equation}
\label{eq:amd-wf}
|\Phi_{AMD}(Z)\rangle
=\hat{A}\prod_{K=1}^{A}\varphi_K %(i)
\end{equation}
where $\hat{A}$ is the full antisymmetrization operator.  Each single-particle state $\varphi_k$ is a product of a Gaussian function and a spin-isospin state
\begin{equation}
\langle{\bf r}|\varphi_k\rangle =
\exp\Bigl[-\nu\Bigl({\bf r}-\frac{{\bf Z}_k}{\sqrt{\nu}}\Bigr)^2\Bigr]
\otimes\chi_{\alpha_k}.
\end{equation}
The spin and isospin of each nucleon are kept fixed, $\alpha_k=p\uparrow,p\downarrow,n\uparrow$ or $n\downarrow$.  The width parameter $\nu$ of the 
Gaussian function is chosen to be $\nu=1/(2.5\ {fm})^2$ in almost
all applications.  
%The wave functions are usually not normalized in AMD and the single-particle states $\varphi_i$ are not orthogonal to each other.  However, the non-normalized Slater determinant  (\ref{eq:amd-wf}) is a proper many-body state of fermions as long as the single-particle states are linearly independent.  
It should be noticed that the single-particle states $\varphi_k$ are not orthogonal to each other, however, as long as they are linearly independent, the Slater determinant  (\ref{eq:amd-wf}) is a proper fermionic many-body state. 
The latter 
%Thus the many-body state $|\Phi_{{AMD}}(Z)\rangle$ 
results parametrized in terms of the Gaussian centroids $Z=\{{\bf Z}_1,{\bf Z}_2,\ldots,
{\bf Z}_A\}$, which are complex vectors.  
%The time-evolution of these variables may be determined by the time-dependent variational principle, from which we obtain an equation of motion
The time-dependent variational principle allows to determine the following 
equations of motion:
\begin{equation}
\label{eq:amd-eq-of-motion}
\frac{d}{dt}{\bf Z}_k=\{{\bf Z}_k,\mathcal{H}\}_{{PB}},
\end{equation}
with the Poisson bracket (PB) suitably defined for the non-canonical variables $Z$ \cite{ono2002,ono2004ppnp,OnoPPNP2019}.  
 The above equation accounts for the wave packet propagation in the mean field.
The Hamiltonian $\mathcal{H}$ is the expectation value of the Hamiltonian operator $H$ and a suitable technique is adopted to subtract the spurious zero-point kinetic energies of fragment center-of-mass motions \cite{ono1992,ono1993}.  
Similarly to the transport approaches discussed above, an effective interaction is employed in AMD, such as the Gogny force 
or the Skyrme force.  

 A stochastic collision integral, describing hard
two-body scattering between nucleon packets and preserving the
antisymmetrization is explicitely included in the description.

As in QMD models, in AMD the width parameter $\nu$ is always kept fixed. 
On the other hand, in the fermionic molecular dynamics (FMD) model 
proposed by Feldmeier \cite{feldmeier1990}, the width parameters $\nu_k$ of the 
individual wave packets are treated as time-dependent.
This generalization allows one to get the exact quantum-mechanical solution 
for free particle propagation and for the spreading  of single-particle states. 
However, the fixed width choice looks more suitable to describe fragmentation
events, thanks to the localization of the nucleon wave packet \cite{FFMD}, as 
shown already by the very first applications of the AMD approach to the simulation
of heavy ion collisions \cite{ono1992,ono1992prl}.  
%Before the first application of AMD, Feldmeier independently proposed the fermionic molecular dynamics (FMD) \cite{feldmeier1990}, in which the width parameters $\nu_i$ for individual wave packets are also treated as time-dependent parameters which take complex values.  This allows the exact evolution of free particles and the spreading of single-particle states in expanding systems.  In AMD, on the other hand, the width parameter $\nu$ is always a fixed parameter, 
%as already explained  in the previous subsection for QMD, in order to describe fragment formation.  
However, further extensions of AMD also include the possibility to consider
deformation of the wave packects, which actually can be represented as a
superposition of many Gaussian wave functions, 
%However, when the AMD wave function is applied to nuclear structure problems, the wave function is extended by treating $\nu$ as a variational parameter, by allowing deformation of the wave packets, by superposing many wave functions, and so on.  
see Ref.~\cite{kanada2012} for a review.

A recent upgrade of the AMD model considers the possibility to include explicit
light cluster production, as an extension of the nucleon correlations induced
by the collision integral \cite{OnoJPG2013,OnoCim2016}. 
The implementation of light clusters, up to mass A=3, was actually 
introduced first in the context of a BUU approach, leading to the
pBUU model \cite{DanielNPA1991,DanielPRC1992}.
 
%Some of the mean-field and molecular dynamics models have been improved
%e.g. by considering light clusters explicitly, which can be regarded as incorporating
%quantum mechanical features mentioned above in practical manners. 

%These attempts of transport models will be reviewed
%in this article with emphasis on the description of the formation of fragments and clusters.

%\section{Models with explicit cluster correlations}

\subsection{\label{RBUU}Relativistic transport approaches} \label{qhd}

%\markright{Chapter \arabic{section}: qhd}

Within a fully relativistic picture, 
equilibrium and dynamical properties of nuclear systems at the hadronic 
level 
are well described by the the Quantum Hadro Dynamics (QHD) effective field model 
\cite{WaleckaAP83,SerotANP16,SerotIJMPE6}.

Such a framework provides 
consistent results for the nuclear structure of finite 
nuclei \cite{SharmaAP231,RingPPNP37,TypelNPA656}, for the NM
Equation of State and liquid-gas 
phase transitions \cite{MuellerPRC52}. and for 
collective excitations \cite{MaNPA686,GrecoPRC67} and 
nuclear collision dynamics \cite{GiessenRPP56,KoJPG22}. 

%Relativistic 
%Random-Phase-Approximation ($RRPA$) 
%theories have been developed to study the nuclear collective response 
%\cite{DellaPRC44,HorowitzNPA531,MateraPRC49,VretenarNPA621,MaPRC55,MaNPA686}.

%%%%%%%%%%%%%%%%%%%%%%%%%%%%%%%%%%%%  mettere dopo

%We will often derive transparent analytical results. In order to show
%also some quantitative effects of the dynamical contribution of the
%$\delta$-channel we have to fix in some way the corresponding coupling.
%We have used a constant value (see Table \ref{qhdsets}) extracted 
%from the $DBHF$ analysis
%of refs.\cite{JongPRC57,HofmannPRC64}, where it actually appears not strongly
%density dependent in a wide range of baryon densities.
%Some results are also presented with Fock correlations
%explicitly accounted for ($NLHF$ case, see following). 

\subsubsection{QHD effective field theory}

%As a starting point of the relativistic approaches introduced below, 
%We consider the {\it $QHD$} effective field picture of the hadronic 
%phase of nuclear matter \cite{WaleckaAP83,SerotANP16,SerotIJMPE6}. 
In QHD theory, the main dynamical 
degrees of freedom of the system are included by considering the nucleons
coupled to the isoscalar scalar $\sigma$ and vector $\omega$ mesons
and to the isovector scalar $\delta$ and vector $\rho$ mesons.

Hence the Lagrangian density for this approach, including non--linear isoscalar/scalar 
$\sigma$-terms \cite{BogutaNPA505}, is given by:

\begin{eqnarray}
{ L} = {\bar {\psi}}[\gamma_\mu(i{\partial^\mu}-{g_\omega}{V}^\mu
- g_{\rho}{{\bf B}}^\mu \cdot  {\vec {\tau}} ) 
-(M-{g_\sigma}\phi-g_{\delta}{\vec {\tau}} \cdot {\vec {\delta}}~)]\psi +~~~~~~
\nonumber\\
  {1\over2}({\partial_\mu}\phi{\partial^\mu}\phi
- m_s^2 \phi^2) - {a \over 3} \phi^3 - {b \over 4} \phi^4
- {1 \over 4} W_{\mu\nu}
W^{\mu\nu} + {1 \over 2} m_\omega^2 {V}_\nu {{ V}^\nu}+ \nonumber\\
\frac{1}{2}(\partial_{\mu}{\vec {\delta}} \cdot \partial^{\mu}{\vec {\delta}}
-m_{\delta}^2{\vec {\delta}}^2)
- {1 \over 4} {\bf G}_{\mu\nu}
{\bf G}^{\mu\nu} + {1 \over 2} m_{\rho}^2 {{\bf B}}_\nu
{{{\bf B}}^\nu}~~~~~~~~~~~~~~~~~~~
\label{eq.1}    
\end{eqnarray}

where
$W^{\mu\nu}(x)={\partial^\mu}{{V}^\nu}(x)-
{\partial^\nu}{{V}^\mu}(x)~$ and
${\bf G}^{\mu\nu}(x)={\partial^\mu}{{{\bf B}}^\nu}(x)-
{\partial^\nu}{{{\bf B}}^\mu}(x)~.$

Here $\psi(x)$ is the nucleon
fermionic field, $\phi(x)$ and ${{V}^\nu}(x)$ represent neutral scalar
 and
vector boson fields, respectively, whereas ${\vec {\delta}}(x)$ and 
${{{\bf B}}^\nu}(x)$
are the charged scalar and vector
fields and ${\vec {\tau}}$ denotes the isospin matrices.
The coefficients of the type $m_i$ and $g_i$ indicate masses and coupling
constant of the different meson channels and $M$ denotes the nucleon mass.

From the Lagrangian, Eq.(\ref{eq.1}), %with the Euler procedure 
a set of 
coupled equations of motion for the meson and nucleon fields can be 
derived. The basic 
approximation in nuclear matter applications consists in neglecting all the
terms containing derivatives of the meson fields with respect to the their mass
contributions. Then the meson fields are simply connected to the operators
of the nucleon scalar and current densities by the following equations: 
\begin{equation}\label{Eq.4a}
{\widehat{\Phi}/f_\sigma} + A{\widehat{\Phi}^2}
+ B{\widehat{\Phi}^3}~=\bar\psi(x)\psi(x)~\equiv \widehat{\rho_S}
\end{equation}
\vskip -0.5cm
\begin{eqnarray}\label{Eq.4b}
{\widehat{ V}}^\mu(x)~\equiv g_{\omega} V^\mu~=~f_\omega\bar\psi(x)
{\gamma^\mu}\psi(x)~
\equiv f_\omega \widehat j_\mu~,\nonumber\\
{\widehat  {\bf  B}}^\mu(x)~\equiv g_{\rho} {\bf B}^\mu~=~f_{\rho}\bar\psi(x)
{\gamma^\mu} {\vec {\tau}}
\psi(x)~,\nonumber\\
{\widehat{\vec {\delta}}}(x)~\equiv g_{\delta}{\vec {\delta}}~=~f_{\delta}
\bar\psi(x){\vec {\tau}}\psi(x)
\end{eqnarray}
where $\widehat \Phi=g_\sigma\phi$, $f_\sigma = (g_\sigma/m_\sigma)^2$, 
$A = a/g_\sigma^3$, $B = b/g_\sigma^4$, $f_\omega = (g_\omega/m_\omega)^2$, 
$f_{\rho} = (g_{\rho}/m_{\rho})^2$, $f_{\delta} = (g_{\delta}/m_{\delta})^2$.

Exploiting Eqs.(\ref{Eq.4a},\ref{Eq.4b}) for the meson field operators,
it is possible to derive  a Dirac-like equation for the nucleon fields which
contains only nucleon field operators. This  
equation can be consistently solved within a Mean Field Approximation 
(Relativistic Mean Field, RMF), i.e. in a self-consistent Hartree scheme
 \cite{SerotANP16,SerotIJMPE6}.

Some attempts were performed to go beyond this scheme. 
In particular, it is interesting to notice that the inclusion of Fock terms
%\cite{HorowitzNPA399,BouyssyPRC36}
automatically leads to contributions to the various meson exchange 
channels, also
in absence of explicit direct coupling terms. 
A thorough study of these effects
%the Fock contributions in a $QHD$ approach with non-linear
%self-interacting terms 
was performed in \cite{GrecoPRC64}, for asymmetric nuclear matter. 
%has been recently performed \cite{GrecoPRC63}, in
%particular for asymmetric matter \cite{GrecoPRC64}.
 
\subsubsection{Relativistic semi-classical transport equations}
%\addtocontents{toc}{\hspace{0.55cm}\thesubsubsection \hspace{0.12cm}
%Relativistic transport equations with Fock terms}
Within the previous assumptions, 
we move to discuss a kinetic approach which adopts  
the semi-classical approximation. 
%The latter holds when the nuclear medium is supposed
%to be in states for which the nucleon scalar and current densities
%are smooth functions of the space-time coordinates.
This allows one to establish a formal connection to the
transport models described above. 

%As anticipated above,  
%within the $QHD$ model
%we focus our analysis on a description of 
%the many-body nuclear system in terms of one--body dynamics.
In the RMF model, the nuclear system is essentially described at the one-body level, with some correlation effects included, in an effective manner, through (density dependent) coupling constants.
To derive the kinetic equations for the one-body nucleon density matrix, it is useful to introduce, 
in quantum phase-space, the Wigner function %transform of the one-body density matrix 
for the
fermion field \cite{DegrootRelKin,HakimNC6}.
The latter is defined as:
$$[{\widehat F}(x,p)]_{\alpha\beta}=
{1\over(2\pi)^4}\int d^4R~e^{-ip \cdot R}
\langle :\bar{\psi}_\beta(x+{R\over2})\psi_\alpha(x-{R\over2}):\rangle~,$$
where $\alpha$ and $\beta$ are double indices for spin and isospin. 
The brackets denote statistical averaging and the colons indicate  
normal ordering.
The equation of motion can be derived from
the Dirac field equation by using standard procedures
(see e.g.\cite{DegrootRelKin,HakimNC6}). 
%One obtains: 
%\begin{eqnarray}\label{wig}
%{i\over2}{\partial_\mu}[\gamma^\mu{\hat F}(x,p)]_
%{\alpha\beta}+p_\mu[{\gamma^\mu}{\hat F}(x,p)]_{\alpha\beta}-
%M{\hat F}_{\alpha\beta}(x,p) \nonumber\\
%-{g_{\omega}\over(2\pi)^4}\int_R e^{-ip\cdot R}
%<:\bar{\psi}_\beta(x_+){\gamma^\mu_{\alpha\gamma}}\psi_\gamma(x_-)
%{V}_\mu(x_-):> \nonumber\\
%+{g_{\sigma}\over(2\pi)^4}\int_R e^{-ip\cdot R}
%<:\bar{\psi}_\beta(x_+)\psi_\alpha(x_-)\phi(x_-):> \nonumber\\
%-{g_{\rho}\over(2\pi)^4}\int_R e^{-ip\cdot R}
%<:\bar{\psi}_\beta(x_+){\gamma^\mu_{\alpha\gamma}}\psi_\gamma(x_-)
%{\vec{\tau}} \cdot {\bf B}_\mu(x_-):> \nonumber\\
%+{g_{\delta}\over(2\pi)^4}\int_R e^{-ip\cdot R}
%<:\bar{\psi}_\beta(x_+)\psi_\alpha(x_-){\vec{\delta}}(x_-):> \nonumber\\
%=0
%\end{eqnarray}
%with $x_+=x+{R\over2}$ and $x_-=x-{R\over2}$.
%When we insert Eqs.(\ref{Eq.4a},\ref{Eq.4b}) for the meson field operators
%we clearly see the appearance of Hartree and Fock contributions (at the lowest %order in
%density matrices).

%The Wigner function is a matrix in spin and isospin space; 

%Following the treatment of the Fock terms in non-linear $QHD$ introduced 
%in Refs. \cite{GrecoPRC63,GrecoPRC64}, 
Within the Hartree-Fock scheme and in the semi-classical approximation, 
one obtains the following 
 kinetic equations \cite{DellaPRC44,GrecoPRC64}: 
\begin{eqnarray}\label{trans}
&&{i\over2}{\partial_\mu}{\gamma^\mu}{\hat F}^{(i)}(x,p)+\gamma^\mu
p^*_{\mu i}{\hat F}^{(i)}(x,p) - M^*_i{\hat F}^{(i)}(x,p)+\nonumber\\
&&{i\over2}\Delta
\left[\tilde f_\omega j_\mu(x)\gamma^\mu \pm \tilde f_\rho j_{3\mu}(x)
\gamma^{\mu}
- \tilde f_\sigma \rho_S(x) \mp \tilde f_\delta \rho_{S3}(x)\right]
{\hat F}^{(i)}(x,p)=0,
\end{eqnarray}
%In the general case of
%asymmetric $NM$ it is useful to decompose 
where the Wigner function 
has been decomposed into neutron and proton components  ($i=n,p$)
and the upper (lower) sign 
corresponds
to protons (neutrons).
In the avove equation $\Delta={\partial_x} \cdot {\partial_p}$, with $\partial_x$ acting only 
on the first term of the products, and  $\rho_{S3}=\rho_{Sp}-\rho_{Sn}$ and $j_{3\mu}(x)=j^p_{\mu}(x)
-j^n_{\mu}(x)$ 
are the isovector scalar density and baryon current, 
respectively. 
The kinetic momenta and effective masses are defined as:
\begin{eqnarray}\label{psms}
&&p^*_{\mu i}=p_\mu-{\tilde f_\omega}j_\mu(x)\pm \tilde f_\rho{j}_{3\mu}(x)~,
   \nonumber\\
&&M^*_i=M- {\tilde f_\sigma}\rho_S(x) \pm \tilde f_\delta \rho_{S3}(x)~, 
\end{eqnarray}
where the effective coupling functions $\tilde f_i (i=\sigma,\omega, \rho, \delta)$ have been introduced. 
The latter are generally space, i.e. density,
 dependent. Within the Hartree-Fock approximation, 
the effective coupling constants incorporate the Fock term contributions, and get finite values in 
all channels, even in absence of direct contributions. 
%to the channel considered.  
%It is interesting to notice that 
%the expression of Eq.(\ref{psms}) for the effective mass 
%embodies an isospin contribution from Fock terms even without 
%a direct inclusion of 
%the $\delta$ meson in the Lagrangian. 
%As seen from Eqs.\ref{cincoup} 
It is easy to realize that the common RMF approximation (Hartree level) is recovered from the 
Hartree-Fock results by changing the effective coupling functions 
$\tilde f_i$ %%%%%%%%%%%, Eqs.(\ref{cincoup}), 
to the 
the explicit coupling constants $f_i$.  
%Moreover, in practice one can always
%work within the Hartree scheme, but adopting effective, density dependent, 
%coupling functions. 

The explicit effect of two-body correlations can be included also 
within the relativistic covariant framework. 
Eq.(\ref{trans}) can be 
complemented by a two-body collision integral, to explicitely take into
account effects beyond the mean-field picture (RBUU models) \cite{buss2012}.  
%%%%%%%%%%%%%%  fare un cenno alle interazioni relativistiche utilizzate. 

%%%%%%%%%%%%%%%%%%%%%%%%%%%%%%%%%%%%%%%%%%%%%%%%%%%%%%%%%%%%%%%%%%%%%%%

\section{\label{sym_ene}Effective interactions and symmetry energy}

%From the above discussion, we expect BUU-like and stochastic one-body 
%theories to better
%describe genuine mean-field effects, owing to the better mapping of the
%one-body distribution function in phase space.  On the other hand, the effect of fluctuations and correlations is emphasized in QMD-like models.  

In the following we will review results obtained with both classes of models
discussed in Section \ref{theo:intro}. 
A common ingredient is certainly the nuclear effective interaction, from 
which
the nuclear EOS can be derived in the equilibrium limit. 
%Effective interactions (associated with a given EOS) 
%can be considered as an input of the transport code
%and 
The comparison of suitable reaction observables, evaluated in 
simulated transport model events, to experimental data  
would allow one to extract information on relevant 
nuclear matter features. 

For illustrative purposes, let us
consider, as a standard effective interaction, a Skyrme interaction, with the energy density ${\cal E}$ expressed in terms of the isoscalar, $\rho=\rho_n+\rho_p$,
and isovector, $\rho_{3}=\rho_n-\rho_p$,  densities and 
kinetic energy densities ($\tau=\tau_{n}+\tau_{p}, \tau_{3}=\tau_{n}-\tau_{p}$) as \cite{radutaEJPA2014}:
\begin{eqnarray}
{\cal E}&=&\frac{\hbar^2}{2 m}\tau + C_0\rho^2 + D_0\rho_{3}^2 + C_3\rho^{\alpha + 2} + D_3\rho^{\alpha}\rho_{3}^2 ~+ C_{eff}\rho\tau \nonumber\\
&& + D_{eff}\rho_{3}\tau_{3} + C_{surf}(\bigtriangledown\rho)^2 + D_{surf}(\bigtriangledown\rho_3)^2.
\label{eq:rhoE}
\end{eqnarray}
The coefficients $C_{..}$, $D_{..}$ are combinations of the traditional Skyrme parameters \cite{zheAXV2018}.
The nuclear mean-field potential $U_q$ (q = n,p), which enters the transport 
equations,  is consistently derived from the energy functional ${\cal E}$.
 %The strength of the momentum dependence is linked
%to the effective mass value.}
%In particular, the terms with coefficients $C_{eff}$ and $D_{eff}$ are the momentum dependent contributions to the nuclear 
%effective interaction.  
%{\color{red} To obtain simplified Skyrme interactions without momentum dependence, one can simply set  $C_{eff}=D_{eff}=0$.
%%%%%%%%%%%%%%%%%%%%%% mettere prima
%The Coulomb interaction is also considered in the calculations \cite{zhengPRC2016}.
%The integration of Eq. (\ref{vlasov}) is based on the test-particle (t.p.) method \cite{wong}. 
%%%%%%%%%%%%%%%%%%%%%%%%%%%%%%%%%%%%%%%%

The isoscalar section of the energy functional is usually fixed requiring that 
the saturation properties of symmetric nuclear matter, with 
a compressibility modulus around $K \approx 200-250 MeV$, are reproduced.

Many discussions will concentrate on   
the nuclear symmetry energy $E_{sym}$, that we define starting from 
the expression of the energy per nucleon: %at zero temperature:
$E(\rho,\rho_3)/A \equiv E(\rho)/A + {E_{sym}(\rho) \over A} (\rho_3/\rho)^2
 + O(\rho_3/\rho)^4 +..$.

$E_{sym}/A$ gets a
kinetic contribution directly from basic Pauli correlations and a potential
part, $C_{pot}(\rho)$,  from the highly controversial isospin dependence of 
the effective interactions. At zero temperature: 
\begin{equation}
\frac{E_{sym}}{A}=\frac{E_{sym}}{A}(kin)+\frac{E_{sym}}{A}(pot)\equiv 
C_{sym}(\rho) = \frac{\epsilon_F}{3} + {C_{pot}(\rho)} 
\end{equation}

The coefficient $C_{sym}(\rho)$ can be written as a function of the Skyrme coefficients:
\begin{equation}
C_{sym}(\rho) = \frac{\varepsilon_F}{3} + D_0\rho + D_3\rho^{\alpha+1} ~+ 
\frac{2m}{\hbar^2}\left(\frac{C_{eff}}{3} + D_{eff}\right)\varepsilon_F\rho,
\end{equation}
with $\varepsilon_F$ denoting the Fermi energy at density $\rho$ and $m$ the nucleon mass.
It is often convenient to expand the symmetry energy $C_{sym}(\rho)$ around its value
at the saturation density $\rho_0$: 
\begin{equation}
\label{sym_Tay}
C_{sym}(\rho) = S_0 + {L\over 3}\left({{\rho-\rho_0}\over \rho_0}\right) + 
{K_{sym}\over 18}\left({{\rho-\rho_0}\over \rho_0}\right)^2,
\end{equation}
where $S_0$ (ofted denoted also as $J$) indicates the symmetry energy value at
normal density, whereas $L$ and $k_{sym}$ are related to first and second
derivative, respectively.

As an example, we illustrate here some results associated with the recently introduced SAMi-J Skyrme effective interactions \cite{SAMi-J}, which provide a variety of trends for the nuclear symmetry energy.  
The corresponding parameters are determined from a 
fitting protocol which accounts for the following properties: binding energies and charge radii of some doubly magic nuclei - which allow the SAMi-J family to predict a reasonable saturation density ($\rho_0=0.159$ fm$^{-3}$), energy per nucleon $E/A (\rho_0) =-15.9$ MeV and incompressibility modulus ($K = 245$ MeV) of symmetric nuclear matter -; some selected spin-isospin sensitive Landau-Migdal parameters \cite{caoPRC2010};  the neutron 
matter EOS of Ref.\cite{wiringaPRC1988}.

\begin{figure}[h]
\includegraphics[width=20pc]{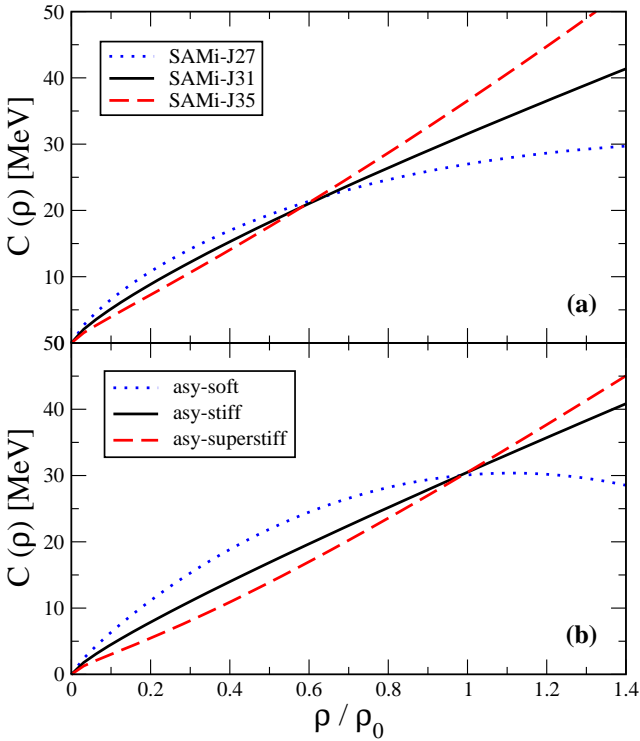}\hspace{4pc}%
\begin{minipage}[b]{17pc}\caption{\label{isoEOS}
(Color online) (a) Three effective parameterizations of the symmetry 
energy, as given by the Skyrme SAMi-J family \cite{SAMi-J}.
(b) Parametrizations corresponding to momentum independent interactions: 
asystiff (dotted line), asysoft (full line) and
asysuperstiff (dashed line) \cite{baranPR}.
}
\end{minipage}
\end{figure}

Let us consider three SAMi-J parametrizations: SAMi-J27, SAMi-J31 and SAMi-J35 \cite{SAMi-J}. 
Since, as mentioned above, Skyrme interactions are usually fitted in order to reproduce the main features of selected nuclei, 
%for the three parametrizations 
the symmetry energy coefficient takes, for the three parametrizations, the same value, $C_{sym}(\rho_c) \approx 22$ MeV, 
at the density $\rho_c \approx 0.6\rho_0$, 
which can be considered as the average density of medium-size nuclei. 
%Thus the curves representing the density dependence of $C(\rho)$ cross each other at $\rho = \rho_c$, 
%i.e., below saturation density, see the vertical dashed line 
Then the value, J, of the symmetry energy at saturation density is different in the three cases, being equal to 27 MeV (SAMi-J27), 31 MeV (SAMi-J31)
and 35 MeV (SAMi-J35), respectively. The values of the slope parameter
 $\displaystyle L = 3 \left. \rho_0 \frac{d C_{sym}(\rho)}{d \rho} \right\vert_{\rho=\rho_0}$ are equal to $L = 29.9$ MeV (SAMi-J27), $L = 74.5$ MeV (SAMi-J31)  and $L = 115.2$ MeV (SAMi-J35).  
The corresponding density dependence of $C_{sym}(\rho)$ is shown in Fig.\ref{isoEOS}(a).
It would be interesting to explore the effects of these interactions, which
provide a nice reproduction of nuclear ground state and structure properties, also in transport simulations \cite{ZhengPLB,ZhengPRC,BurrelloPRC}.      
  %\begin{figure   

For a comparison,  Fig.\ref{isoEOS}(b)
illustrates
three different parameterizations of $ C_{sym}(\rho)$ %\equiv E_{sym}/A $
corresponding to momentum independent 
Skyrme interactions widely employed in the literature: the asysoft,
the asystiff and asysuperstiff respectively, 
see \cite{baranPR} for a detailed description.
The latter interactions are mainly fitted on nuclear matter properties, 
thus the symmetry energy curves crosses at normal density ($\rho = \rho_0$).
The sensitivity of the simulation results can be tested against these 
different choices. 
We also mention that some transport models employ the following parametrization
for the potential part of the symmetry energy: 
$C_{pot}(\rho) \approx 17.5(\rho/\rho_0)^{\gamma_i} $. In this case, the stiffness
of the symmetry energy is obviously determined by the exponent $\gamma_i$.

 It should be noted that the Skyrme mean-field potential 
%consistently derived from Eq.(\ref{eq:rhoE}), 
exhibits a quadratic momentum
dependence, which can be considered as a good approximation in low momentum regions \cite{bao_PLB2015}. 
%Still concentrating on the isovector terms of the nuclear 
%effective interaction, a rather important ingredient is also
In asymmetric NM, 
the momentum dependence of the neutron/proton mean-field potentials
leads to the splitting of neutron and proton effective masses.

According to the strength of the momentum dependent terms, the SAMi-J 
interactions lead to 
an effective isoscalar nucleon mass $m^*(I = \rho_3/\rho = 0) = 0.67~m$ and a
neutron-proton effective mass splitting 
$m^*_n - m^*_p = 0.023~mI$ MeV at saturation density.
This small splitting effect is associated with a quite flat momentum dependence of the symmetry potential. 
We note that a steeper decrease at high momenta is suggested from the optical model analysis 
of nucleon-nucleus scattering data performed in \cite{bao_PLB2015}. 
This feature should not impact nuclear structure properties and low energy reactions, where 
one mainly 
explores the low-momentum region of the symmetry potential (i.e., the region
below and around the Fermi momentum).

\begin{figure}
\centering
\includegraphics[scale=1.,width=.75\textwidth]{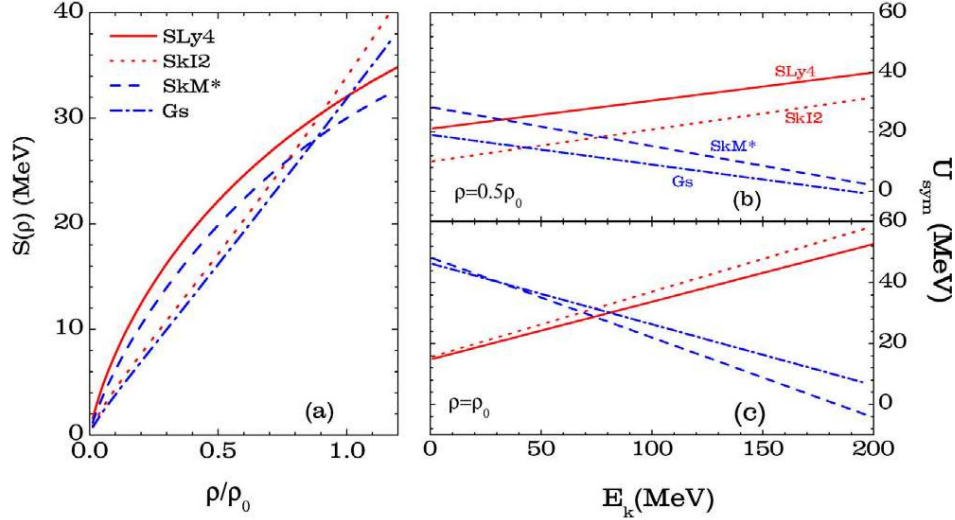}
\caption{\label{fig1_MD}
(Color online) (a) Density dependence of symmetry energy, (b) energy dependence of the Lane potential at $\rho = 0.5 \rho_0$ and (c) $\rho = \rho_0$, 
for the Skyrme parametrizations SLy4 (solid line), SkI2 (dot line), SkM* (dash line) and Gs (dot dash line).
Taken from \cite{ZhangPLB2014}.
}
\end{figure}

%In asymmetric NM, this leads 
%to the splitting of neutron and proton effective masses.
A recent illustration of several possible combinations of 
symmetry energy and effective mass splitting trends is found in Ref.\cite{ZhangPLB2014}, still in the context of Skyrme interactions. 
Fig.\ref{fig1_MD}
represents four Skyrme parametrizations which lead to a soft or a stiff
behavior of the symmetry energy.   
The right panel of the figure shows the energy dependence of the 
Lane potential $U_{sym}$, which is related to the difference between neutron and
proton mean-field potentials, at two fixed density values. 
One can see that the parameters of the Skyme interactions can be tuned
in such a way that, for a given symmetry energy trend, on can obtain
a decreasing (increasing) trend, with the energy, of the Lane potential,
corresponding to proton effective mass smaller (greater) than neutron effective
mass.  

% Fig.1 articolo Yingxun, PLB 732 (2014), Coupland et al., PRC94 (2016)
New perspectives towards a more general formulation of the density and
isospin dependence of the nuclear EOS  
are provided by metamodeling \cite{Margueron2018}.  This is  
a flexible approach that can interpolate continuously between
existing EOS and allow a more global analysis of the relation between
experimental constraints and EOS features.  
It would be interesting to follow the same philosophy for the effective
interactions employed in transport models.

The discussion about symmetry energy and effective masses is quite relevant
also within a relativistic framework.
Effective interactions which are commonly employed in the relativistic
frame are the so-called $NL\rho$ (where the isovector-scalar $\delta$
coupling is set to zero) and $NL\rho\delta$ interactions, see
Ref.\cite{baranPR} for more details. 
Actually one observes  that the two effective couplings,
vector and scalar, in the isovector channel influence in a different way
the static (symmetry energy) and dynamic (collective response, reaction 
observables) properties
of asymmetric nuclear matter. 
Hence reaction dynamics studies 
can be useful also to solve the
open problem of the determination of the (scalar- and vector-) 
isovector coupling, in connection 
to symmetry energy and effective mass splitting. 
In particular, the contributions to the scalar-isovector channel are mainly
coming from correlation effects \cite{GrecoPRC64}, thus it would
be desirable to employ, within the QHD-RMF framework, effective coupling constants
derived from microscopic 
Dirac-Brueckner-Hartree-Fock (DBHF) calculations. 
Several attempts 
have been performed, 
see \cite{JongPRC57,HofmannPRC64} for instance, %,SchillerEPJA11,MaPRC66},
 but the results still exhibit some degree of model dependence.

\section{\label{Fermi-central}Reaction dynamics at medium energy: central collisions}
\begin{figure}
%\vskip -3.cm
\centering
%\begin{tabular}%{c}
\includegraphics*[scale=0.97]{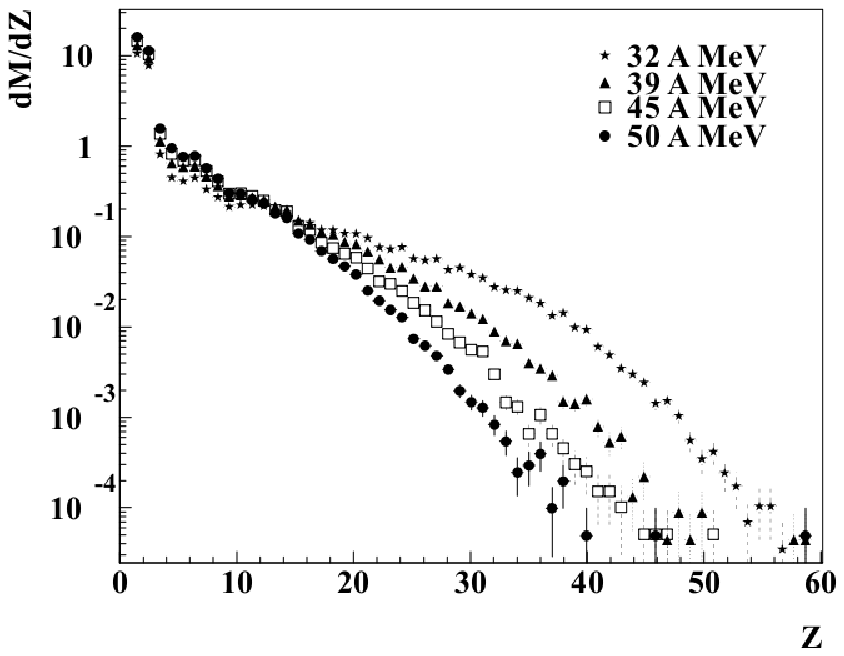}
\includegraphics*[scale=0.97]{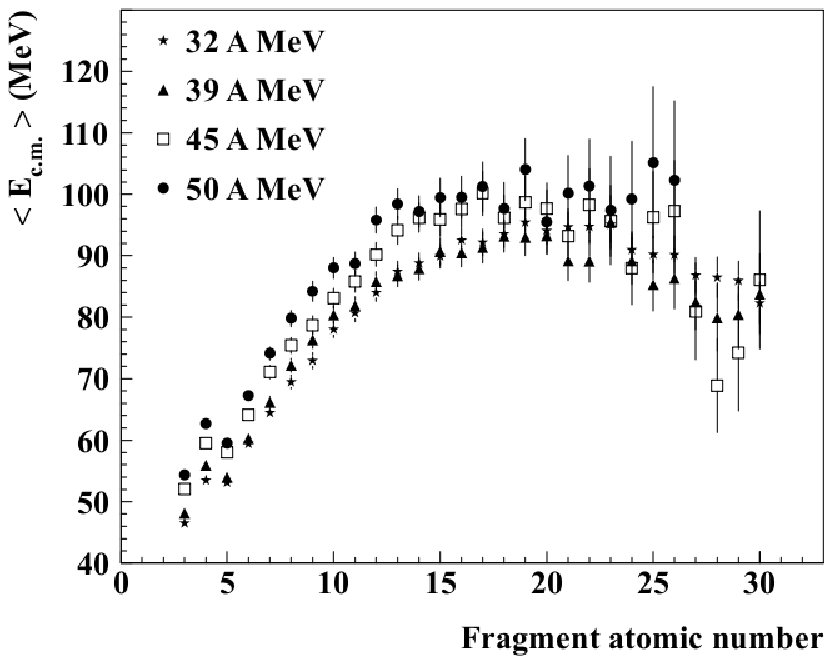}
%\end{tabular}
\caption{Left panel: Charge distributions of fragments produced in central
collisions of Xe + Sn at four incident energies: 32, 39, 45, and
50 A MeV.
Rigt panel: Average center of mass kinetic energy of the fragments
produced in the same reactions, 
%in central collisions of Xe +
%Sn reactions, at four incident energies: 32, 39, 45, and 50 A MeV 
as a function
of their atomic number. The statistical error bars are shown.
Adapted from \cite{Hudan2003}, with kind permission of the APS.
}
\label{hudan}
\end{figure}

In the Fermi energy regime (30-60 $AMeV$), 
different reaction mechanisms are explored, according to the reaction centrality, in heavy ion collisions, 
ranging from (incomplete) fusion and deep-inelastic binary processes, up to 
fragmentation of the projectile-target overlap region (the neck region) and multifragment production. 
In very central collisions, the degree of stopping is such as to lead to the 
formation of a unique composite source 
\cite{BorderiePPNP2008,BorderiePLB2018,BorderiePPNP2019,OnoPPNP2019} 
with a temperature in the range
of T$\approx$ 3-5 MeV, which eventually breaks up into
many pieces, as a result of thermal effects and of the compression-expansion dynamics. 
Indeed, as a quite interesting feature, this process is also accompanied 
by the development of a radial flow, which characterizes the kinematical
properties of the reaction products \cite{Hudan2003,Bonnet2014}. 
These features are summarized in Fig.\ref{hudan}, which represents the
evolution of fragment charge distribution and kinetic energies, as measured
for the system $^{129}$Xe + $^{nat}$Sn at different beam energies.  
From the left panel one can appreciate how the fragmentation process evolves
with the energy transferred to the system, with smallest fragments produced
at the highest beam energy.  
The right panel of the figure shows clearly that, excluding from the analysis
the heaviest
fragments, the mean kinetic energy per nucleon exhibits an almost linear
increase with the fragment charge. This trend can be interpreted as an
evidence of the presence of a radial collective flow, leading to 
similar radial velocities for all fragments (and thus to a kinetic 
energy scaling with the fragment mass).  

From the experimental observations outlined above, one can already figure out
that the occurrence of surface and/or volume instabilities may play a central role in the description of the reaction path. 
Simulations of the multifragmentation dynamics based on 
transport models have allowed, through the comparison with experimental 
data, to shed light on the compression-expansion dynamics, yielding
independent information on the nuclear matter compressibility
\cite{NapolitaniEPJ}.
We illustrate here a selection of simulation results related to central 
collisions at Fermi energies. Emphasis will be put on the features
of the fragmentation process, in connection to the treatment employed to
describe the many-body dynamics, and on aspects related to isospin transport
and isotopic features of the reaction products.

\subsection{\label{pre-eq}Pre-equilibrium nucleon and cluster emission}

During the first stage of the reaction, hard two-body scattering plays an essential role 
and pre-equilibrium emission is observed, i.e. 
nucleons and light particles are promptly emitted
from the system. 
One may expect that two- and higher order correlations play
an important role in determining mass, isotopic properties and kinematical
features of the particles emitted. 
Moreover, it is easy to realize that this stage influences significantly also 
the following evolution of the collision. Indeed, the amount of particles and energy
removed from the system affects the properties of the composite source that eventually
breaks up into pieces. Hence, when discussing multifragmentation mechanisms, a detailed
analysis of this early emission is in order.
Moreover, pre-equilibrium emission, involving energetic particles, is particularly
sensitive also to the momentum dependence of the nuclear effective interaction. 

To get an insight into the theoretical description of the reaction dynamics, 
it is quite instructive to compare the results
obtained, for selected collisions at Fermi energy, within the scheme of
the AMD and SMF models (employing similar 
momentum dependent interactions and n-n cross section parametrizations 
in the two cases). 
This analysis was performed in Ref.\cite{ColonnaPRC2010}, %Akira-comparison
%PHYSICAL REVIEW C82, 054613 (2010)
for the neutron-poor $^{112}Sn+^{112}Sn$ and neutron-rich
$^{124}Sn+^{124}Sn$ reactions, at 50 MeV/nucleon.
As far as the early emission of 
%the number of neutrons and protons contained in
nucleons and light clusters is concerned, %(with  mass number $A\leq 4$),
one observes that these particles (with  mass number $A\leq 4$) 
leave the system mostly in the
time interval between $\approx$ 70 -120 fm/$c$ %and $\approx 120$ fm/$c$
(see also Fig.\ref{explosion}).    
%{\bf and can be attributed to evaporative processes.}
%are essentially free nucleons and light clusters (with mass number $A\leq 4$).
%In fact, at later times, intermediate mass fragments (IMF) start to appear and the light cluster 
%emission rate decreases. 
%Let us focus at the moment on this early nucleon and light cluster emission. %the so-called pre-equilibrium emission.     
%In charge-asymmetric systems, 
A striking difference between the two models
concerns the amount of these emitted particles, that is larger in the SMF case,
though the average kinetic energy of this emission is similar in the two models
(being 20.72 MeV/nucleon in SMF and  21.95 MeV/nucleon in AMD) \cite{ColonnaPRC2010}. 
%%%%%%%%%%%%%%%%%%%%  sistemare qui
Moreover, mainly unbound nucleons are emitted   
in the SMF case.  
However, it should be noticed that, in AMD, also light fragments, 
with mass number $5 \leq A\leq 15$,  %emitted at later times, 
leave the system  during the early stage of the reaction, on equal footing as light particles. This is not observed in SMF. 
%%[[ONO:
%%Is this about nucleon emission?  Then MeV/nucleon => MeV]].
%% There are also particles with A<5.
%%, in favour of bigger particles.
%%In fact, 
%This more abundant emission in SMF is %of very light particles in SMF is 
%compensated by a larger emission of
%%On the other hand, %as it will be discussed in the following, 
%light fragments, with mass number $5 \leq A\leq 15$,  %emitted at later times, 
%in AMD, which leave the system  on equal footing as light particles.  
Thus, as a matter of fact, 
the total amount of mass which escapes from the system 
is close in the two models. 
%it is observed that 
%%inspite of the different nucleon emission mechanism
%%and clustering effects, 
%%%that leads to a different mass distribution
%%%of 
%the total amount of emitted nucleons belonging to particles 
%with  $A \leq 15$ (including free nucleons) %, the total emitted mass
%is close in the two models. 
%As a consequence, we will also get a 
%similar production of
%intermediate mass fragments (IMF), with charge  $A > 15$, as it 
%will be discussed in the following.
As a consequence, we also expect a similar global amount of mass going into 
the production of
intermediate mass fragments (IMF), with mass number $A > 15$, as it 
will be discussed in the following.

%NO These features are illustrated in Fig...%figura articolo Akira
As argued in Ref.\cite{ColonnaPRC2010},
the difference observed  %amount of emission 
between the two models %, in connection to 
for single nucleon and light
particle emission  can be ascribed to the fact that 
clustering effects and many-body correlations 
are more efficient in AMD, due to the nucleon localization,
reducing the amount of mass that goes
into very light reaction products. 
Some effects may also be connected to a different compression-expansion
dynamics in the two models, as we will discuss below.
Clustering effects are under intense investigation nowadays, in a variety of
contexts, including astrophysical environments \cite{Burrello2015},
and it deserves much attention to get new insights from 
multifragmentation events \cite{OnoPPNP2019,OnoCim2016,Daniel,Natowitz2010}.

\subsubsection{\label{pre-iso}Isospin effects}
\begin{figure}
\centering
\includegraphics[scale=1.2]{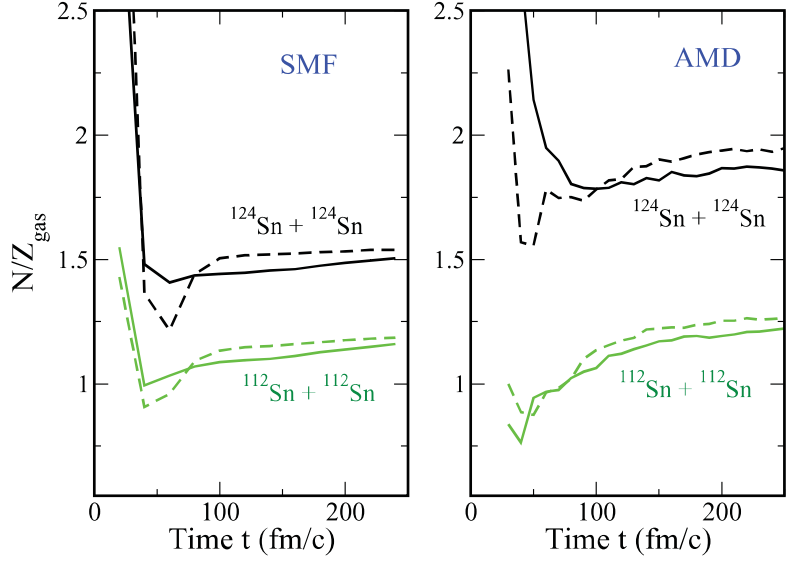}
\caption{\label{fig1_Akira}
(Color online) Time evolution of the N/Z content of the pre-equilibrium emission, as obtained in SMF (left panel) and AMD (right panel) models, 
for the $^{112}$Sn + $^{112}$Sn and $^{124}$Sn + $^{124}$ Sn reactions. 
Solid lines, asystiff interaction; dashed lines, asysoft.
Taken from \cite{ColonnaPRC2010}.}
\end{figure}

A lot of interest in the last years has been directed also to the isotopic
composition of the pre-equilibrium emission.  This appears as 
a rather interesting  
feature, which is expected to be particularly sensitive to the 
isovector channel of 
the effective interaction and, namely, to the symmetry energy.   
Because the symmetry energy is density dependent, this kind of
observables are also rather useful to track the reaction dynamics itself. 
Moreover, also isotopic features are expected to be sensitive, generally
speaking, to 
correlations and clustering effects. 
The comparison between AMD and SMF results, for 
$^{124}$Sn + $^{124}$Sn and $^{112}$Sn + $^{112}$Sn reactions at 50 MeV/A
 is shown in Fig.\ref{fig1_Akira}, when employing
a soft or a stiff symmetry energy parametrization. 
%%%%%%%%%%%%%%%%  mettere fig.12 dell'articolo con Akira
One can observe that the more abundant pre-equilibrium emission %(particles
%with $A<4$)
obtained in SMF is associated with a lower N/Z value, with respect to the AMD results,
especially for the neutron-rich system. 
Both effects could be connected to the more pronounced 
clustering effects in AMD and/or the
different density conditions explored along the reaction path. Indeed, 
if cluster production is favored, protons are more likely bound in it.
%As a general feature, one can observe that the N/Z content of the pre-equilibrium emission is more abundant in the asysoft case, reflecting the fact that
%these particles are mostly originating from low-density regions, where the
%symmetry energy is higher in the soft case. 

%%%%%%%%%%%%%%%%%%%%%%%  sistemare questo pezzo
Some general considerations emerge from the results displayed in 
the figure: 
when comparing the two reactions, it is seen that neutron (proton) emission is more
abundant in the neutron-rich (poor) systems, as expected.   Moreover, 
the figure also shows that a larger (smaller) number of neutrons (protons)
is emitted in the asysoft case, as compared to the asystiff case. 
This is consistent with the  larger repulsion of the symmetry potential for the soft parametrization below normal density.   
%This is observed in both models and for both reactions.
Hence this result confirms that 
pre-equilibrium particles leave the system 
%are mostly emitted  
from regions that are at sub-normal density, 
%(i.e. during the expansion phase), 
where the symmetry
energy is higher in the soft case (see Fig.~3).
Inspite of these common features, one can easily notice 
that the quantitative differences between the AMD and SMF 
result obtained for a given parametrization  are larger than the
differences given by the two parametrizations.
This observation highlights the important impact of the description
of the many-body dynamics, namely the interplay between mean-field and
correlation effects, even on isospin observables. 
A correct description of clustering effects looks crucial 
in order to extract reliable
information on the low-density symmetry energy behavior from pre-equilibrium
observables.

\begin{figure}
%\vskip -1.cm
\centering
\includegraphics[scale =1.3]{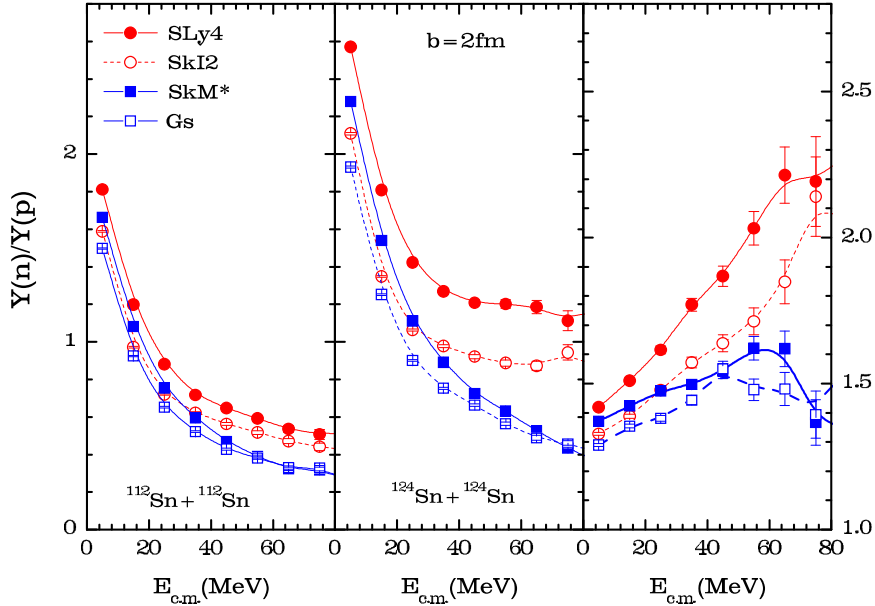}
\caption{\label{fig2_MD}
(Color online) Left panel: Neutron-proton ratio Y(n)/Y(p) as a function of the kinetic energy for $^{112}$Sn + $^{112}$Sn at b= 2 fm with angular cuts
$70^0 < \theta_{c.m.} < 110^0$; Middle panel: the same quantity, 
 for $^{124}$Sn + $^{124}$Sn. 
Right panel: Double ratios, DR(n/p), as a function of kinetic energy. 
The results shown on the figure are for SLy4 (solid circles), SkI2 (open circles),SkM* (solid squares) and Gs (open squares).
Taken from \cite{ZhangPLB2014}.}
\end{figure}

An insight into effects related to the momentum dependence 
of the nuclear interaction is got by looking at the
N/Z ratio versus the kinetic energy of the pre-equilibrium 
emitted particles. 
As an illustrative example, we show here the results discussed in 
Ref.\cite{ZhangPLB2014}, obtained with the ImQMD model, 
see Fig.\ref{fig2_MD}.
%%%%%%%%%%%%%%%%%%%%%  mettere Fig.3 di Physics Letters B 732 (2014) 186–190
The figure displays the ratio of the the yields of neutrons and protons, 
for the same reactions considered above (at 50 MeV/nucleon), as a function of the kinetic energy. 
Four Skyrme effective interactions are employed, 
corresponding to different symmetry energy parametrization (stiff or soft) 
and different splitting of the proton/neutron effective masses
(see also Fig.\ref{fig1_MD}).
One observes that, whereas the N/Z ratio of the particles emitted at 
low kinetic energies is governed
by the low-density symmetry energy behavior (being larger in the soft case), the trend at 
large kinetic energy reflects the effective mass splitting sign. 
Higher N/Z values (see especially the case of the 
$^{124}$Sn + $^{124}$Sn system, middle panel) 
are associated with parametrizations 
having $m^*_n < m^*_p$.  Indeed the latter case corresponds to a more
repulsive symmetry potential for neutrons. 
This effect is particularly evident on the right panel, which represents
the double ratio DR(n/p), i.e. the ratio between the N/Z
content of the pre-equilibrium emission obtained in the two systems 
considered 
(the neutron-rich $^{124}$Sn + $^{124}$Sn and the neutron poor 
$^{112}$Sn + $^{112}$Sn).
\begin{figure}[h]
%\vskip -5.cm
\centering
\includegraphics[scale=1.]{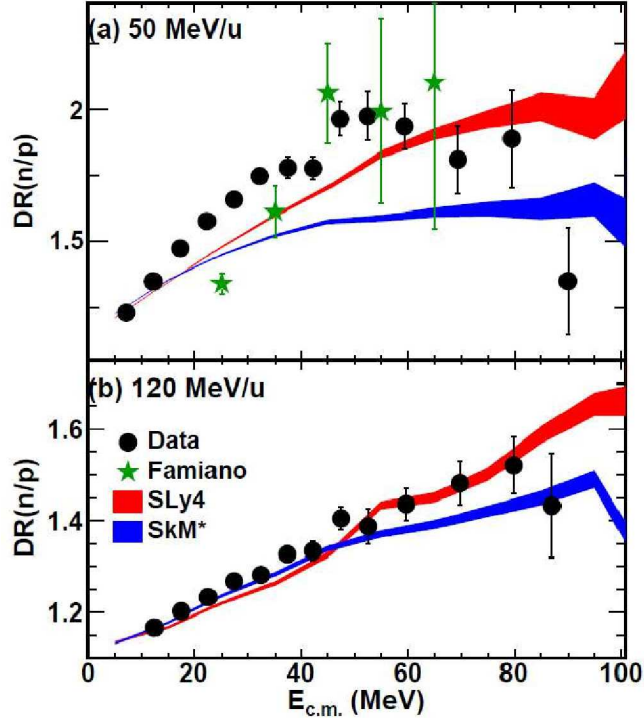}
\caption{\label{fig_coup}
 (Color online) Neutron to proton double ratio, for Sn + Sn collisions at 
beam energies $E_{beam}/A$=50 MeV (a) and 120 MeV (b).
Experimental data are confronted with ImQMD transport calculations employing 
two different Skyrme interactions. 
Reprinted from \cite{CouplandPRC94}, with kind permission of the APS.}
\end{figure}  
There have been recent attempts to compare predictions related to pre-equilibrium
observables to experimental data, aiming at extracting information
on symmetry energy and effective mass splitting sign at once.
We report here, as an example, the results of Ref.\cite{CouplandPRC94},
where simulations are still performed 
with the ImQMD model, employing the
``soft'' parametrizations of Fig.\ref{fig1_MD}. 
The double ratio DR(n/p),
%i.e. the ratio between the N/Z
%content of the pre-equilibrium emission obtained in the two systems 
%considered before
%(the neutron-rich $^{124}$Sn + $^{124}$Sn and the neutron poor 
%$^{112}$Sn + $^{112}$Sn) 
evaluated for the same systems considered above  
%The analysis is done for the same systems considered before, 
at two beam energies, 50 and 120 MeV/nucleon, is represented in 
Fig.\ref{fig_coup}. 
%coupland fig.2 da  Coupland et al., PRC94 (2016), 011601
The comparison 
%of the double ratio, i.e. the ratio between the N/Z
%content of the pre-equilibrium emission obtained in the two systems 
%$^{124}$Sn + $^{124}$Sn and $^{112}$Sn + $^{112}$Sn reactions, 
to the 
corresponding experimental data seems to favour the $m_n^* < m_p^*$ scenario. 
As already discussed above, in this case an increasing trend of the double 
n/p ratio with the kinetic energy is expected (see Fig.\ref{fig2_MD}).

%%%  for pre-equilibrium:   articolo comparison con Akira,
%% Napolitani, Colonna,  PRC (2017) (figure finali)
% articolo Yingxun, PLB 732 (2014), Coupland et al., PRC94 (2016)
%% articolo Joseph PRC72, 064609 (2005)
%% citare Bao-An, PPNP 99 (2018) 29

%%%%%%%%%%%%%%%%%%%%%%%%%%%%%%%%%%%%%%%%%%%%%%%%%%%%%%%%%%%%%%%%%%%%%%%%%%%%
\subsection{\label{comp}Multifragment emission}%as it will

\begin{figure}[h]
%\vskip 1.cm
\centering
\includegraphics{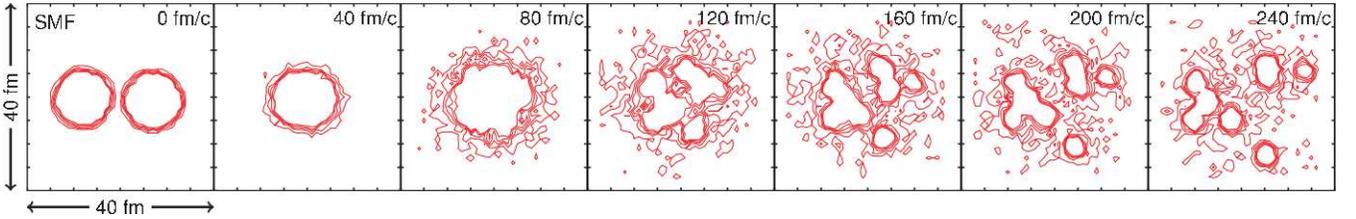}
\caption{\label{explosion}Contour plots of the density projected on the reaction plane, calculated with the SMF model, for the
central reaction $^{112}$Sn + $^{112}$Sn at 50 MeV/nucleon, at several times (fm/c). The lines are drawn at projected densities beginning at 0.07 fm$^{−2}$
and increasing by 0.1 fm$^{−2}$. The size of each box is 40 fm.
Taken from \cite{ColonnaPRC2010}.}
\end{figure}

As a result of the initial compression/expansion dynamics and/or thermal effects,
the composite sources formed in central heavy ion collisions %at Fermi energies 
may reach low density values, attaining the co-existence zone of the
nuclear matter phase diagram. 
A qualitative illustration of the corresponding reaction path is
given in Fig.\ref{explosion} for the system $^{112}$Sn + $^{112}$Sn at 50 MeV/nucleon. 
%For instance, an excited system that expands under
%the conditions of thermal equilibrium could perform a phase transition 
%staying close to the liquid branch of the co-existence line \cite{mononuclear,M%oretto}.
%%%%modif aggiungere altre referenze  
%However, due to the Coulomb instabilities, the
%limiting temperature, that a nucleus can sustain as a compact configuration, 
%may be lower than the critical temperature \cite{Nato}.

In this situation, as a possible scenario (see the recent review, 
Ref.\cite{BorderiePPNP2019}), the system
%is brought inside the co-existence zone of the nuclear matter phase diagram and
undergoes a spontaneous phase separation, breaking up into several fragments, 
 as a consequence of the occurrence of 
 mean-field spinodal instabilities \cite{Colonna1994,chomaz2004}. 
This scenario is supported by simulations performed with stochastic mean-field
models \cite{Frankland2001}. 

\begin{figure}
\centering
\includegraphics[scale=1.2]{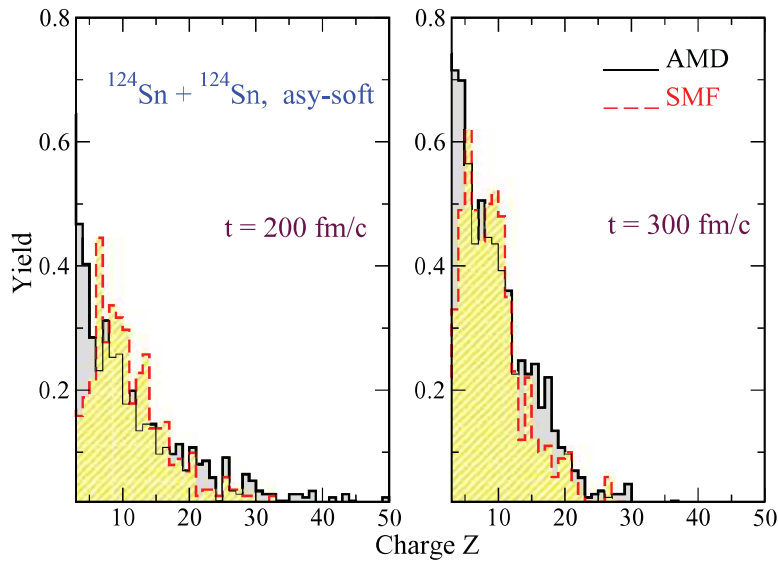}
\caption{\label{fig_8_akira}
(Color online) Charge distribution as obtained in AMD (solid histogram) and in SMF (dashed histogram), for the reaction
$^{124}Sn+^{124}Sn$ 
at the time instant t=200 fm/c (left) and t=300 fm/c (right).
Taken from \cite{ColonnaPRC2010}.}
\end{figure}

It should be noticed, however, that in the initial high density phase nucleon correlations 
are expected to be rather large, owing to the huge amount of
two-body nucleon-nucleon collisions.  Hence some memory of these high density correlations
could be kept along the fragmentation process,
%According to the theoretical description outlined above, 
even assuming that clusters emerge essentially from the occurrence of mean-field instabilities. In any case, fluctuations of the one-body density induced 
by two-body scattering provide, at least, the initial 
seeds for the nucleon assembly into clusters.  
%many-body correlations still play an essential role because they provide the seeds for the nucleon assembly into clusters.  
As already discussed in the case of pre-equilibrium observables, 
an insight into the interplay between mean-field and correlation effects
can be got comparing the results of extended 
BUU-like models, including fluctuations, and molecular dynamics models. 
In the following we come back to the comparison performed between AMD and SMF calculations in Ref.\cite{ColonnaPRC2010}, briefly discussing the main results.   
%In this paper we have undertaken a quantitative comparison of the features observed in
%heavy-ion fragmentation reactions at Fermi energies, as predicted by two transport 
%models: SMF and AMD. 
%As far as observables of experimental interest are concerned,
%one significant  discrepancy  between the two models is connected to the amount of
%pre-equilibrium emission, i.e.\ the energetic particles that leave the system a%t the early
%stage of the reaction.
As already pointed out in subsection \ref{pre-eq}, in the AMD approach clustering effects appear to be more relevant, 
reducing the amount of free nucleons emitted, compared to SMF, in favor of  
a richer production of primary light IMFs. 
However, it is quite interesting to notice that the yield of sizeable primary IMFs ($Z>6$) is rather close in the two models, likely keeping the fingerprints of low-density mean-field dynamics. 
This is shown in Fig.\ref{fig_8_akira}, %figura confronto SMF-AMD
which compares the charge distributions obtained %in the two models 
for the reaction $^{124}Sn+^{124}Sn$ at 50 MeV/A. 
%%%%%%%%  mettere figura 8 dell'articolo con Akira
Looking more in detail, in SMF fragments with charge around $Z = 10$ are slightly more abundant, while in AMD
the tail at larger $Z$ (around 20) is more pronounced. 
The shape of the SMF charge distribution is closer to the expectations of spinodal decomposition
\cite{chomaz2004}.  However, it should be noticed that
these differences are likely smoothened by secondary decay effects.  In fact, both models are 
able to fit experimental IMF charge distributions reasonably well \cite{ono2002,Frankland2001}.   
A good reproduction of experimental data is
provided also by QMD calculations, see Ref.\cite{Zbiri2007}. 
A more refined analysis, based on event-by-event fragment correlations
would be needed to disentangle among possible different fragmentation scenarios
\cite{BorderiePLB2018}.

Major differences among models are connected to fragment kinematical properties: 
for instance, in SMF fragment kinetic energies are smaller, compared to AMD, by about 20$\%$.
%see Fig.\ref{fig_11_akira}
%%figura confronto SMF-AMD per le energie cinetiche.  
%%%%%%%%%%%%%%%%%  mettere figura 11 dell'articolo con Akira
%\begin{figure}
%\centering
%\includegraphics{fig11_akira.ps}
%%\includegraphics{fig-others/rios2011-2dfis.pdf}
%\caption{\label{fig_11_akira}
%(Color online) IMF average kinetic energy as a function of the charge Z, as obtained in AMD and SMF models at t=200 fm/c (dashed lines) and at t=600 fm/c 
%(solid lines), for the same reaction of Fig.\ref{fig_8_akira}. 
%The solid and hatched gray areas mark the change during this time interval for AMD and SMF, respectively.
%}
%\end{figure}
These observations corroborate the scenario of a faster fragmentation process in
AMD, while in SMF the system spends a longer time as a nearly 
homogeneous source
at low density, thus quenching radial flow effects before fragment formation sets in \cite{Bonnet2014,Frankland2001}. 
%emitting a larger amount of nucleons prior to fragment formation.
%Moreover, SMF kinetic energies are smaller than observed experimentally for
%similar systems \cite{Bonnet2014,Frankland2001}, see also Fig.\ref{hudan}. 
This delay in the fragmentation process, probably associated with the approximate treament of fluctuations
in SMF, could be overcome in upgraded stochastic mean-field 
models, introduced more recently \cite{Napolitani2013,Hao2019}.
We review here some results obtained with the BLOB model.
 
\begin{figure}[b!]
\centerline{
	\includegraphics[angle=0, width=.85\textwidth]{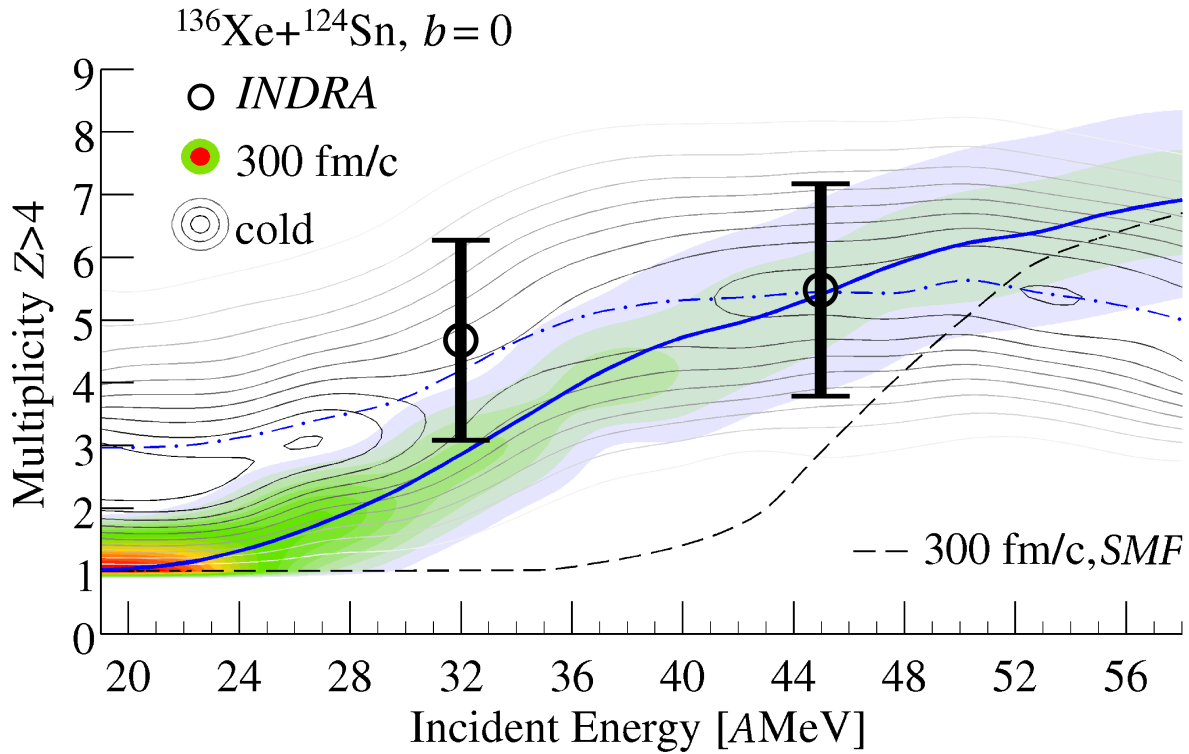}
}
\caption{(Color online)
	BLOB simulation: evolution of IMF ($Z>4$)  multiplicity, as a function of incident energy for the system $^{136}$Xe$+^{124}$Sn and a selection of central collisions at 300fm/c (colour shades) and for the cold system (grey contours). Corresponding mean values are indicated for the BLOB simulation 
(full blue line for primary fragments and dot-dashed line for final fragments) and for a SMF calculation (black dashed line, primary fragments).
%(adapted from ref.~\cite{Napolitani2013}\cite{clusters}).
	Corresponding experimental data from Indra experiments~\cite{Moisan2012,Ademard2014} are added for comparison, with average (symbols) and 
variance (bars) of the multiplicity distributions. Taken from  
\cite{clusters}. 
%a gaussian fitted to the multiplicity distributions (bars).
}\label{fig_frag_threshold}
\end{figure}

The latter was conceived with the purpose of including fluctuations in full 
phase space, thus improving the treatment of fluctuations and correlations, %with a large amplitude, 
but preserving, at the same time, mean-field features such as the 
proper description  of spinodal instabilities at low density \cite{Napolitani2017}. 
%Landau dispersion relation for unstable modes
%which would grow in nuclear matter for the corresponding employed nuclear interaction.
%%%%%%%%%%%%%%%%%%%%%%%%%%%%%  testo preso da Udo
The improvement introduced by the BLOB approach is primarily providing a correct sampling of the fluctuation amplitude in full phase space, yielding a 
faster fragmentation dynamics and also a consistent description of the threshold toward multifragmentation.
Fig.~\ref{fig_frag_threshold} 
%%%%%%%%%%%%%%%%%%%%%%%  mettere Figura 2 di Udo
shows results of simulations performed for the system $^{136}$Xe$+^{124}$Sn, at several incident
energies, analysed for central impact parameters at the time t = 300fm/c. 
It is observed that fragmentation events start competing with the predominant 
low-energy fusion mechanism already at around 20 MeV/nucleon of beam energy. The multiplicity of the primary 
IMF with $Z>4$, evaluated at 300 fm/c (full blue line), tends 
to grow with the beam energy, however a maximum at around 45 MeV/nucleon 
is observed
when considering cold fragments, i.e. after the secondary
de-excitation stage has been taken into account.  (The cooling of the hot system is undertaken by the use of the decay model Simon~\cite{Durand1992}.) %PN[moved to this place]
On the other hand, calculations performed with SMF, also shown on the figure, 
tend to overestimate the
energy threshold toward multifragmentation. 
%Two experimental points, from INDRA~\cite{Moisan2012,Ademard2014} shown in the figure indicate that the BLOB simulation is quantitatively consistent when compared to the cold distribution. %PN[added]
%PN2
%
%	--- FIGURE 2, Fragmentation threshold 
%

In the figure, two experimental points from INDRA~\cite{Moisan2012,Ademard2014} indicate the IMF multiplicity extracted from the analysis of central 
compact sources; they can be compared to the calculated cold distribution and indicate that the BLOB simulation, performed employing a soft 
EOS (with compressibility K = 200 MeV), is quantitatively consistent.
\begin{figure}
\centering
\includegraphics[scale=1.2]{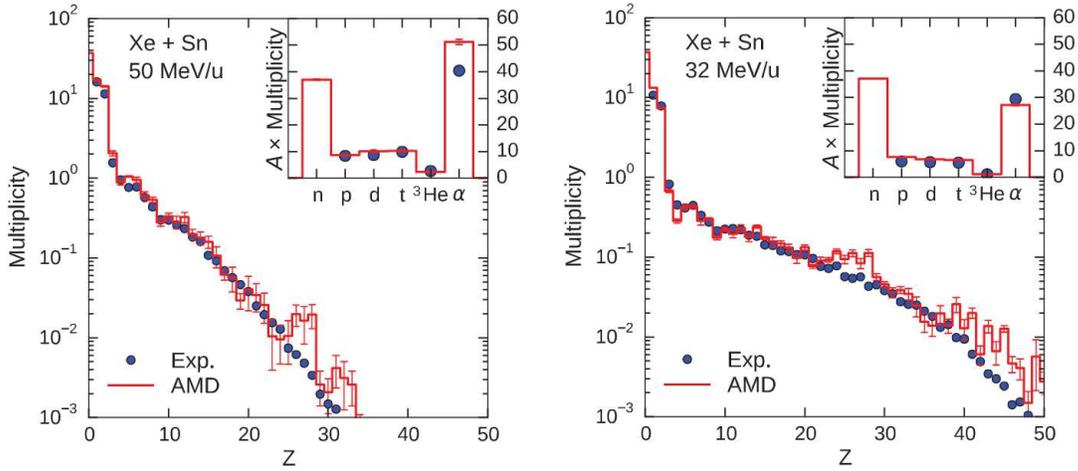}
\caption{\label{Ak_IWM16}
(Color online) Fragment charge distribution in central Xe+Sn collisions 
at the incident
energies of 50 (left) and 35 (right) MeV/nucleon, calculated by AMD with cluster correlations.
The inset shows the multiplicities of light particles multiplied by the mass number. The INDRA
experimental data are taken from ref.\cite{Hudan2003}. Taken from
\cite{OnoCim2016}.}
\end{figure}
%PN2
%Additional details on the comparison between theory and 
%experiment~\cite{Moisan2012,Ademard2014} are provided in fig.~\ref{fig_M_spectra}, where the difference between SMF and BLOB is investigated around the multifragmentation threshold (32$A$MeV) and in full multifragmentation regime (45$A$MeV).
%%An example of comparison between theory and 
%%experiment~\cite{Moisan2012,Ademard2014} is given in fig.~\ref{fig_M_spectra}, where the difference between SMF and BLOB models is tested around the multifragmentation threshold (32$A$MeV) and in full multifragmentation regime (45$A$MeV).
%One can observe that the BLOB model, employing a soft 
%EOS (compressibility K = 200 MeV), is suitable to give a reasonable reproduction of the IMF multiplicities measured by INDRA, 
%even at the lowest beam energy.   
% %The cooling of the hot system (300fm/c) is undertaken by the use of the decay model Simon~\cite{Durand1992}. %PN[moved away]
%%%%%%%%%%%%%%%%%%%%%%%%%%% mettere fig.3 di Udo
The faster BLOB dynamics also leads to a better description of 
fragment kinetic energies, as recently pointed out in 
Ref.\cite{Napolitani2019} in the case of 
fragmentation events emerging from mass-asymmetric reactions at Fermi energies,
of recent experimental interest. 

New improvements have been reported also in the case of the latest version
of the AMD model, which includes a refined treatment of light cluster 
dynamics \cite{OnoJPG2013,OnoCim2016}.
Fig.\ref{Ak_IWM16} reports a comparison of 
the calculated charge distribution,
for central collisions (with impact parameter $0 < b < 2~fm$),  with the INDRA 
data, for the system  $^{129}$Xe + $^{nat}$Sn at 32 (right panel) and 50 (left panel)
MeV/nucleon.    
It is interesting to see that the inclusion of cluster correlations leads
to a quite good reproduction  of the whole charge distribution spectrum, including protons
and very light clusters \cite{OnoCim2016}.

%
%
%	--- FIGURE 3, multiplicity data
%
%\begin{figure}[t!]
%\centerline{
%	\includegraphics[angle=0, width=.7\textwidth]{fig3_Mspecra_Gauss.eps}}
%\caption{
%	BLOB and SMF simulations: IMF ($Z>4$) multiplicity distributions for the systems $^{136}$Xe$+^{124}$Sn at 32 and 45$A$MeV, for a selection of central collisions at 300fm/c and after secondary decay
%are compared to experimental data from Indra, with average multiplicity and variance given in Refs.~\cite{Moisan2012,Ademard2014}. 
%The experimental distributions correspond to the bars shown in fig.~\ref{fig_frag_threshold}. From \cite{clusters}%PN2
%}\label{fig_M_spectra}
%\end{figure}

%%%%%%%%%%%%%%%%%%%%%%%%%%%%%%%%%%%%%%%%%%%%%%%%%%%%%%%%%%%%%%%%%%%%%%%%%%%%

%Unstable isoscalar modes can be generated, allowing to develop bifurcations in the dynamical evolution of the system and mechanisms of amplification like the spinodal instability.
\subsubsection{Isotopic features}
\begin{figure}[h]
\includegraphics[width=20pc]{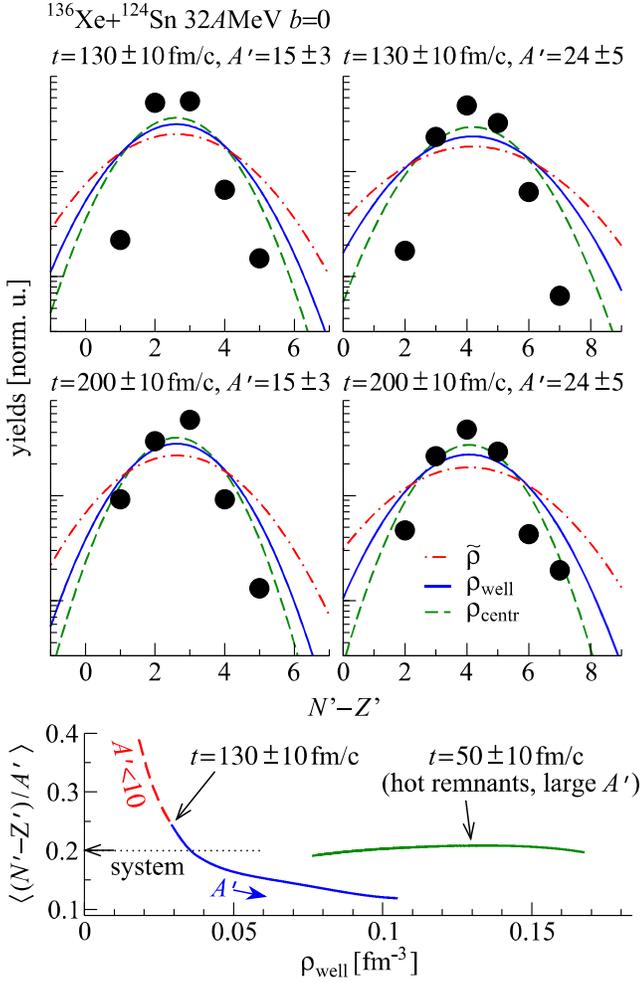}\hspace{3pc}%
\begin{minipage}[b]{17pc}\caption{\label{fig_PRC_2017}
(Color online)
BLOB simulation of head-on $^{136}$Xe+$^{124}$Sn collisions at 32AMeV. (upper row) Isotopic distribution (full dots) in potential ripples containing 
N′ neutrons and Z′ protons for cluster-forming configurations with masses aroundA′=15 (left) and A′=24 (right) at t=130 fm/c. The lines correspond to analytic distributions (see text), corresponding to the density extracted from different portions of potential ripples. (middle row) The same as in the upper row, but at t=200 fm/c. (bottom) Average isospin content measured in potential ripples at two different times, as indicated on the figure, as a function of
the ripple density $\rho_{well}$. Larger density values correspond to 
prefragments of larger mass. The dot-dashed line with an arrow indicates the asymmetry of the projectile-target composite system.
Taken from \cite{Napolitani2017}.}
\end{minipage}
\end{figure}

It is quite well established that, in neutron-rich systems, the fragment formation mechanism also keeps the fingerprints of the isovector channel of the
nuclear effective interaction, in connection to the symmetry energy term
of the EOS \cite{bao-an2008,EPJA_Colonna}. 
 Indeed one observes that the clusters (liquid drops) 
which emerge from the low-density nuclear matter have a lower $N/Z$ ratio, 
with respect to the surrounding nucleons and light particles.
This effect, the so-called isospin distillation (or fractionation), is connected to the density derivative of the symmetry energy and leads to the minimisation of the system potential energy \cite{Xu2000,Larionov1998,baranPR}.   
Fluctuations act on both isoscalar and isovector degrees of freedom.
On top of the average isospin distillation mechanism, isovector 
fluctuations generate the isotopic distribution of the fragments observed.
Combining the two features, namely the average fragment N/Z and the 
corresponding variance, it should be possible to grasp the behavior of the
low-density symmetry energy, see the analyis in Ref.\cite{Ono2004}. This has been done also resorting to 
the so-called isoscaling analysis \cite{Xu2000,Kohley2014}.   
As an illustrative example, Fig.\ref{fig_PRC_2017} 
%%%%%%%%%%%%%%%%%%%% mettere   figura dell'articolo PRC 2017  ~\ref{fig_phase_trans_b} 
shows the action of the isovector terms on the system  $^{136}$Xe+$^{124}$Sn 
(at 32 MeV/A), looking at fragment isotopic features. %{\bf define ! stiff or soft ?}
Results obtained with BLOB, employing an asystiff EOS, are displayed
\cite{Napolitani2017}.   
  %\begin{figure   

%\begin{figure}[h]
%%\vskip -2.45cm
%\centering
%\includegraphics[scale=0.9]{prova1.ps}
%%\includegraphics{fig-others/rios2011-2dfis.pdf}
%\caption{\label{fig_PRC_2017}
%BLOB simulation of head-on $^{136}$Xe+$^{124}$Sn collisions at 32AMeV. (upper row) Isotopic distribution (full dots) in potential ripples containing 
%N′ neutrons and Z′ protons for cluster-forming configurations with masses aroundA′=15 (left) and A′=24 (right) at t=130 fm/c. The lines correspond to analytic distributions [see Eq.(36)], corresponding to the density extracted from different portions of potential ripples (see text). (middle row) The same as in the upper row, but at t=200 fm/c. (bottom) Average isospin content measured in potential ripples at two different times, as indicated on the figure, as a function of
%the ripple density $\rho_{well}$. Larger density values correspond to 
%prefragments of larger mass. The dot-dashed line with an arrow indicates the asymmetry of the projectile-target composite system.
%\cite{Napolitani2017}}

%\end{figure}

To explore the behavior of the system as a function of the local density, 
ripples of the mean-field potential, which can be considered
as pre-fragments, are identified at several time instants along the
dynamical evolution of the nuclear reaction. 
Then isospin effects are investigated as  
a function of the average local density and for different ripple sizes. 
Different sets of potential-concavity sizes are indicated by extracting a corresponding mass $A'$.
As one can see from the bottom panel of the figure, 
from an initial situation where the system is close to saturation density and the average isospin is determined by the mixing of target and projectile nuclei, densities drop to smaller values and the isospin distribution extends over a large range.
In particular, the smaller is the local density, the larger is the 
neutron content measured in corresponding sites, as neutrons favour the most volatile phase.
%Such ordering has been intensively studied within the SMF approach~\cite{baranPR}. %\PN2[HAS been]
%However, this second isovector observable can only be exploited if isoscalar fluctuations have the large amplitude required to disassemble the system into clusters and fragments.

%
The top panel of the figure shows the isotopic distribution obtained, at two different times, for fragments in two different mass ranges $A'$.  
Considering that, at the instants considered, thermal equilibrium is attained, 
the distribution of the 
difference $\delta = (N' - Z')$
with respect to its average is expected to behave as: $P(\delta) \approx
exp[-(\delta^2/A')C_{sym}(\rho_{well})/T]$, thus reflecting the temperature T 
and bringing information on the symmetry energy value 
at the corresponding 
fragment density $\rho_{well}$ (full blue lines on the figure).
Temperatures of the order of T = 3 MeV are extracted from the simulations.  
One can observe that the BLOB calculations become closer to the analytical predictions
at the largest time instant considered, though they remain narrower.  
This indicates that, whereas the BLOB fluctuation treatment looks efficient in 
out-of-equilibriums and/or unstable situations, providing the seeds for fragment formation, 
some refinements are still needed to properly
describe the fluctuations characterizing the following (thermodynamical) 
interaction phase among formed fragments.  
These fluctuation issues are a crucial point to be 
checked in transport codes, in order to link fragment isotopic features 
to the trend of the symmetry energy \cite{ColonnaPRL2013}.

%%%%%%%%%%%%%%%%%%%%%%  predizioni di AMD sono corrette ? 

%%%%%%%%%%%%%%%%%%%%%%%%%%%%%%%%%%%%%%%%%%% Conclusions
%relative abundance of ``liquid'' and
%and to the isotopic properties of particles and fragments.
 
%Moreover, a larger amount of nucleon and light particles is emitted in SMF.
%While the difference in the primary charge distributions could be probably compensated
%by the secondary decay process, 
\subsection{Concluding remarks on multifragmentation}
The results reviewed above indicate that the pre-equilibrium and fragmentation 
dynamics is quite sensitive to the subtle interplay between mean-field and 
many-body correlation effects.  This may lead to different predictions of the transport
models which are commonly employed to simulate heavy ion collisions, 
according to the different approximations adopted to go beyond the mean-field picture.
Whereas closer results are obtained for 
IMF multiplicities and charge distributions,  
light cluster and isospin observables exhibit a larger sensitivity to the
treatment of the many-body dynamics. 
In particular, one observes that composition and  
$N/Z$ ratio of the pre-equilibrium emission, %and primary IMF's 
as predicted by BUU-like and QMD-like models, can be different.  
As discussed above in the specific case of the comparison between AMD and SMF models, 
for a fixed parametrization of the symmetry potential   
the isotopic content of the emitted light particles appears model-dependent, 
being systematically lower in SMF, that also gives a
more abundant emission. 
As a consequence, though the IMF charge distributions may exhibit 
some similarities, IMFs are predicted neutron-richer in SMF than in AMD. 
%Hence 

The latter observations lead to the conclusion that 
isospin observables not only reflect the features of the isovector channel of 
the nuclear effective interaction, but they 
are largely affected by the global reaction dynamics. 
This could be simply expected from the fact that the
%Since the 
symmetry potential is density dependent, 
%thus
%%isotopic properties should 
%%one expects that 
%isospin observables may keep the fingerprints of 
%the density regions spanned along the reaction path (which may vary 
%among the different model descriptions). 
%However, 
however
the impact of the nuclear dynamics on these observables 
looks generally more intricate, 
also revealing the presence of correlations and clustering effects. 
One may conclude that the discrepancies between transport model predictions
should be ascribed essentially
to differences in the description of 
compression-expansion dynamics and many-body correlations, which 
affect isoscalar, as well as isovector properties, of the reaction products. 
From this point of view, %one could try to exploit
isospin observables could be also exploited 
as a tracer of the reaction dynamics,  
to probe %the corresponding 
pre-equilibrium stage and fragmentation path.

Finally, it would be quite appealing to extend the comparison of
pre-equilibrium and fragmentation features 
discussed above to other approaches recently introduced 
to deal with the nuclear many-body dynamics, 
namely approaches incorporating explicit light cluster 
production \cite{OnoCim2016} and new stochastic 
models \cite{Napolitani2017,Hao2019}.
\vskip 1.cm

%%%%%%%%%%%%%%%%%%%%%%%%% mettere dopo

\section{\label{Fermi-periph}Reaction dynamics at medium energy: 
semi-peripheral collisions}
\begin{figure}[h]
%\vskip -1.cm
\centering
\includegraphics{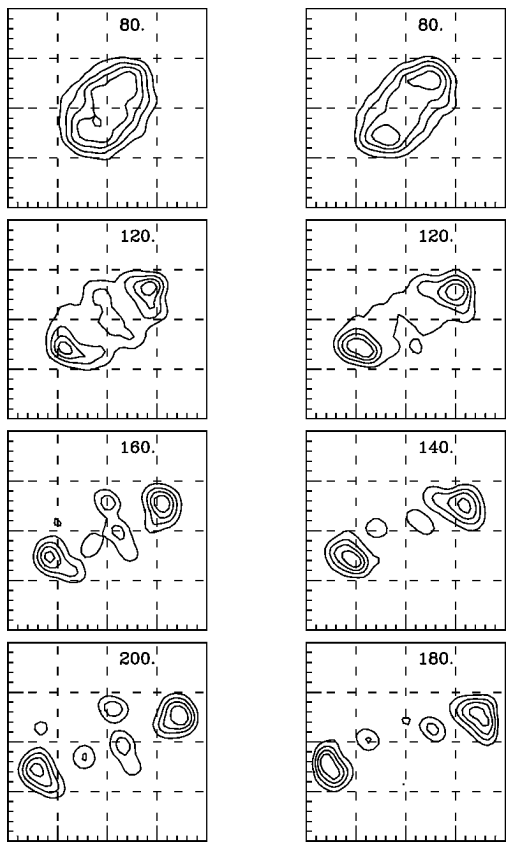}
\caption{\label{explosion_neck}Contour plots of the density projected on the reaction plane, calculated with SMF, for the
reaction $^{112}$Sn + $^{112}$Sn at 50 MeV/nucleon, 
at several time instants (fm/c) indicated on the panels. The lines are drawn at projected densities beginning at 0.07 fm$^{−2}$
and increasing by 0.1 fm$^{−2}$. The size of each box is 40 fm.
Left column: impact parameter b = 4 fm.  Right column: b = 6 fm. 
Taken from \cite{Baran2002}.}
\end{figure}

%\section{Survey of results for semi-peripheral collisions}
% neck dynamics, 
% focus on isospin effects
% comparison with Chimera
%%%%%%%%%%%%%%%%%%%%%%%% preso da Udo
In the previous Section, we focused our attention on 
multifragmentation processes 
observed in central collisions
at Fermi energies. 
%was discussed in terms of a liquid-gas phase transition in a composite
%two-component system.
%A possibility to explain the kinetics of this phase transition 
%is to relate it to spinodal decomposition in two-component nuclear matter, 
%leading to fragment formation accompanied by isospin distillation. 
%%For semiperipheral collisions, 
At semicentral impact parameters, the mechanism changes from fusion/fragmentation to 
predominant binary channels (which give rise to deep-inelastic or quasi-fission processes); 
along this transition an intermediate mechanism appears 
where a low-density neck region, from which fragments can eventually emerge, 
is produced between projectile-like (PLF) and target-like (TLF) 
fragments.
A qualitative illustration of the mechanism 
is given in Fig.\ref{explosion_neck}.

In addition to volume instabilities, surface effects (and related instabilities)
are particularly important for the description
of the configuration paths encountered in low- and Fermi-energy heavy ion collisions. 
%and also in this case a stochastic treatment of the reaction dynamics is
%in order 
As far as the description of low-energy nuclear processes is concerned, it is worth mentioning that  
stochastic extensions of time-dependent 
quantum approaches have been recently proposed \cite{Lacroix2016,Lacroix2012,Simenel2012}. Within this scheme, quantum and/or thermal fluctuations are injected in the
initial conditions, leading to a spread of dynamical trajectories and
corresponding variances of physical observables.  Interesting results have been
obtained for  spontaneous fission of 
superheavy systems \cite{Tanimura2017}. 
One could envisage
a smooth transition from quantum models to the semi-classical
treatments discussed here. In particular, at Fermi energies we still
expect thermal and/or mean-field fluctuations to influence dynamical
fragment emission from the neck region and from PLF/TLF sources. 

\begin{figure}
\centering
\includegraphics{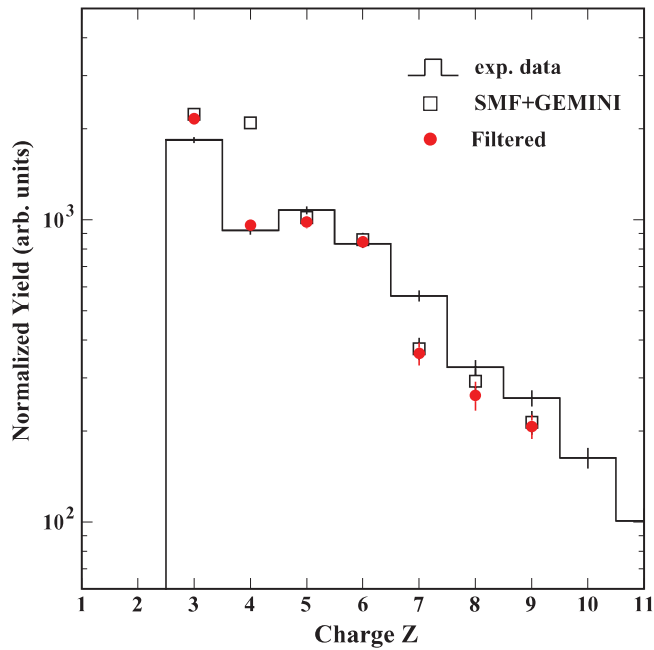}
\caption{\label{fig_en6}
(Color online) Experimental charge  distribution  of  dynamically  emitted  
IMFs detected in the  $^{124}$Sn + $^{64}$Ni reaction at 35 MeV/nucleon.  The  squares
correspond to the SMF+GEMINI calculations. Red circles: filtered 
calculated data. The large effect of the detector filter on the charge Z = 4 
is due to the unbound $^8$Be isotope that is present in the calculation, but filtered 
because it is not identified in the data analysis.
Taken from \cite{Defil2012}.}
\end{figure}

The neck fragmentation with a peculiar intermediate
mass fragment ($2<Z<20$) distribution and an entrance	%PN[$3<Z>20$ --> $3<Z<20$]
channel memory was observed experimentally and
predicted by various transport models \cite{baranPR,Lionti2005,DiToro2006,
Udo,Vesel,Souza,Monto,Barli,Milazzo2001,Kohley2011,Defil2012,Baran2012,Rizzo2014,Jedele2017,Manso2017}.
An interplay between statistical and
dynamical emission mechanisms can be expected; 
however it is possible to identify clear dynamical signatures, such as 
the appearance of hierarchy effects 
of the IMF size vs. transverse velocity, which keep track of the cluster formation time scale
and of the related impact of many-body correlations.  
Moreover, the analysis of the IMF relative velocities with respect to PLF and TLF
reveals that a large fraction of fragments emitted at mid-rapidity
cannot be associated with 
statistical emission from PLF and TLF sources, since they deviate from the
Viola systematics \cite{Defil2012}. 
The dynamical nature of these fragments
has been probed also through the comparison with transport models. 
An example is given in %within the context of the SMF model.  
Fig.\ref{fig_en6}, which shows the charge distribution of the IMFs emitted
at mid-rapidity, for the system $^{124}$Sn + $^{64}$Ni at 35 MeV/nucleon. 
Experimental data \cite{Defil2012} are compared to SMF simulations,
which predict a prompt dynamical emission.  
The model is able to reproduce also fragment velocity distributions. 
%(see Ref.\cite{Defil2012}).
The good agreement supports the interpretation of the neck emission in terms of 
the occurrence of volume/surface instabilities, according to the
scenario offered by given transport models.   

%This latter, when it is not reabsorbed by the projectile- or target-like nucleus, separates as one individual fragment or disintegrates into few IMFs~\cite{Lionti2005,DiToro2006}.
%The large density gradients induced by the neck formation offer favoured conditions for studying isospin migration, and many investigations have been carried out with the SMF approach~\cite{baranPR,Rizzo2008_1}.
Isospin effects are expected also in semi-peripheral reactions. 
Indeed, 
the low-density neck region may trigger an isospin migration from
the PLF and TLF regions, which have normal density. 
 Therefore,
the isospin content of the neck fragments could reflect the
isospin enrichment of the mid-velocity region. 
%Many investigations have been carried out with the SMF approach~\cite{baranPR,Rizzo2008_1}. 
Moreover, new interesting correlations between
fragment kinematical features and isotopic properties,
which can provide clues in searching for the most sensitive
observables to the symmetry energy, are noted and are still
under intense scrutiny. 

For more
peripheral collisions, the binary mechanism is accompanied, for
N/Z-asymmetric
entrance channel combinations, by isospin diffusion, that drives the
system toward charge equilibration. %\cite{betty2009,Rizzo2008_1}.
This mechanism has been widely exploited to probe the low-density
behavior of the symmetry energy \cite{bao_prl2005,betty2009,
Rizzo2008_1,Gali,Sun2010}. 
%More exclusive analyses from new running or planned
%experiments will certainly impose severe restrictions on various
%models and parameterisations concerning the latter quantity.

In the following 
%the main mechanisms responsible for isospin transport
%will be discussed more in detail and 
we will focus our discussion mainly on isospin dynamics, which has 
been the object of more recent investigations,  and 
we will review a selection of results related to
isospin observables typical of semi-peripheral and peripheral collisions
at Fermi energies. 
%The purpose of this article is to perform a detailed
%investigation of the fragmentation dynamics at the transition
%from semicentral to semiperipheral collisions, a region not
%much studied until now and which can be identified in
%modern experiments. 

%%%%%%%%%%%%%%%%%%%%%%%%  togliere questa parte  (era da Udo)

%%%%%%%%%%%%%%%%%%%%%%%%%%%%%%%%%%%%%%%%%  fin qui

\subsection{Isospin transport at Fermi energies}

As anticipated above, 
reactions between charge asymmetric systems are charecterized by  
a direct isospin transport in binary events (isospin diffusion).
This process occurs through the low density neck region and thus it is 
expected to manifest a sensitivity to
the low density behavior of $C_{sym}$, 
see Refs.\cite{tsang2004,bao_prl2005,Rizzo2008_1,Gali,betty2009,Sun2010}.
Moreover, it is now quite well established that a large part of the reaction
cross section measured in the Fermi energy range goes
into the neck fragmentation channel, with IMFs directly
emerging from the interacting zone on short
time scales \cite{DiToro2006,Defil2012}. 
Fragments are still formed in a dilute
asymmetric matter but always in contact with the regions of the
projectile-like and target-like remnants, almost at normal densities.
This may favor the neutron enrichment of the low-density 
neck region (isospin migration).   

%Results on these mechanisms, obtained with the
%SMF model, are discussed below. 
The main role  of the isospin degree of freedom in the collision dynamics 
%of a nuclear reaction
can be easily understood at the hydrodynamical limit,  considering the behavior
of neutron and proton 
chemical potentials as a function of density $\rho$ and 
asymmetry 
$\beta \equiv I = \rho_3/\rho$ \cite{isotr05}. The proton/neutron currents can be expressed as
\begin{equation}
{\bf j}_{p/n} = D^{\rho}_{p/n}{\bf \nabla} \rho - D^{\beta}_{p/n}{\bf \nabla} \beta ,
\end{equation}
with $D^{\rho}_{p/n}$ the drift, and
$D^{\beta}_{p/n}$ the diffusion coefficients for transport dynamics \cite{isotr05}. 
%%%%%%%%% beta dovrebbe essere I
Of interest for the study of isospin effects %here 
are the differences of currents 
between protons 
and neutrons which have a simple relation to the density dependence of the 
symmetry energy
\begin{eqnarray}
D^{\rho}_{n} - D^{\rho}_{p}  & \propto & 4 \beta \frac{\partial C_{sym}}
{\partial \rho} \,
 ,  \nonumber\\
D^{\beta}_{n} - D^{\beta}_{p} & \propto & 4 \rho C_{sym} \, .
\label{trcoeff}
\end{eqnarray}
From these simple arguments, one can realize that 
the isospin transport due to density gradients (isospin migration) 
is ruled by the slope of the symmetry energy, or the symmetry pressure, 
while the 
transport due to isospin concentration gradients (isospin diffusion) 
depends on
 the symmetry energy value. 
Hence transport phenomena in nuclear reactions appear directly linked to 
two relevant properties (direct value and first derivative) of the symmetry 
energy of the nuclear EOS. 

%Thus the low
%density behavior of the symmetry energy is concerned yet. 
%A neck of density below normal density develops between the two 
%heavy residues, the evolution of which is driven by the motion of the 
%spectators. 
%During this phase isospin is transferred to the neck due to the density 
%difference 
%between the neck and the residues; this effect is called isospin migration,
% which 
%leads to a more neutron-rich neck. In addition in collision systems with
% different asymmetry isospin is transported through neck due to the 
%asymmetry gradient
% leading to an equilibration of 
%the isospin of the residues, which has been called isospin diffusion. 
%Thus in asymmetric
% systems there is  a competion of isospin migration and diffusion.  

%In peripheral collisions 
%discussed here, 
%residues of about normal density are in contact with the neck region of 
%density below 
%saturation. At such low densities a stiff iso-EOS has a smaller 
%value but 
%a larger slope compared to a soft iso-EOS. Correspondingly we expect 
%opposite effects of 
%these models on the migration and diffusion of isospin.

\subsubsection{\label{equi}Charge equilibration in peripheral collision dynamics}
In semi-peripheral and peripheral reactions, 
transport simulations allow one to investigate 
the asymmetries of the various parts 
of the interacting system in the exit channel:
emitted particles, PLF 
and TLF fragments, and, for ternary (or higher
multiplicity) events, IMFs.
\begin{figure}[h]
\centerline{
	\includegraphics[angle=0,scale=1.]{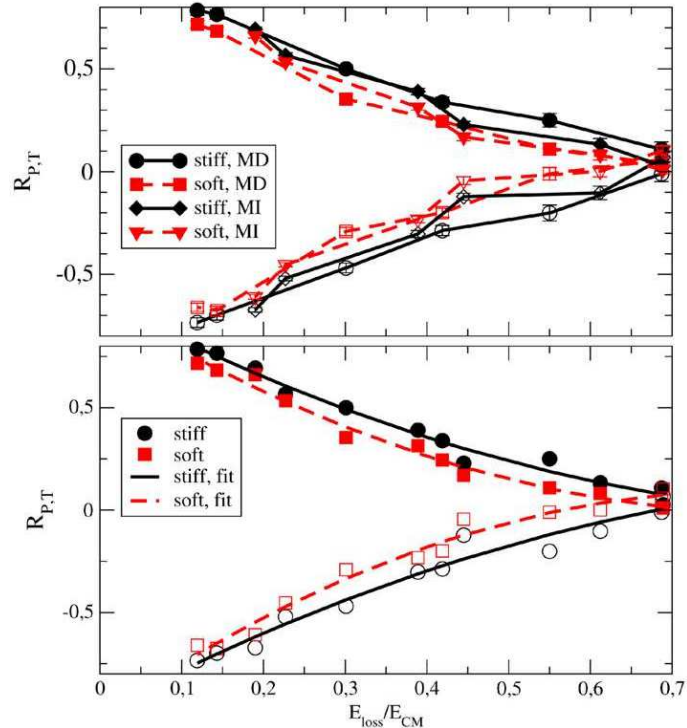}}
%\centering
%\includegraphics{fig_joseph.eps}
%\includegraphics{fig-others/rios2011-2dfis.pdf}
\caption{\label{fig_jos} (Color online)
Projectile (P, upper curves in each panel) and target (T, lower curves in each
panel) imbalance ratio as a function of relative energy loss. Upper panel: separately for stiff (solid) and soft (dashed) asyEOS, and for MD (circles and squares) and MI (diamonds and triangles) interactions, in the projectile region (full symbols) and the target region (open symbols). Lower panel: quadratic fit to all points for the stiff (solid), respectively soft (dashed) asyEOS. 
Taken from \cite{Rizzo2008_1}.}
\end{figure} 
 In particular, one can  study  the
 so-called isospin transport ratio (or imbalance ratio), which is defined as:
%%% Hermann
\begin{equation}
%R^x_{P,T} = \frac{2x^M-(x^H+x^L)}{(x^H-x^L)} .
R^x_{P,T} = \frac{2(x^M-x^{eq})}{(x^H-x^L)}~,
\label{imb_rat}
\end{equation}
with $x^{eq}=\frac{1}{2}(x^H+x^L)$.
 Here, $x$ is an isospin sensitive quantity
that has to be investigated with respect to
equilibration \cite{Rami2000,tsang2004}.   
One can directly consider the asymmetry 
$I = (N-Z)/A$,
but also other quantities, such as isoscaling coefficients, ratios of 
light
fragment's yields, etc, can be of interest \cite{tsang2004,betty2009,Colonna_IMF2008,
ColonnaPRL2013}. %WCI  
The indices $H$ and $L$ refer to symmetric reactions
between 
heavy  ($n$-rich) and light ($n$-poor)  systems, while $M$ refers to the
mixed reaction.
$P,T$ denote the rapidity region, in which this quantity is measured, in
particular the
PLF and TLF rapidity regions. Clearly, this ratio is $\pm1$ in
the projectile
and target regions, respectively, for complete transparency, and oppositely
for complete
rebound, whereas it is zero for complete equilibration.

%\end{minipage}
%\end{figure}  %\begin{figure}

%\begin{figure}[t]
%%\vskip 1.0cm
%\centering
%%\begin{picture}(0,0)
%%\put(135.0,0.0){\mbox{\includegraphics[angle=-90,width=7.0cm]{isotr11.ps}}}
%%\put{20.0,0.0}{mbox{\includegraphics[width=7.0cm]{isotr9.eps}}}
%%\end{picture}
%%\includegraphics[width=7.0cm]{erice3a.eps}
%%\hskip 0.5cm
%\includegraphics[width=7.5cm]{erice3b.eps}
%\caption{ 
%Left Panel.Imbalance ratios for $Sn + Sn$ collisions for 
%incident energies
%of 50 (left) 
%and 35 $AMeV$ (right) as a function of the impact parameter. Signatures of 
%the curves: 
%iso-EOS stiff (solid lines), soft (dashed lines); MD interaction (circles),
% MI interaction (squares); projectile rapidity ( full symbols, upper curves ),
% target rapidity ( open symbols, lower curves ).
%Right Panel. Imbalance ratios as a function of relative energy loss for both 
%beam energies. 
%Upper: Separately for 
%stiff (solid) and soft (dashed) iso-EOS, and for MD 
%(circles and squares) and MI 
%(diamonds and triangles) interactions, in the projectile region (full symbols)
% and the target region 
%(open symbols).
%Lower: Quadratic fit to all points for the stiff (solid), resp.
% soft (dashed) 
%iso-EOS.}
%\label{imb_eloss}
%\end{figure}

%%%%%%%%%%%%%%%%%%%  è utile definire l'energia di simmetria in espansione
%%%%%%%%%%%%%%%%%%%  di Taylor
\begin{figure}[h]
\centering
\includegraphics{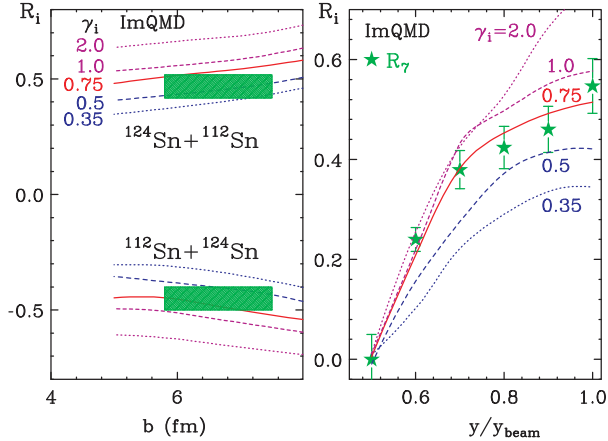}
\caption{\label{fig_prl}
(Color online).  Left panel: Comparison of experimental isospin transport ratios (shaded regions) to ImQMD results (lines), as a function of impact parameter for different values of the parameter $\gamma_i$, related to the symmetry energy
behavior. Right panel: Comparison of experimental isospin transport ratios obtained from the yield ratios of A = 7 isotopes (star symbols), as a function of the rapidity, to ImQMD calculations (lines) at b = 6 fm.
Reprinted from \cite{betty2009}, with kind permission of the APS.}
\end{figure}    
The centrality dependence of the isospin transport ratio
has been widely explored in experiments as well as in theory
\cite{tsang2004,bao_prl2005,Rizzo2008_1,Gali,betty2009,Sun2010}.
Many investigations have been concentrated on
Sn + Sn collisions at 35 and 50 MeV/nucleon \cite{tsang2004}. Some examples are given below. 
Fig.\ref{fig_jos} shows the results of SMF 
calculations employing Momentum Dependent (MD) or Momentum Independent (MI)
Skyrme interactions \cite{Rizzo2008_1}.  The transport ratio, evaluated considering, as the
$x$ observable of Eq.(\ref{imb_rat}), 
the N/Z
ratio of primary PLF and TLF,  is reported as a function of the
kinetic energy loss. The latter quantity is taken as an estimate of the dissipation
degree reached along the collision dynamics and thus of the contact time between the reaction partners.
The energy dissipation 
between projectile and target 
%(that also influences the corresponding reached
%degree of equilibration \cite{Zhang2015}) 
is mainly governed by the isoscalar features of the
nuclear interaction (including effective mass details) \cite{Zhang2015}.
On the other hand, for the same amount of dissipated energy,  
the sensitivity of the isospin transport ratio to the symmetry energy
is nicely isolated (see the lines on the bottom panel of Fig.\ref{fig_jos}).  
%see that, within the representation Fig.\ref{fig_jos}, the sensitivity
%of the isospin transport ratio to the symmetry energy is nicely isolated. 
More equilibration, i.e. a smaller $R$ ratio, is obtained in the case of an 
asysoft parametrization, which is associated with a higher symmetry energy
value below normal density (see Fig.3).   
Quite sizeable isospin equilibration effects characterize the semi-central events (corresponding to large dissipated energy).    

%Also for this mechanism, it is quite interesting to compare results obtained within different model prescriptions of the
%reaction dynamics.  
Calculations performed with the ImQMD model, together with a thorough
comparison to the $MSU$ experimental data, are reported in \cite{betty2009}.  
Corresponding results are shown in Fig.\ref{fig_prl}, % fig Yingxun
for the transport ratio as a function of the impact parameter (left
panel) and of the fragment rapidity (right panel).
Several parametrizations of the symmetry energy, denoted by the parameter 
$\gamma_i$ (see the discussion in Section \ref{sym_ene}) are employed.  
The predictions of the SMD and ImQMD models
 look comparable for peripheral collisions, where a low degree of charge
equilibration is observed in both cases, however  
SMF predicts a larger equilibration at semi-central impact
parameters. 
This can be expected on the basis of the faster dynamics, 
i.e. the larger transparency characterizing QMD-like models \cite{Aich1991}, 
with respect to mean-field ones. 
We observe again that not only isospin observables hold a sensitivity to the 
density behavior of the symmetry energy, but they also trace the 
many-body reaction dynamics. 

It is worth noticing that the isospin transport ratio 
is also affected by isoscalar features of the nuclear effective interaction
\cite{Rizzo2008_1,Zhang2015}.  
Also clustering effects may impact this observable, as pointed out, for
instance, in Ref.\cite{Daniel}. 
%%%%%%%%%%%%%%%%   Danielewicz + Coupland

The information on the stiffness of the symmetry energy, 
extracted from the comparison  performed in Ref.\cite{betty2009},
see Fig.\ref{fig_prl} (left panel),
points to a value of the slope $L$ in
the range between 40 and 80 MeV. 
%(for a symmetry energy coefficient around 30 MeV at saturation).
With respect to the model dependence discussed above, this conclusion should
be rather robust because the analysis is performed at semi-peripheral 
impact parameters, where the predictions of different models are closer 
to each other.  

%%%%%%%%%%%%%%%%%%%%%%%%%%%%%  citare Bao-An ! 

%%%%%%%%%%%%%%%%%%%%%%   citare Galichet
%We report here a new analysis
% which appears experimentally more selective \cite{isotr07}. 

%\vskip -1.0cm
\subsubsection{Neck fragmentation at Fermi energies: isospin dynamics}

As discussed above, 
in presence of density gradients, as the ones occurring 
when a low-density neck region is formed between the two reaction
partners,  the isospin transport
is mainly ruled by the density derivative of the symmetry energy,  
see Eqs.(\ref{trcoeff}),
and so
we expect a larger neutron flow toward
 the neck clusters for a stiffer symmetry energy around saturation  
\cite{baranPR} (see Fig.3). 
\begin{figure}[h]
\centering
\includegraphics[angle=0, width=.5\textwidth]{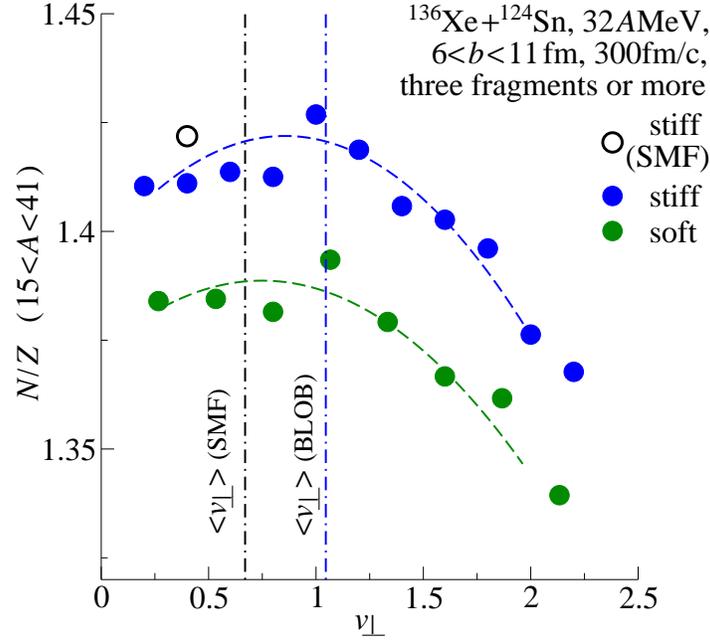}
\caption{\label{fig_udo} (Color online)
BLOB simulations: Isotopic content of neck fragments at 300fm/c, identified in the 
impact parameter region of maximum production, 
as a function of the transverse velocity component with respect to the PLF-TLF 
axis for a stiff and a soft form of the symmetry energy potential. 
Taken from \cite{clusters}.}
\end{figure}    
As predicted by some transport models, this mechanism (isospin migration) 
appears as a good candidate to explain the neutron enrichment of the
neck region. 
The effect is illustrated in Fig.\ref{fig_udo},   
%%%% Figura da Udo
where the neutron enrichment of the neck primary fragments, simulated with
BLOB,  
is shown as a function
of their transverse velocity, for the reaction $^{136}$Xe$+^{124}$Sn at 32 MeV/A.
Two different parametrizations of the nuclear symmetry energy are employed
(asysoft and asystiff).  
Two interesting features emerge from these results: 
i) the neutron migration effect is larger at smaller transverse velocity, 
i.e. for fragments which stay for a longer time in contact with 
the PLF/TLF regions (thus experiencing a more effective isospin migration); ii) neutron-richer
fragments are obtained in the asystiff case, confirming  the sensitivity
of the effect to the symmetry energy derivative.   
These features, related to primary fragment emission, 
could be blurred by secondary decay, thus it would be desirable to
build observables that should be less affected by the de-excitation stage.  
A possibility would be to consider ratios of fragment yields or isotopic features.
%which should be more robust against secondary decay. 
For instance, one can adopt 
the ratio of the asymmetry of the IMFs ($\beta_{IMF}$) to that of the
residues ($\beta_{res}$). 
%The results
% correspond to $b=6fm$ semiperipheral
%events, plotted  here  as a function of the initial isospin asymmetry of the
%colliding system.
%%For $b=8fm$ the behaviour is not very different,
%%except that the error bars are considerably larger. 
For symmetric projectile/target reactions, 
this quantity %between the asymmetry of IMF's and residues  
can be roughly estimated analytically
on the basis of simple energy balance considerations.
%In fact, isospin migration is due to the fact that the neck region
%has lower density with respect to the residues and the symmetry 
%energy is decreasing with density.
By imposing to get a  maximum  (negative) variation of 
$C_{sym}(\rho)$  when transfering the neutron richness from
PLF and TLF towards the neck region, one obtains \cite{Rizzo2008_1}: 
%with respect to $\Delta\beta$ yields the ratio of asymmetries
\begin{equation}
\frac{\beta_{IMF}}{\beta_{res}} 
%= \frac{\beta+\Delta\beta}{\beta-\Delta\beta}
= \frac{C_{sym}(\rho_R)}{C_{sym}(\rho_I)} 
%= 1 +  \frac{E_{sym}(\rho_R)-E_{sym}(\rho_I)}{E_{sym}(\rho_I)}
\end{equation}
%\be
%\frac{\beta_{IMF}}{\beta_{Res}} \approx \frac{\beta + \Delta\beta}{\beta}
%%= \frac{\beta+\Delta\beta}{\beta-\Delta\beta}
%= 1 + \frac{E_{sym}(\rho_R)- E_{sym}(\rho_I)}{E_{sym}(\rho_R) + E_{sym}(\rho_I)}
%\ee
%The minimum of the variation $\delta E$ corresponds to:
%$$\delta\beta/\beta = \left( E_{sym}(\rho_R) - E_{sym}(\rho_{IMF})\right) /
%\left( E_{sym}(\rho_R) + E_{sym}(\rho_{IMF})\right)$$
%From this simple argument, one can see that the neutron enrichment of the
%neck region is proportional to the system asymmetry  $\beta$.
\begin{figure}[h]
%\centering
\vskip 0.5cm
\hskip 1.5cm
\includegraphics[width=15pc]{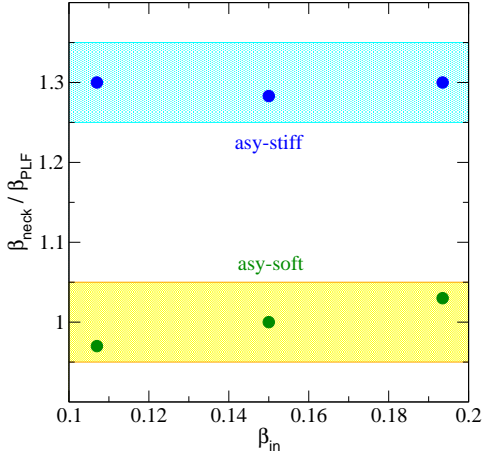}\hspace{3pc}%
\begin{minipage}[b]{17pc}\caption{\label{nzphi}
%\caption{
%Top panel: asymmetry of IMF's (circles) and PLF-TLF (squares), as a function of the
%system initial asymmetry, for two Asy-EOS choices: asystiff (full lines) and
%asysoft (dashed lines).  
%Bottom panel: 
Ratio between the neck IMF and the PLF/TLF asymmetries, in Sn + Sn reactions at 
50 $AMeV$, as a function of the system
initial asymmetry. SMF simulations. 
The bands indicate the uncertainty in the calculations. 
Taken from \cite{EPJA_Colonna}, with kind permission of 
the European Physical Journal (EPJ).}
%\vskip -2.0cm
%\label{nzphi}
%\end{center}
\end{minipage}
\end{figure}    
From this simple argument the ratio between IMF and residue (PLF and TLF) 
asymmetries should
depend only on symmetry energy properties and, in particular, on the different  
symmetry energy values associated with residue and neck densities ($\rho_R$ and
$\rho_I$, respectively),
%between the residue and the neck regions, 
as appropriate 
for isospin migration.  
%symmetry energy properties and, in particular,
%on the ratio of the symmetry energies of the residue and the
%neck regions. 
It should also be larger than one, more so for the asystiff than
for the asysoft EOS.
%{\bf Check!!} 
Results obtained with SMF, employing the asystiff and asysoft parametrizations 
(see Fig.3), 
are represented in Fig.\ref{nzphi}, for Sn + Sn systems with different 
initial asymmetry, $\beta_{in}$, at 50 MeV/nucleon of beam energy. 
One can see that this ratio %of IMF over residues asymmetry
is nicely dependent on the asyEOS only (being larger in the asystiff case) 
and not on the system considered.
Assuming that final asymmetries are affected in a similar way by secondary evaporation, 
%in the case of neck and PLF fragments, 
one could directly compare the
results of Fig.\ref{nzphi} to data.  However, some words of caution are 
needed, because of 
the different size/temperature conditions
of the neck region with respect to PLF and TLF sources.
%some caution is needed and  
%de-excitation effects should be carefully checked with the help of
%suitable decay codes. 

%After having discussed some theoretical predictions, 

\begin{figure}[h]
\centering
\includegraphics{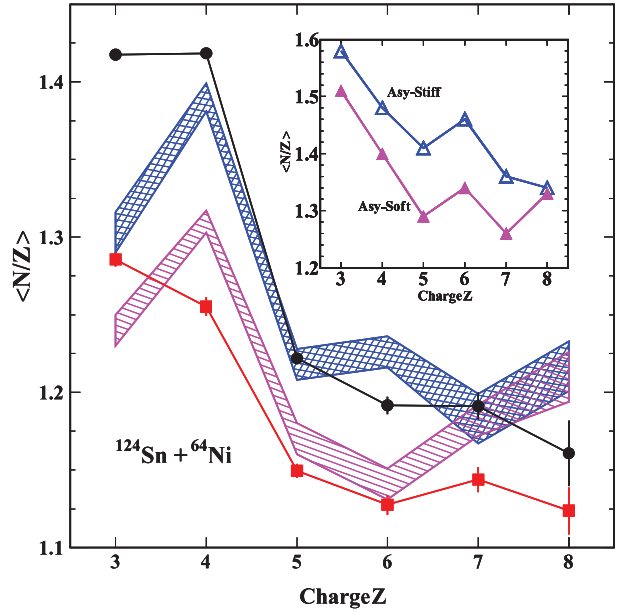}
\caption{\label{fig_en4}
(Color online) Experimental $<N/Z>$ distributions, as a  function  of  the 
charge Z, for  statistically  emitted  (solid squares - red line)  and  dynamical  emitted  fragments  (solid  circles - black line),  for  the reaction $^{124}$Sn +
$^{64}$Ni. Blue hatched area: SMF-GEMINI calculation for  dynamical  emitted  particles  and  asystiff  parametrization;  magenta hatched area: asysoft parametrization. The $<N/Z>$ of primary IMFs  as  a  function  of  the  atomic  number Z, as obtained in SMF calculations, is plotted in the inset for the two 
parametrizations. 
The hatched zone indicates the error bars in the calculations. 
Taken from \cite{Defil2012}.
}
\end{figure}    
We now move to review some 
recent experimental evidences related to the neutron enrichment of 
the neck fragments \cite{Defil2012,yzhang2017,Jedele2017,Manso2017}. 
The isotopic features of the neck emission have been investigated 
in experimental analyses performed 
on Sn+Ni data at $35~AMeV$
by the Chimera Collab.\cite{Defil2012}
% see figure \ref{nzphi} right panel.
The experimental observation that these fragments are neutron-rich appears compatible 
with the isospin migration mechanism predicted by stochastic mean-field
models. 
Provided that
charge distribution and kinematical properties of the neck
fragments are also consistently described, this observable 
can probe the density dependence of the symmetry energy, 
as shown in Fig.\ref{fig_en4}. 
First of all one can see that, in the data analysis \cite{Defil2012},
the ``dynamical'' emission (black line), i.e. 
the fragments which are directly emitted at mid-rapidity, exhibits a
larger N/Z than the ``statistical'' emission from
PLF/TLF sources (red line). SMF simulations, including statistical
de-excitation, (the bands on the figure) 
indicate that an asystiff parametrization of the symmetry energy 
leads to a better agreement with the data.  
Within the same data set, a strong correlation between neutron enrichement and fragment alignement (which enforces a 
short emission time) has been evidenced.
The comparison with the simulations point  
to 
%only with 
a stiff behavior of the symmetry energy ($L\approx 75~ MeV$). 
We notice that this result is compatible with the constraints
extracted from the analysis of the isospin transport ratio discussed in the
previous subsection.
  
\begin{figure}[h]
\centering
\includegraphics[scale=0.7]{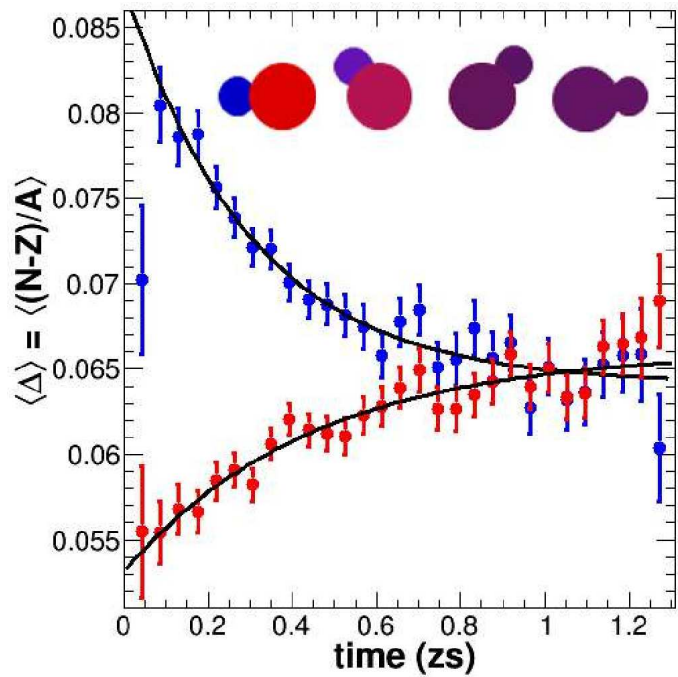}
\caption{\label{fig_sh2} (Color online)
Measured neutron excess of both the heavy (red) and light (blue) 
daughters produced in binary splits of a
deformed excited projectile-like fragment as a function of the duration of the equilibration. Data shown is
for $^{70}$Zn + $^{70}$Zn at 35 MeV/u, $Z_H$ = 12, $Z_L$ = 7. The cartoon across the top suggests the neutron-rich (blue) - proton-rich (red) equilibration as the system rotates. Taken from \cite{Sherry_PPNP}.
}
\end{figure}  
New experimental studies have been devoted to probe the isotopic content
of the fragments emitted at mid-rapidity, in connection to kinematical features
and emission time scales \cite{Jedele2017,Manso2017}. 
\begin{figure}[h]
\centering
\includegraphics[scale = 1.]{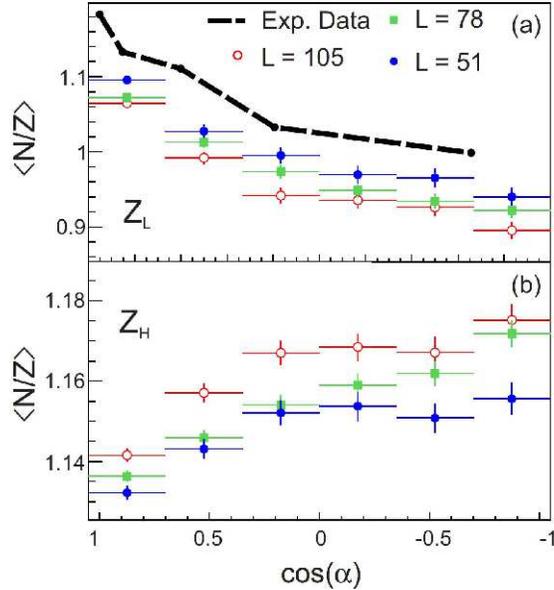}
\caption{\label{fig_sh1} (Color online)
Top panel: Composition of the light fragment as a function of its orientation relative to the projectile-like
fragment for reactions of Zn+Zn at 45 MeV/u. Black: measured data. Color points: CoMD model calculations for varying forms of the asymmetry energy. 
Bottom panel: The CoMD model also shows how the composition of the
heavy fragment changes in proportion and in the opposite direction to the composition of the light fragment.
Reprinted from \cite{Stiefel2014}, with kind permission of the APS.
}
\end{figure}    
In particular, the main goal was to explore the interplay between the 
isospin migration/diffusion
effect and the time scales characterizing the fragment emission
dynamics.  Fig.\ref{fig_sh2} illustrates the features of  
the two fragments (of charge $Z_L$ and $Z_H$) emitted on short time scales from deformed 
PLF sources, in semi-peripheral reactions at typical Fermi energies. 
Emission times are extracted from 
the light fragment emission angle, according to the estimated PLF 
intrinsic angular momentum.  
It is seen that the light fragments are neutron-richer when 
%are promply emitted, i.e. 
their orientation with respect to the (heavier) 
projectile-like fragment corresponds to a small angle, i.e. they are 
promptly emitted,
along a rather aligned configuration.  
The latter observation would be consistent with the fact that the light 
fragments, when promptly
emitted, still reflect the neutron enrichment of the
neck region.   On the other hand, larger emission angles would correspond to 
a longer
emission time, i.e. to a significant rotation of the light fragment around the heavy one,
that would lead to isospin equilibration, prior to the emission, among the
two objects.  This would correspond to an increase (decrease) 
of  the N/Z of 
the light (heavy) fragment, towards equilibration.   

 In Fig.\ref{fig_sh1} the experimental data are compared to the results of
the CoMD model \cite{Papa2001}.   
It is interesting to see that the model is able to reproduce the 
decreasing (increasing) trend of the N/Z of the light (heavy) 
fragment with the emission time, i.e. the emission angle. 
The calculations also exhibit
a sensitivity to the symmetry energy parametrization employed (denoted
by the slope parameter $L$ on the figure). 
However, the model predicts neutron-richer heavy fragments, with
respect to the light ones (compare top and bottom panels),  
contrarily to what is observed in the data, 
see Fig.\ref{fig_sh2}, and expected on the basis of isospin migration
arguments. 
A deeper discussion about the predictions of different 
transport models 
is tackled in the next subsection.

%--- citare MacIntosh e l'importanza di un'analisi combinata (FAZIA)--- 
%, fig.\ref{nzphi} right panel \cite{isotr07}. 
%This represents an 
%evidence in favor of a relatively large slope ($L\approx 70~ MeV$) 
%around saturation. 
%We note a recent confirmation from structure data,
%i.e. from monopole resonances in Sn-isotopes \cite{garg_prl07}.   

%%%%%%%%%%%%%%%%%%%%%%%%%%%%%%%%%%%%%%%%%%%%%%%

%%%%%%%%%%%%%%%%%%%%%%%%%%%%   Da EPJA
\subsection{Comparison between the predictions of different transport models}
%As illustrated above, 
%a detailed investigation of isospin equilibration was undertaken
%within transport codes based on the molecular dynamics approach \cite{betty2009}.
%QMD-like models, with nucleons represented as individual wave packets of fixed compact shape, are expected  to be particularly  
%well suited to describe  the correlations associated with 
%the exit channel of fragmentation events. 
%However 
A closer inspection of the main mechanisms governing 
energy dissipation and isospin transport 
can be made by comparing the predictions of different transport models. 
\begin{figure}[h]
%\vskip 2.cm
\centering
\resizebox{0.7\textwidth}{!}{%
  \includegraphics{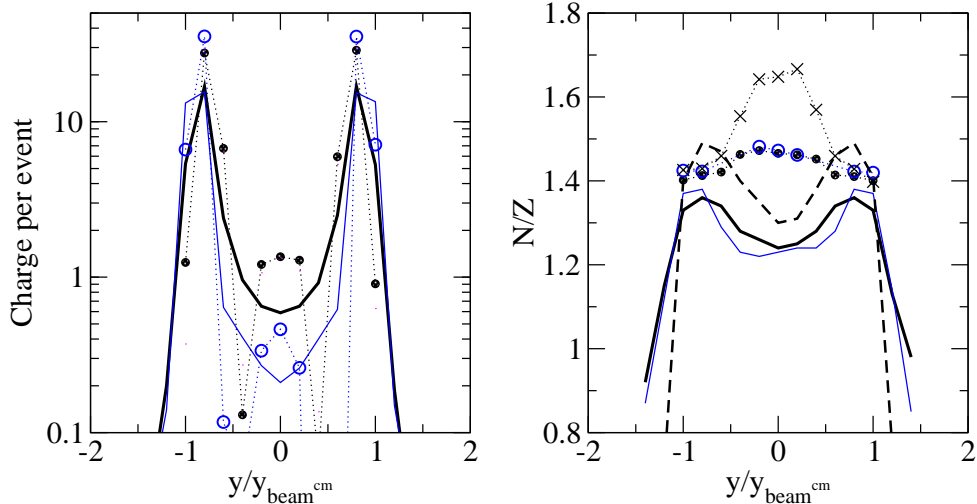}
}
%\includegraphics[width=30pc]{fig_nn2012.eps}%\hspace{4pc}%
%\end{center}
%\begin{minipage}[b]{14pc}
\caption{\label{confro} (Color online)
Left panel: Average total charge per event, associated with IMFs, as a function of the reduced 
rapidity, obtained in the reaction $^{124}$Sn + $^{124}$Sn at 50 MeV/u.
Results are shown for ImQMD calculations at b = 6 fm (thick black line) 
and b = 8 fm (thin blue line)
and for SMF calculations at b = 6 fm (full circles) and b = 8 fm (open circles). 
A soft interaction is considered for the symmetry energy.
Right panel: N/Z of IMFs as a function of the reduced rapidity. Lines and symbols are
like in the left panel. 
Results corresponding to a stiff asyEOS are also shown for ImQMD (dashed line) and
SMF (crosses), for b=6 fm. Taken from \cite{EPJA_Colonna}, with kind permission
of the European Physical Journal (EPJ).}
%\end{minipage}
\end{figure}  %\begin{figure}  
%would be rather interesting, 
%also by making a comparison with the predictions of other transport models. 
%(surely leads to some approximation in the description...) 
%one expects that the description of mean-field
%effects is approximated, 
%On the other hand, fluctuations and correlations should be well
%accounted for, especially in
%the exit channel of multifragmentation events.
As shown by Figs.\ref{fig_jos} and \ref{fig_prl}, which illustrate 
charge equilibration for $Sn + Sn$ reactions
at 50 MeV/nucleon, 
for the most central reactions the ImQMD code predicts a quite different behavior with respect to SMF: indeed
the trend of the isospin transport ratio versus the impact paramter 
is flatter in ImQMD.  
This behavior would suggest that, even in the case of central collisions, the contact time
between the two reaction partners remains rather short, 
inhibiting charge equilibation. Then energy dissipation is mainly due
to nucleon correlations and particle emission, rather than to mean-field effects. In other words, the faster (and more explosive)  
dynamics predicted by ImQMD simulations
would lead to the lower degree of isospin equilibration observed.
It should be noticed that the stronger impact, in molecular dynamics approaches, of many-body correlations on the fragmentation
path was evidenced also in the case of central multifragmentation reactions,
see the discussion concerning the comparison between 
AMD and SMF approaches in Section \ref{comp} \cite{ColonnaPRC2010}.

To explore in deeper details the neck fragmentation dynamics, 
results concerning IMF ($Z>2$) properties, obtained with the SMF and ImQMD codes, are compared in Fig.\ref{confro}.
In the left panel, the average total charge per event, bound im IMFs, is plotted
as a function of the reduced rapidity, for the reaction $^{124}$Sn + $^{124}$Sn at 50 MeV/u and 
impact parameters b = 6 and 8 fm. 
It appears that in ImQMD a larger number of light IMFs, distributed
over all the rapidity range between PLF and TLF, are produced.
On the other hand, mostly binary or ternary events are observed in SMF, with light IMFs
located very close to mid-rapidity. 
%Then the different reaction dynamics predicted by the two codes
%may explain the different results seen 
These observations are consistent with the results obtained for
charge equilibration, 
especially
in semi-peripheral reactions (b $\approx$ 4-6 fm).
In fact, a more abundant cluster production goes in the direction of 
reducing isospin equilibration, as also discussed in Ref.\cite{Daniel}
in the context of the pBUU model. 
The fast ImQMD fragmentation dynamics inhibits nucleon exchange and charge equilibration,
though the energy loss can be sizeable, owing to particle and light 
cluster emission.
%leads to less charge equilibration, while 
On the other hand, in the SMF model, 
 mean-field effects, acting over longer time intervals, have a stronger 
impact on dissipation, leading to more 
equilibration.

Results on the fragment neutron content are illustrated in the right panel of Fig.\ref{confro},
that shows the global N/Z of IMFs as a function of the reduced rapidity. 
As already discussed above,
SMF calculations clearly predict a larger N/Z for IMFs produced at mid-rapidity,  
with respect to PLF and TLF regions (isospin migration). 
The effect is particularly pronounced in the case of the asystiff parametrization. 
On the contrary, ImQMD calculations predict a minimum of the N/Z ratio at mid-rapidity, probably caused by the abundant neutrom emission from the hotter
neck region, suggesting that isospin migration toward the neck is not present.    
We notice that the observation of 
mid-velocity (light) fragments 
with smaller or comparable N/Z with respect to the heavier ones, is consistent
with the CoMD results presented in Fig.\ref{fig_sh1}. 
Interestingly enough, 
isospin migration effects towards IMFs at mid-rapidity are absent also in 
pBUU calculations including explicit cluster correlations, 
whereas standard BUU simulations (i.e. without clusters) predict it
 \cite{Daniel}.
In perspective, it would be interesting to investigate 
isospin transport effects also in 
the recently upgraded AMD model, including explicit 
cluster production \cite{OnoCim2016}. 
%The reasons of these difference  need to be further investigated. 
From the above discussions it emerges that, 
in addition to the expected sensitivity 
to the symmetry energy, isospin diffusion and, to an
even larger extent, isospin migration are quite affected by the treatment of many-body dynamics. 
From the experimental point of view, 
it would be interesting to study both isospin transport phenomena 
(isospin equilibration and migration) within the same data set\cite{Chbihi}.
This should allow to further probe, through the comparison with transport simulations, the underlying reaction mechanisms and validate the current 
constraints on the symmetry energy trend at sub-normal density.

%%%%%%%%%%%%%%%%%%%%%%%%%%%%%%%%%%%%%%  spostarlo su Conclusions !!!
%\subsection{Concluding remarks}

\section{
Collision dynamics at relativistic energies
}
%When increasing the beam energy (up to $\approx$ 1 GeV/A), 
For beam energies in the range of 0.1 - 1 GeV/nucleon, 
the heavy ion 
reaction dynamics is characterized by a larger degree of stopping in 
central collisions \cite{OnoPPNP2019}, with also high density and temperatures values
explored, which may lead to the full disassembly of 
the system into nucleons and light clusters. 
Cluster and IMF production holds also at semi-central/peripheral 
impact parameters, together with the detection of some remnants 
of PLF and TLF fragments.
%%%%%%%%%%%%%%%%%%%%% From Science -- Danielewicz, Lacey, Lynch
Nuclear collisions in this energy range are the ideal tool to investigate
the  relationship  between  pressure,  density and temperature 
characterizing the
EOS of dense nuclear matter. 
We stress again that the latter is a rather important object, which also 
governs the compression achieved in supernovae and neutron stars,  as  well  as  their  internal  structure.
The  pressure that results from the high densities reached  
during  collisions  at relativistic energies strongly  influences 
the motion of the ejected matter. Thus sensitive observables can be devised to probe the 
high density EOS \cite{Dan2002}.
%thus providing the sensitivity to the EOS that is needed to constrain it (10–19). 
%     W. Reisdorf, H. G. Ritter
%, Annu. Rev. Nucl. Part. Sci. 47, 663 (1997)
%%%%%%%%%%%%%%%%%%%%%%%               citare i Kaoni come segnale di EOS e 
%%%%%%%%%%%%%%%%%%%%%%%               short-range correlations (presentazione
%%%%%%%%%%%%%%%%%%%%%%%               di Che-Ming)
 
Also in this energy regime,
the collective response of the system deserves strong attention. 
Collective flows are very good candidates to probe the reaction
dynamics since they are 
expected to be 
quite sensitive to the features of the nuclear effective interaction, 
including the momentum
dependence of the mean field, see \cite{DanielNPA673,baranPR}.
On the other hand, stopping observables look more sensitive to the details
of the in-medium n-n cross section \cite{Dan2002,OnoPPNP2019}. 

%The flow observables can be seen
%respectively as the
%first and second coefficients from the Fourier expansion of the
%azimuthal distribution \cite{OlliPRD46}:
%$\frac{dN}{d\phi}(y,p_t)=1+V_1cos(\phi)+2V_2cos(2\phi)$, 
%where $p_t=\sqrt{p_x^2+p_y^2}$ is the transverse momentum and $y$
%the rapidity along beam direction. 
The transverse flow, 
$V_1(y,p_t)=\langle \frac{p_x}{p_t} \rangle$,
where $p_t=\sqrt{p_x^2+p_y^2}$ is the transverse momentum and $y$
the rapidity along the beam direction, 
provides information on the anisotropy of 
nucleon (or particle) emission on the reaction plane.
Very important for the reaction dynamics is also the elliptic
flow,
$V_2(y,p_t)=\langle \frac{p_x^2-p_y^2}{p_t^2} \rangle$.
%It measures the competition between in-plane and out-of-plane emissions. 
 The sign of $V_2$ indicates the azimuthal anisotropy of the emission:
on the reaction
plane ($V_2>0$) or out-of-plane ($squeeze-out,~V_2<0$).

The flow mechanisms, and their dependence on the nuclear EOS, can be 
qualitatitely explained as it follows:   
After the initial compression phase, 
the spectator nucleons (i.e. the nucleons which remain outside the overlap 
area) initially  block  the  escape  of  the compressed region along trajectories in the reaction plane, thus forcing the matter to flow in  directions  perpendicular to the reaction plane. At a later stage,  after  these  spectator  nucleons have moved away, the particles from the squeezed central region preferentially escape along trajectories on the reaction plane, which are no longer 
blocked.   The in-plane emission becomes dominant at higher incident energies (about 5 GeV/nucleon), because the spectator nucleons move quickly away. 
Thus, along the reaction path, nucleon emission first occurs out of plane, 
then spreads into all directions in the transverse plane and 
finally it favors the reaction plane. 
 It is clear that this evolution mainly reflects the interplay between 
the  time  scale  associated with the blockage induced by the spectator matter and the time scale related to the pressure and corresponding flow of the compressed matter, the latter being directly connected to the EOS. 
In fact, a more  repulsive  mean  field potential  leads  to higher pressures and to a quicker expansion flow, with the spectator matter still present around. 
On the other hand, a softer mean field generates a  slower expansion and preferential  emission  on  the  reaction  plane,  after  the spectators have passed.
  The  sideways  deflection  or  transverse flow  of  the  spectator  
fragments  (the $v_1$ observable) 
%occurs  primarily while 
%the spectator fragments are adjacent to the compressed region.
is mainly generated by the pressure exerted from the compressed region on 
the spectator matter and it is positive at relativistic energies. 
It acquires negative values for reactions at beam energies below the
so-called balance energy (around 50 MeV/nucleon), because of the dominance of attractive
mean-field effects.

On the other hand, the elliptic flow reflects
the blockage of the spectators on the participant zone.
%In general, larger  deflections  are  expected  for  more  repulsive  mean  fields,  which  generate  larger pressures.  
%and  conversely,  smaller  deflections are expected for less repulsive ones.

\begin{figure}
\centering
\includegraphics{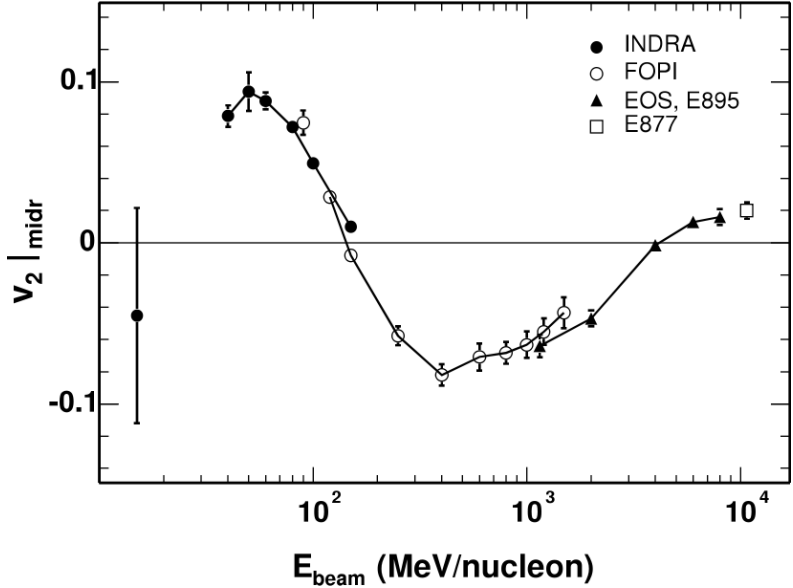}
\caption{\label{fig_uni}
Elliptic flow parameter $v_2$ at mid-rapidity 
for $^{197}$Au+$^{197}$Au collisions at intermediate impact
parameters (b = 5.5-7.5 fm), as a function of incident energy. 
The filled and open circles represent
the INDRA and FOPI data for Z = 1 particles, the triangles represent the EOS and
E895 data for protons, and the squares represent the E877 data for all charged particles. Figure reprinted
from \cite{Andronic2006}, with kind permission of the European Physical
Journal (EPJ), 
where the references to the data are also given.
%Andronic, A.; Lukasik, J.; Reisdorf,W.; Trautmann,W. Systematics of stopping and flow in Au + Au collisions.
%Eur. Phys. J. A 2006, 30, 31–46.
}
\end{figure}

A systematics of elliptic flow measurements,  
for Au + Au collisions at intermediate impact parameters, 
is shown in Fig.\ref{fig_uni}, as a function of the beam energy.
It is interesting to see how the elliptic flow evolves from 
positive values (observed at beam energies up to 100 MeV/nucleon), indicating that
particle emission occurs mainly in plane because of the moderate
pression effects, to negative values (up
to energies of the order of 1 GeV/nucleon), corresponding to out-of-plane 
emission.  As expected on the basis of the arguments presented above, 
the elliptic flow becomes
positive again at high energy (beyond 1 GeV/nucleon) \cite{Andronic2006,Andronic2005,Pinke1999,Braun1998} 
 
In the beam energy interval under consideration, the flow  observables  
reflect the behavior of the  EOS  at  the central densities, in the range of 
2 to 5  $\rho_0$ predicted by transport model simulations.  
From a thourough analysis of transverse and elliptic flows,
based on the comparison between model predictions and experimental data, 
robust constraints on the nuclear matter compressibility
have been derived \cite{Dan2002}, pointing to a value between 200 and 300 MeV. 
%%%%%%%%%%%%%%%%%%% preso da Hermann --- Universe
%\cite{DanielNPA673}.

From the discussion above, it is clear that 
reactions at relativistic energies 
(0.1-1 GeV/nucleon ) allow one to generate high momentum particles and 
to probe regions
of high baryon and (in the case of neutron-rich systems) isovector density during the
reaction dynamics.
In the theoretical description of these collisions,
%it is important to consider 
the momentum dependence of the nuclear effective interactions, leading
to isoscalar effective mass
and neutron/proton effective mass splitting, is rather important.  
%If also the isovector
%component of the interaction is momentum dependent, one observes different
%effective masses (i.e. effective mass splitting) for neutrons and protons.  
The problem of the precise determination of the momentum dependence in the isovector
channel of the nuclear interaction \cite{BaoNPA735,rizzoPRC72}
is still controversial and it would be extremely
important to get more definite experimental information, 
 looking at observables which may also be sensitive to the mass splitting,
see Ref.\cite{LiPPNP2018} for a recent review.
%To this purpose it is interesting to investigate observables related to
%particles with high momenta, formed in high density regions, where
%n/p mass splitting becomes relevant.

%We discuss here some preliminary results for reactions induced by $^{132}Sn$
% beams on $^{124}Sn$ targets at $400AMeV$ \cite{vale08}. 

%\section{EOS at supra-saturation density: collective flows} 
\subsection{Collective flows and isospin effects}

%However, in order to conclude on the properties of the effective
%interaction (asystiffness and MD), it would be desirable to combine 
%the analysis of several obervables. 
%For instance, in the considered beam energy range, 
%the N/Z content of the particle emission  looks 
%particularly sensitive just to the sign of neutron/proton mass splitting, rathe%r than to the asystiffness
%\cite{vale08}. 
%Recent experimental analyses look very promising in this direction  \cite{Russotto2016}.

Several investigations, based on both BUU-like and QMD-like models,
have focused on particle emission and collective dynamics
characterizing heavy ion reactions at incident energy around 
0.5 GeV/nucleon.  One of the goals of these analyses was to constrain the
symmetry energy behavior at supra-saturation densities.  
In particular, the difference between neutron and proton flows, that is mainly ruled
by the balance between the Coulomb repulsion (acting for protons) and the symmetry
potential (repulsive for neutrons in neutron-rich systems),  
appears as a suitable observable to 
%investigate the supra-saturation behavior of 
%the symmetry energy, as well as the effective mass
%behavior 
probe the isovector channel of the nuclear effective interaction
\cite{Li2002,Paolo,Coz,vale08}.

Transport codes are usually implemented with 
%a $BGBD-like$ \cite{GalePRC41,BombaciNPA583}   mean field 
%with a 
different $(n,p)$ density and momentum dependent interactions, see 
for instance \cite{BaoNPA735,rizzoPRC72}. 
This  allows one to probe different symmetry energy parametrizations and 
also to follow the dynamical
effect of opposite n/p effective mass ($m^*$) splitting while keeping the
same symmetry energy behavior 
\cite{Rizzo2008_1,ZhangPLB2014}.

%We discuss here some preliminary results for reactions induced by $^{132}Sn$
% beams on $^{124}Sn$ targets at $400AMeV$ \cite{vale08}. 
As an example, we discuss below some results obtained for
%Let us consider 
semicentral ($b/b_{max}$=0.5) collisions of
 $^{197}$Au+$^{197}$Au at 400 MeV/nucleon.
%For central collisions 
Transport models predict that 
in the interacting zone baryon densities around
$1.7-1.8 \rho_0$ can be reached in a transient time of the order of 15-20 fm/c. 
The system 
is quickly expanding and the freeze-out time is around 50 fm/c.
A rather abundant particle emission is observed 
over this time scale \cite{Li2002,vale08}. 
%A detailed study of neutron and proton elliptic flow appear rather
%useful to probe isospin aspects of the reaction dynamics. 
%In fig.6 we present the  $(n/p)$ and $^3H/^3He$ yield ratios at freeze-out,
%for two choices of Asy-stiffness and mass splitting, vs. transverse
%momentum (upper curves) and kinetic energy (lower curves). In this way we can
%separate particle emissions from sources at different densities.
%We note two interesting features: i) the curves are crossing at
%$p_t \simeq p_{projectile}= 2.13 fm^{-1}$; ii) the effect is not much dependent
%on the stiffness of the symmetry term. The crossing nicely corresponds
%to a source at baryon density $\rho \simeq 1.6 \rho_0$, 
% \cite{ditoroAIP05,rizzoPRC72}. These data seem to be suitable
%to disentangle $Iso-MD$ effects.
%Now we
%present some qualitative features of the dynamics in heavy ion
%collisions in higher energy regions, of large interest for the $RIA$ 
%facility, related to the splitting of nucleon 
%effective masses.
\begin{figure}[h]
\vskip0.3cm
\centering
\resizebox{0.7\textwidth}{!}{%
\includegraphics{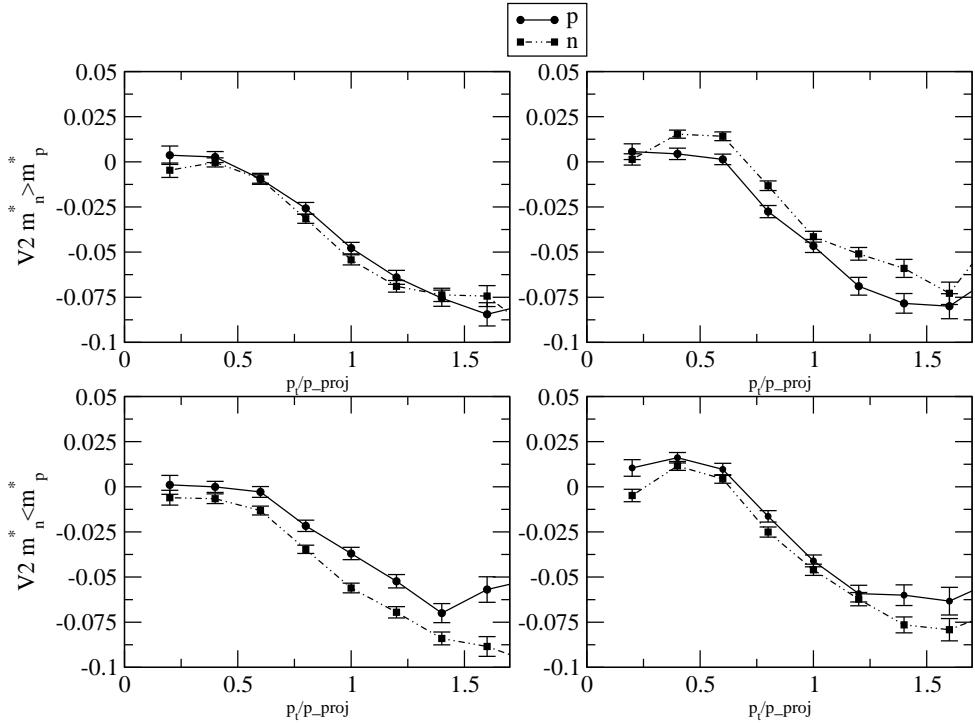} } 
%\begin{minipage}[b]{14pc}
\caption{\label{v2ypt400}
%Transverse momentum dependence of the difference between proton 
%and neutron $V_2$ flows, at mid-rapidity, in a 
%semi-central
%reaction Au+Au at 400 $AMeV$. Taken from Refs.\cite{vale08,Erice08}.
Proton (thick) and neutron (thin) $V_2$ flows in a 
semi-central
reaction Au+Au at 400 $AMeV$.
%Top Panels:
%Rapidity dependence. 
%Bottom Panels:
Transverse momentum dependence 
for particles emitted at mid-rapidity, $\mid y_0 \mid <0.3$.
Upper curves for $m_n^*>m_p^*$, lower curves for
the opposite splitting $m_n^*<m_p^*$. Left: asystiff. Right: 
asysoft. Taken from \cite{vale08}. }
%\end{minipage}
\end{figure}  %\begin{figure}
Figure \ref{v2ypt400}, %it is clear how at, mid-rapidity, 
represents the elliptic flow of emitted neutrons and protons, as obtained
in BUU-like calculations \cite{vale08}, for different asy-stiffness
and effective mass splitting choices. 
We are now exploring density regions above normal density, thus we expect a larger
neutron repulsion in the asystiff case, corresponding to the larger symmetry energy value (see figure 3). 
Indeed Fig.\ref{v2ypt400} shows that 
a larger (negative) neutron $v_2$, close or even larger than
the proton $v_2$, is obtained in the asystiff case
(compare left and right panels). 
Moreover, one can see that the  $m^*_n < m^*_p$ case favors the neutron repulsion,
leading to a larger squeeze-out for neutrons (compare top and bottom panels). 
In particular, in the asysoft case (right panels) we observe an inversion of the 
neutron/proton squeeze-out trend when using the $m^*_n > m^*_p$
parametrization or considering the  $m^*_n < m^*_p$  splitting.  
%at mid-rapidity for the two effective mass-splittings. 

We conclude that a rather interesting interplay exists between the effects
linked to symmetry energy and effective mass splitting: 
a larger (smaller) neutron effective mass may compensate the larger (smaller) neutron repulsion
corresponding to the asystiff (asysoft) case.  
In fact, the $m^*_n < m^*_p$ case,
with the asysoft EOS, yields very similar results of the $m^*_p < m^*_n$ case with the 
asystiff EOS.

This interplay between the effects of several ingredients of the nuclear
effective interaction needs to be further 
investigated \cite{LiPPNP2018}. 
The $m^*_p < m^*_n$ option
seems supported by optical model analyses \cite{bao_PLB2015}.

\begin{figure}[h]
\vskip0.2cm
\centering
\includegraphics[scale=0.6]{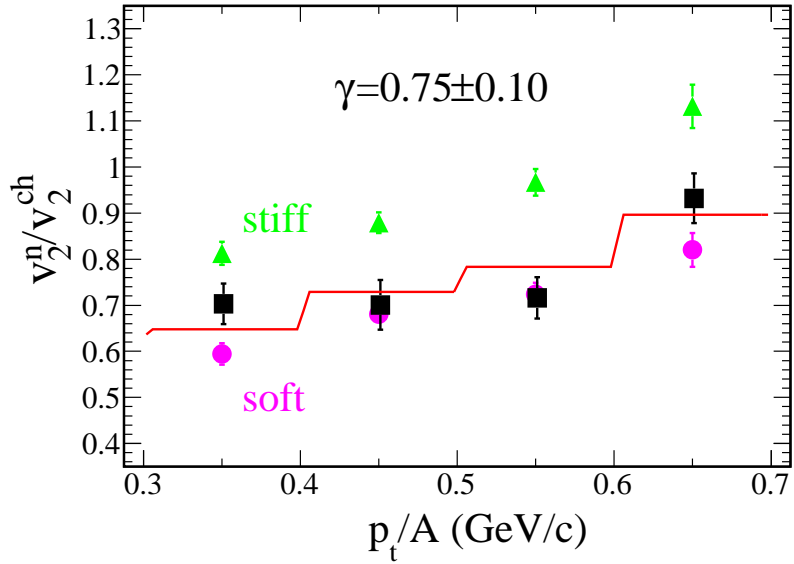}
\caption{\label{fig_paolo1}
(Color online) Elliptic flow ratio of neutrons over all charged particles for central ($b < 7.5$ fm) collisions of $^{197}$Au+$^{197}$Au at 400 MeV/nucleon, as a function of the transverse
momentum per nucleon $p_t$/A. 
%evaluated with a fraction
%of $80\%$ for the second step of timing corrections (see
%Sec. IVA).
The black squares represent the experimental
data, the green triangles and purple circles represent the
UrQMD predictions for stiff ($\gamma$ = 1.5) and soft ($\gamma$ = 0.5)
power-law exponents of the potential symmetry term, respectively. The
solid line is the result of a linear interpolation between the
predictions, weighted according to the experimental errors, 
%of the included four bins in $p_t$/A, 
and leading to the indicated
$\gamma$ = 0.75 $\pm$ 0.10.
Taken from \cite{Russotto2016}.}
\end{figure}

Reaction dynamics at relativistic energies and collective flows 
have been analyzed also in 
QMD-like models 
(TuQMD and UrQMD codes, \cite{Paolo,Coz,Cozma2013}, 
employing different parametrizations of the symmetry energy (asysoft or 
asystiff, identified by the exponent $\gamma$), and a fixed effective mass splitting option (of the   
$m^*_p \leq m^*_n$ type).
The calculated ratio of neutron and proton flows is displayed in 
 Fig.\ref{fig_paolo1}, where also 
%the FOPI-LAND and
the ASY-EOS experimental data  are plotted \cite{Paolo}
(see \cite{Russotto2016} and references therein). 
%The comparison of transport calculations (QMD-like codes \cite{Paolo,Coz,Cozma2013}) to FOPI-LAND and
%ASY-EOS data \cite{Russotto2016} has recently allowed to extract new constraints on the symmetry
%energy slope: L = 72 $\pm$ 13 MeV.  

\begin{figure}[h]
\centering
\includegraphics{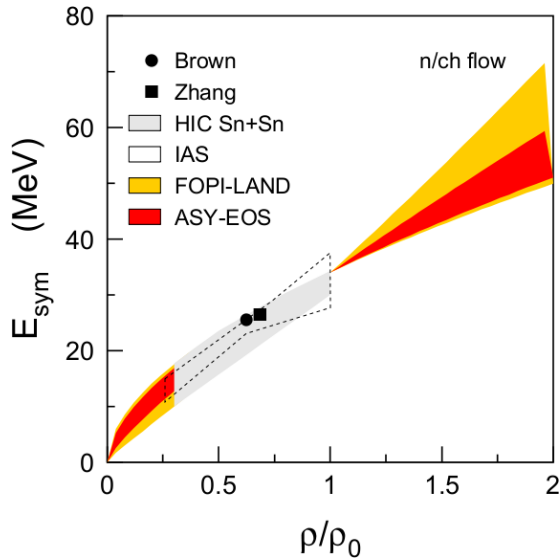}
\caption{\label{fig_paolo2}
(Color online) Constraints deduced for the density
dependence of the symmetry energy from the ASY-EOS data \cite{Russotto2016} in
comparison with the FOPI-LAND result of Ref.\cite{Paolo}, as a function
of the reduced density $\rho/\rho_0$. The low-density results of
Refs.\cite{betty2009,DanLee2014,Brown2013,ZZhang2013} are given by the symbols,
the grey area (Heavy Ion Collisions), and the dashed contour (IAS). For clarity,
the FOPI-LAND and ASY-EOS results are not displayed
in the interval $0.3 < \rho/\rho_0 < 1.0$.
Taken from \cite{Russotto2016}.
}
\end{figure}

The latter analysis has recently allowed to extract new constraints, 
emerging from high density dynamics, 
on the symmetry
energy slope: L = 72 $\pm$ 13 MeV.  
A global picture of the symmetry energy behavior, 
obtained combining the results discussed above and the 
low-density constraints deriving from reaction mechanisms at
Fermi energies and structure effects
 is given in 
Fig.\ref{fig_paolo2}. In particular, the low-density behavior represented
in the figure is evinced from the isospin diffusion analysis
discussed in Section \ref{equi} \cite{betty2009} and from constraints associated with the energy of the Isabaric Analog State (IAS) in several nuclei \cite{DanLee2014}.   
The black points correspond to constraints
coming from structure properties (ground state features of
doubly magic nuclei \cite{Brown2013} 
and neutron skin thickness of heavy nuclei \cite{ZZhang2013}). 

%%%%%%%%%% FOPI-Land [5]  P. Russottoet al., Phys. Lett.B 697, 471 (2011).
New analyses, aimed at probing symmetry energy and effective mass splitting
at once, could possibly couple  
%Possibly, coupling 
the flow information to the study of 
other observables in heavy ion reactions. 
%more definite conclusions could be reached. For instance,  
For instance, the N/Z content of particle emission  looks 
more sensitive just to the sign of the effective mass splitting, rather than to the asy-stiffness
\cite{vale08}, as also observed for the energetic particles emitted in 
reactions at Fermi energies \cite{rizzoPRC72,ZhangPLB2014}, see
Section \ref{pre-iso}.  
It would be interesting the combine observables related to particle flows and yields. 
Recent experimental analyses look very promising in this direction  \cite{Russotto2016}.
%seem to be suitable
%to disentangle $Iso-MD$ potentials \cite{Paolo,Coz}.
%The mass-splitting effect is large at high $p_t$ (Bottom Panels), again in a 
%mid-rapidity selection, 
%as expected for particle emitted from higher density regions.
%The results are also slightly depending on the Asy-stiffness,
% with large neutron $squeeze-out$ effects in the Asystiff case.
Owing to the difficulties in
measuring neutrons, one could also consider the difference between
light isobar (like $^3$H vs. $^3$He) flows and yields. 
We still expect to observe symmetry energy and effective mass splitting effects \cite{vale08}. It should also be noticed that,
as already discussed for reactions at Fermi energies, the treatment of 
cluster correlations could have a significant impact on the nucleon and
light particle observables discussed here. %flow and yield observables. 

\subsection{Meson production}
A quite interesting observable to probe both reaction dynamics and isospin
effects is meson production, arising from inelastic n-n collisions.
In particular, processes related to n-n scattering strongly depend on the energy and density
conditions reached during the initial compression phase and, as such, 
bear precious information on the nuclear EOS.  
 Indeed, kaon production has been one of the most useful observables to determine the EOS of symmetric nuclear matter \cite{Fuchs}. 
%As discussed in Ref.\cite{Xiao2009}, on the basis of BUU calculations,  
Turning to isospin effects, 
isotopic meson ratios, such as the pion $\pi^- / \pi^+$ ratio, are expected to
reflect the N/Z content of the high density regions explored along the
reaction dynamics, which is connected to the symmetry energy behavior 
\cite{Li2002,Xiao2009}.
%The latter in turn depends on the symmetry
%energy trend (being larger in the asysoft case). 
%Indeed, 
A simple explanation of this relation is as it follows: 
The isovector channel of the nuclear interaction is responsible
for the different forces felt by neutrons and protons in the medium. 
This difference has an impact on the N/Z content of the nucleons that escape
from the system and, consequently, of the matter which remains kept inside the
interacting nuclear system and may increment particle production 
through inelastic n-n scattering. 
% Thus  the ratios of the newly produced particles 
%can  serve  as  indicators  of  the  symmetry  energy  in  the  high-density  phase. 
For instance, 
pions  are  produced  predominantly  via  the $\Delta$ resonances $NN
\rightarrow N\Delta$ and  the  subsequent  decay
$\Delta\rightarrow N\pi$.  The ratio of the isospin partners $\pi^-/\pi^+$
can thus serve as a probe of the high-density symmetry energy.

\begin{figure}[h]
\centering
\includegraphics{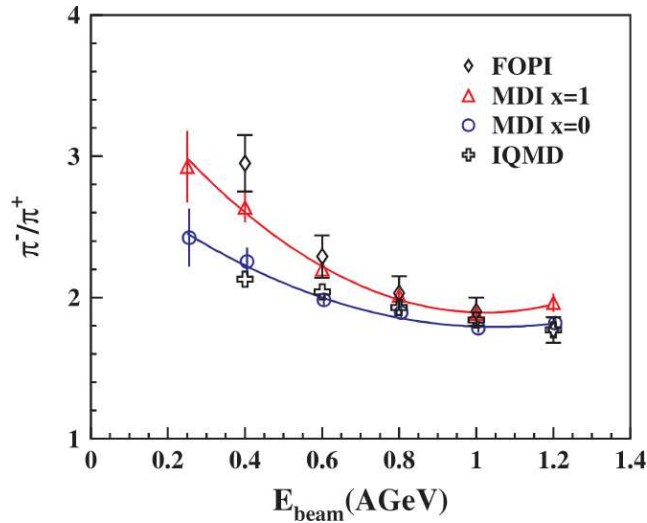}
\caption{\label{fig_xi} (Color online)
Excitation function of the $\pi$ ratio
in the most central Au + Au collisions.
A soft symmetry energy corresponds to the parameter $x$ = 1 (red line), whereas
$x$ = 0 denotes a stiff parametrization (blue line).  
Reprinted from \cite{Xiao2009}, with kind permission of the APS.
}
\end{figure}

An illustration of this effect is given in Fig.\ref{fig_xi}, 
which represents the results of BUU calculations 
for $^{197}$Au + $^{197}$Au collisions, see Ref.\cite{Xiao2009}.
One 
observes that a larger  $\pi^- / \pi^+$ ratio is obtained for 
the soft parametrization of the symmetry energy.  In this case, owing to
the lower neutron repulsion at high density, see Fig.3, the interacting nuclear
system remains neutron-richer, thus favoring $\pi^-$ production.  
More recent calculations, performed with Relativist BUU (RBUU) or QMD codes,  
have shown that emission threshold effects, which act differently
for  $\pi^+$  and $\pi^-$, may reverse the trend of the pion ratio, with respect 
to the effects discussed just above %linked to the neutron richness of the high density regions 
\cite{Ferini2006,Cozma2016,Ko2015}.  
%%%%%%%%%%%%%%%%%%%%%%% dal JPG
This can be illustrated within the relativistic framework introduced
in Section \ref{RBUU}. 
The ``threshold effect'' stems from the fact that nucleon and 
$\Delta$ masses are
modified in the medium. The unknown self-energies of the $\Delta$’s are usually specified
in terms of the neutron and proton ones by the use of Clebsch-Gordan coefficients for
the isospin coupling of the $\Delta$′s to nucleons \cite{BaoPRC2003,Ferini2005}.
These in-medium modifications
are isospin dependent and they influence the phase-space available for meson
production in a n-n collision, 
because the difference between
the invariant energy in the entrance channel $s_{in}$ and the production threshold
$s_{th}$ changes.
This effect is, of course, present in general for all meson production, but, just for the sake of illustration, we consider here one specific inelastic channel:  $nn\rightarrow p\Delta^-$, mainly responsible
for $\pi^-$ production. According to Eqs.(\ref{psms}), 
the invariant energy in the entrance channel and the
threshold energy are given, respectively, by
\begin{equation}
\sqrt{s_{in}}/2 = [E_n^* + \Sigma^0] \rightarrow_{p\rightarrow 0} [m^*_N + \Sigma^0_\omega
+ \Sigma^0_\rho + \Sigma_\delta] > m^*_N + \Sigma^0_\omega  
\end{equation}
\begin{equation}
\sqrt{s_{th}} =  [m^*_p + m^*_{\Delta^-} +  \Sigma^0_p
+ \Sigma^0(\Delta^-)] =   [m^*_N + m^*_{\Delta} +  2\Sigma^0_\omega]
\end{equation}
where $m∗_N$ , $m∗_\Delta$ are the isospin averaged values for nucleon
and $\Delta$ masses.  
The last equality for the threshold
energy derives from the prescription for the $\Delta$ self-energies, which leads to an exact
compensation of the isospin-dependent parts, thus the threshold $s_{th}$ is not modified by the isospin dependent 
self-energies. 
In general, in a self-consistent many-body calculation
higher order effects can destroy this exact balance, 
%But this is not so important for our
%qualitative discussion, 
however compensating effects are expected anyway in $s_{th}$. On the
other hand, the energy available in the entrance channel, $s_{in}$, is shifted in an explicitely
isospin dependent way by the in-medium self-energy $\Sigma^0_\rho + \Sigma_\delta > 0$. In particular, the vector
self-energy gives a positive contribution to neutrons that increases the difference $s_{in} - s_{th}$, thus
enlarging the cross section of the inelastic process, owing to the opening up of
the phase-space.  This happens especially close to the production threshold, since the intermediate $\Delta$ resonance will
be better probed.
A similar modification but opposite in sign is present in $s_{in} - s_{th}$ 
for the  $pp \rightarrow n\Delta^{++}$
channel that therefore is suppressed by the isospin effect on the self-energies. In conclusion,
owing just to threshold effects, the ratio $\pi^-/\pi^+$ is predicted to 
increase with the stiffness of the symmetry energy $C_{sym}$; 
indeed a stiffer trend leads to 
a larger  ($\Sigma^0_\rho + \Sigma_\delta$) self energy.
%%%%%%%%%%%%%%%%%%%%%%%  mettere un commento su come si comporta l'energia
%%%%%%%%%%%%%%%%%%%%%% di simmetria negli approcci relativistici.  

RBUU calculations contain isospin contributions also in the mean field propagation, 
but the results look dominated by the threshold effect discussed above, in particular at lower energies \cite{Ferini2006}.

\begin{figure}[h]
%\vskip -2.cm
\centering
%\begin{tabular}%{c}
\includegraphics*[scale=1.2]{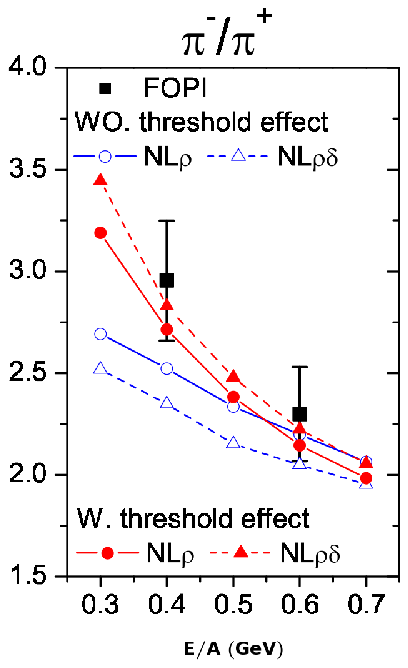}
\includegraphics*[scale=1.2]{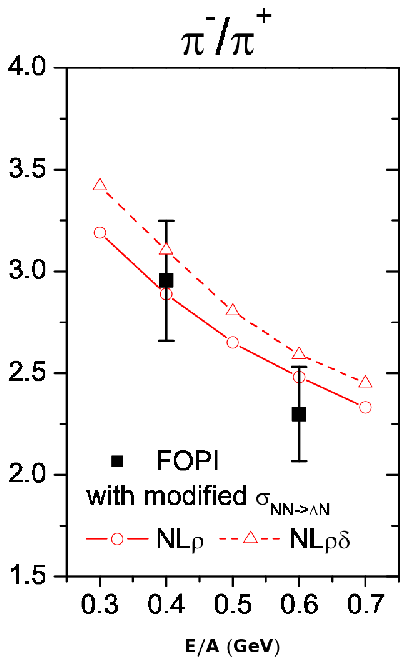}
%\end{tabular}
\caption{Left panel: $\pi^−/\pi^+$+ ratio 
as a function of the collision energy with
and without the threshold effect in Au+Au collisions at impact 
parameter of 1.4 fm from the $NL\rho$ and $NL\rho-\delta$  models.
Experimental data are from the FOPI Collaboration \cite{FOPI}.
Rigt panel: $\pi^−/\pi^+$ ratio as a function of the collision energy
obtained with the threshold effect and the density-dependent
 production cross section in Au+Au collisions at impact
parameter of 1 fm for both the $NL\rho$ and $NL\rho-\delta$  models. 
Adapted from \cite{Ko2015}, with kind permission of the APS.
}
\label{che}
\end{figure}

\begin{figure}[h]
\centering
\includegraphics[scale=1.2]{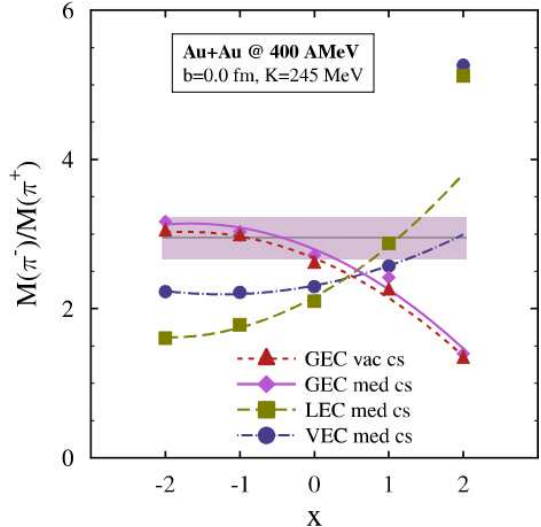}
\caption{\label{fig_coz}
The $\pi^-/\pi^+$  ratio
for the vacuum 
(“vac cs”) and in-medium (“med cs”)
cross-sections scenarios for the case of global energy conservation (GEC) as a function of the asyEOS stiffness parameter $x$.
 Results for the LEC and VEC approximations making use of in-medium modified cross-sections (``med cs'') are also displayed.
 The experimental FOPI results \cite{FOPI} are represented 
by horizontal bands. Taken from \cite{Cozma2016}.}
\end{figure}

Threshold effects on pion production are illustrated in Fig.\ref{che}, 
which reports the results of RBUU calculations employing $NL\rho$ and 
$NL\rho-\delta$ models \cite{Ko2015}. 
Calculations are compared to the FOPI data  \cite{FOPI}. 
On the left panel, one can observe that
neglecting threshold effects the pion ratio is larger in the softer 
$NL\rho$ case, in agreement with the BUU calculations discussed above
\cite{Xiao2009}. On the other hand,  
the trend is reversed when threshold effects are turned on. Indeed in the
latter case a larger pion ratio is obtained for the $NL\rho-\delta$ parametrization,
which corresponds to a stiffer symmetry energy \cite{Ferini2006,Ko2015}. 
The right panel shows calculations that include threshold effects and 
also a density-dependent production cross section.  In this case a
good reproduction of the experimental data can be obtained not only for the
pion ratio (shown on the figure), but also for the separate  
$\pi^-$ and $\pi^+$ yields \cite{Ko2015}.
%%%%%%%%%%%%%%%%%%%%%%%%%%%%%% referenza \cite{FOPI}
A further illustration of the importance of effects related to the
threshold mechanism, namely global
energy conservation (GEC) in inelastic n-n collisions, is provided below, 
%in Fig.\ref{fig_coz}, also 
in the context of QMD-like 
calculations \cite{Cozma2016}.
 Fig.\ref{fig_coz} reports the pion ratio, obtained in central Au + Au collisions at 400 MeV/nucleon, as a function of the parameter $x$ which identifies the
stiffness of the symmetry energy parametrization employed, 
ranging from negative (stiffer) to positive (softer) values.  
Comparing the GEC results to the other options, one can nicely see that enforcing energy conservation leads to 
a reversal of the trend of the pion ratio with respect to the
symmetry energy stiffness.
The figure also shows the effects induced by in-medium modification
of the production cross section.    

\begin{figure}[h]
\centering
\includegraphics[scale=1.2]{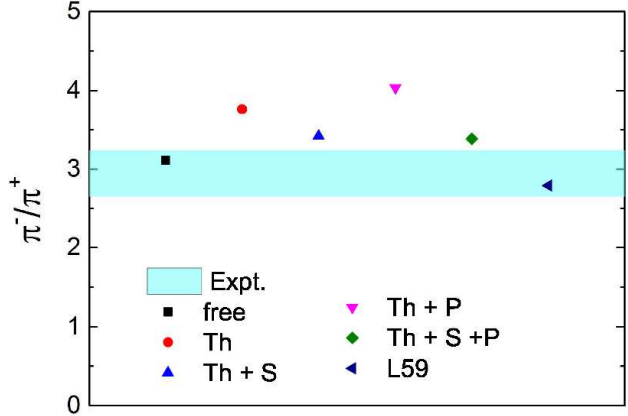}
\caption{\label{che2}
(Color online) The $\pi^-/\pi^+$  ratio in Au + Au collisions
at impact parameter of 1.4 fm and energy of E/A = 400 MeV
from the $NL\rho$ model in different cases (see text for details). 
Experimental data \cite{FOPI} from the FOPI collaboration
are shown as the cyan band.
Reprinted from \cite{Zhang2017}, with kind permission of the APS.}
\end{figure}

As far as threshold effects are concerned, some other physical aspects 
should be considered.
One is the fact that the self-energies usually employed in 
relativistic calculations are not explicitly
momentum dependent, as studies of the optical potential would require. Thus
we come back to the importance of the isospin momentum dependence 
already discussed for collective flows (see previous subsection). 
Because we neglect the dependence of the self-energies on momentum, the difference
$s_{in} - s_{th}$ might increase too strongly with the energy, i.e. the temperature of 
the composite nuclear system. 
%The problem is directly
%related to the strong energy dependence of the optical potential in RMFT. 
It is likely that a more realistic calculation, 
including such a momentum dependence, 
would reduce the threshold
effects discussed above. 
In addition, a fully consistent treatment should include an optical
potential also for the pions. 
More generally, open problems in pion production include the treatment of  
%possible effects associated with 
the width of the $\Delta$ resonance (off-shell effects),
the density dependence of the $\Delta$ production cross section
in the medium 
%(see the
%right panel of Fig.\ref{che}, in this case also the total pion yield
%is reproduced) 
and the pion potentials in the nuclear 
medium \cite{Hong2014,Guo2015,Feng2017,Zhang2017,Cozma2017}. 
%Another effect that is certainly present is the modification of the pion spectral
%function which in an asymmetric medium becomes isospin dependent. This has been
%recently pointed out in Ref.[142] using a pion interaction in the nuclear medium given
%by chiral perturbation theory. 
The latter effect, i.e. the inclusion of the pion potential,  seems to go in the direction of 
smaller pion ratios. 
%Therefore it would be important to consider such pion in medium interaction in the
%transport models.
%, even if it means to go beyond the simple quasi-particle approximation [143].  
%Cassing W, 2009Eur. Phys. J.ST168 3
%%%%% check !!!!
A global representation of threshold and pion potental effects is given 
in Fig.\ref{che2}, which display RBUU results  
\cite{Zhang2017}.  
Adopting the $NL\rho$ interaction (corresponding to a 
symmetry energy slope L = 83 MeV), one observes that the charged pion ratio is increased by threshold (Th) effects, reduced by s-wave pion potential (S), increased by p-wave pion potential (P), finally acquiring a slightly larger value (green point, Th + S + P) than obtained without
any in-medium effects. 
A good reproduction of the FOPI data would require to employ a parametrization
with a smaller 
symmetry energy slope parameter
$L\approx 60$ MeV. 
We notice that the latter would be comparable to the constrained values extracted from nuclear structure and reactions studies discussed above. 
%%%%%%%%%%%%%%%%%%%%%%%%% citare il code comparison

\begin{figure}[h]
\centering
\includegraphics[scale=1.2]{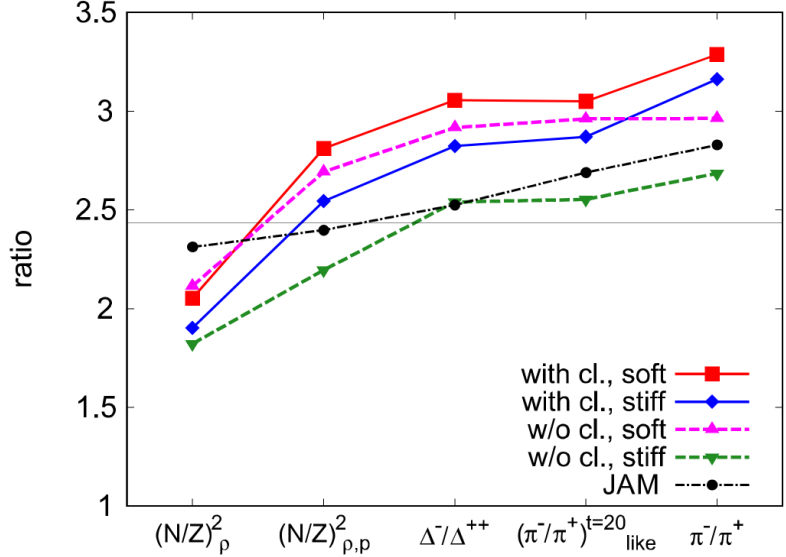}
\caption{\label{fig_Ik}
The nucleon ratios $(N/Z)^2_\rho$
and $(N/Z)^2_{\rho,p}$
%[Eq. (25)] 
in
the high-density region with and without high-momentum condition,
respectively, the $\Delta^-$/$\Delta^{++}$
production ratio, 
%[Eq. (26)], 
the pion-like
ratio at t = 20 fm/c, and the final $\pi^-/\pi^+$
ratio (see Ref.\cite{Ikeno2016} for more details). Each line connects
the ratios for each of the five cases of calculation for central collisions
of $^{132}$Sn+$^{124}$Sn at 300 MeV/nucleon. The horizontal line represents
the $(N/Z)^2_{sys}$
ratio of the total system. The statistical uncertainties in
the final $\pi^-/\pi^+$ ratio are smaller than 0.03.
Reprinted from \cite{Ikeno2016}, with kind permission of the APS.}
\end{figure}

Other interesting effects on pion production are related to many-body dynamics. 
Indeed, as already noted, reactions at relativistic energies
are characterized by a huge amount of light cluster emission.  One may
expect an interplay between clustering effects and meson production. 
Recent AMD calculations, including cluster correlations, have shown that light
cluster emission affects the pion ratios \cite{Ikeno2016}. 
This is illustrated in Fig.\ref{fig_Ik}, which reports results
obtained for central 
$^{132}$Sn+$^{124}$Sn collisions at 300 MeV/nucleon.
Several isospin ratios are plotted on the figure.  The different lines
correspond to AMD calculations with or without cluster correlations, 
for a soft or a stiff parametrization of the symmetry energy,
and to a cascade calculation (JAM, black line).  
One can see that, especially in the asystiff case, the inclusion of 
cluster correlations has a significant impact on the final pion ratios.
Larger ratios are obtained, reflecting the larger N/Z 
of the high density phase (see in particular the high-momentum nucleons),
when clustering effects are turned on. This could be ascribed to the 
fact that the produced light clusters are mainly symmetric, thus
the N/Z of the nucleons which experience inelastic collisions 
and may produce pions increase.    
There is presently an intense activity 
in the field of meson production, also triggered by new devoted experiments \cite{Spirit}. 
It can be envisaged that a stronger sensitivity to the features of the
effective interaction and, in particular, to the symmetry energy,  can
be seen looking at pion ratios as a function of the transverse energy.

It has also been suggested that the ratio of the anti-strange kaon isospin partners, $K^0/K^+$, could be a useful observable to probe
the symmetry energy behavior above normal density.
%the symmetry energy. Indeed, kaon production has been one of the most useful observables to determine the EOS of symmetric nuclear matter \cite{Fuchs}.  
Indeed kaons weakly interact with nuclear matter and thus 
could keep a direct signature of the properties of the dense matter in which 
they are produced \cite{Ferini2006}.  

\section{Conclusions and perspectives}
In this article, 
we have reviewed recent results connected to the rich phenomenology
offered by heavy ion collisions, from medium up to relativistic energies.
Transport theories are essential to tackle the description of the collision
dynamics and to connect the features of the nuclear effective interaction
(and nuclear EOS) to the predictions of reaction observables which can be 
directly compared to experimental data.
In particular, our discussion was focused on the subtle interplay between
mean-field dynamics and many-body correlations in determining the reaction
path, and on isospin effects.  We have shown that theoretical investigations
can allow one to scrutinize the complex nuclear many-body dynamics and, at the
same time, to learn about fundamental nuclear matter properties, of great interest also in the astrophysical context.
 
Selected observables and corresponding constraints which have been extracted over the past years were reviewed.
For reactions at Fermi energies, our attention has been focused on the mechanisms
responsible for light particle and fragment emission.   Several isospin transport 
effects, such as ``isospin migration'' and charge equilibration, have been discussed. 
Whereas the charge distribution of the reaction products emerging from central reactions
provides information on the nuclear matter compressibility and on the role of
many-body correlations (like clustering effects), observables connected to
isospin transport give constraints on the low-density behavior of the symmetry energy,
but also probe the reaction dynamics.
At relativistic energies, the collision dynamics is dominated by nucleon and 
light cluster emission, with a persistence of collective effects, such as
transverse flow (for the spectator matter) and elliptic flow.  Meson production, 
arising from inelastic n-n collisions, also sets in. 
In this regime, neutron/proton collective flows and isotopic meson ratios 
can constrain the high density behavior of the symmetry energy. 

In perspective, it would be highly desirable to reduce the model dependence which presently 
affects the evaluation of some observables. 
In particular, the differences between BUU-like 
and QMD-like approaches need to be understood and accounted for. 
A dedicated program, 
aiming at comparing the predictions of different transport models under 
controlled situations, to understand and eliminate possible sources of
discrepancies, is presently running.
Some clues have already emerged, concerning 
the impact of the Pauli principle preservation along the dynamical evolution
\cite{comp2},  the robustness of some reaction observables \cite{comp1} and
the description of the inelastic processes generating pion production
\cite{comp3}. 
%The synergy between theory and experiments is of paramount importance, in
%order 

Also, it would be quite helpful to investigate several transport
observables at the same time, both within the same
code and between different codes and different models.
The global reaction dynamics could be probed looking at suitable isoscalar observables
such as fragment and particle yields and %kinetic 
energy spectra, as well as at the isotopic features of the reaction 
products (and related observables). 
In particular, 
the study of pre-equilibrium emission, 
which is not influenced  by secondary decay effects,
%should be helpful.  The N/Z content of this emission 
could prove particularly useful in this direction.
%More generally, %to constrain several sensitive transport quantities at once, 
%and conclusively determine the value of the symmetry
%energy at subsaturation density, 
More generally, it would be very interesting to explore %and obtain 
a consistent map of many observables in a
multiparameter physics input space, see for instance the recent 
Bayesian analysis 
of Ref.\cite{Morfouace_new}.
%When the general robustness of the model description 
%has been established, one may undertake
%a deeper dedicated investigation of isospin observables, from which more detail%ed information on
%the symmetry energy and its density dependence can be accessed. 

Finally, to improve our understanding of nuclear dynamics and our 
knowledge of the nuclear EOS, the synergy between theory and experiments is of paramount importance. 
From the experimental 
point of view, new detection systems and analyses, which allow to investigate several 
reaction
mechanisms and observables within the same data set, look extremely promising.
%The large effects pointed out in this review suggest that more work is needed to pin down the input physics in transport
%models other than the symmetry energy.  
%The results discussed in this review suggest that
%isospin observables are sensitive enough to
%the collision dynamics that 

%This is the goal of the transport code comparison project, which has 
%already provided some important clues \cite{comp1,comp2,comp3}. 

%Therefore, the simultaneous analysis of several experimental observables in nuclear
%reactions at Fermi energies should help to shed light on the underlying fragmentation
%mechanism and the corresponding role of mean-field and many-body correlations.  In other words, a check of the global reaction dynamics could be a way to test
%the validity of the approximations employed in the dynamical models devised to
%deal with the complex many-body problem. 

%Indeed, presently it appears that, for several observables,  
%the differences between predictions corresponding to different  
%symmetry energy  parametrizations are smaller than the differences associated with different models.    

%%%%%%%%%%%%%%%%%%%%%%%%%%%%%%%%%%%%%%%%%%%%%%%%%%%%%%%%%%%%%%%%%%%%%%%%% 

\section{Acknowledgements}
Useful discussions with P.Napolitani and H.H.Wolter are gratefully 
acknowledged. This project has received funding from the European Union's Horizon 2020 research and innovation programme under grant agreement N. 654002.

\end{document}